\pdfoutput=1
\documentclass[12pt]{article}

\usepackage{twemojis}
\usepackage{graphicx,psfrag,epsf,color}
\usepackage{xcolor}
\definecolor{darkblue}{rgb}{0.0, 0.0, 0.55}
\usepackage{amsmath,amssymb,amsfonts}
\usepackage{array}
\usepackage{marvosym}
\usepackage{linearA}
\usepackage{cite}
\usepackage{rotating}
\usepackage{slashed,mathtools}
\usepackage{xparse}
\usepackage{stmaryrd}
\usepackage{cancel}
\usepackage{centernot}
\usepackage{stackengine}
\usepackage{hyperref}
\usepackage{marvosym}

\usepackage{tikz}
\usetikzlibrary{shapes, arrows.meta, positioning, patterns, decorations.pathmorphing}
\usepackage{tikz-feynman}
\tikzfeynmanset{warn luatex=false}

\newcommand{\be}{\begin{equation}}
\newcommand{\ee}{\end{equation}}
\newcommand{\bea}{\begin{eqnarray}}
\newcommand{\eea}{\end{eqnarray}}
\newcommand{\bi}{\begin{itemize}}
\newcommand{\ei}{\end{itemize}}
\newcommand{\ben}{\begin{enumerate}}
\newcommand{\een}{\end{enumerate}}
\newcommand{\bt}{\begin{tabular}}
\newcommand{\et}{\end{tabular}}

\newcommand{\klammer}[1]{ \left( #1 \right)  }

\newcommand{\nm}{n_-}
\newcommand{\np}{n_+}
\newcommand{\nms}{\slashed{n}_-}
\newcommand{\nps}{\slashed{n}_+}

\newcommand{\as}{\alpha_s}
\newcommand{\msbar}{\overline{\rm MS}}
\newcommand{\eps}{\epsilon}

\newcommand{\PS}{d\mbox{PS}}
\newcommand{\LQCD}{\Lambda_{\rm QCD}}
\newcommand{\enp}[1]{ \left\llbracket #1 \right\rrbracket}
\newcommand{\dotp}{
    \mathop{
        \mathchoice{\vcenter{\hbox{\LARGE$\cdot$}}}
                   {\vcenter{\hbox{\LARGE$\cdot$}}}
                   {\vcenter{\hbox{\normalsize$\cdot$}}}
                   {\vcenter{\hbox{\small$\cdot$}}}
    }
}

\numberwithin{equation}{section}
\allowdisplaybreaks

\newcommand{\rf}{f^{\text{NLP},\circledR}}

\begin{document}

\begin{titlepage}

\begin{flushright}
{\small
TUM-HEP-1600/26\\
Nikhef 2026-009\\
August 25, 2026
}
\end{flushright}

\vskip0.8cm
\begin{center}
{\Large \bf Next-to-leading power endpoint DIS and PDF factorization}
\end{center}

\vspace{0.2cm}
\begin{center}
{\sc M.~Beneke,$^{a}$ M. Schnubel,$^{b}$ and  R.~Szafron$^c$} \\[6mm]
{\it $^a$Physik Department T31,\\
James-Franck-Stra\ss e~1, 
Technische Universit\"at M\"unchen,\\
D--85748 Garching, Germany
\\[0.2cm]
${}^b$Nikhef, Theory Group,\\ 
Science Park 105, 1098 XG, Amsterdam, The Netherlands\\[0.2cm]
 ${}^c$Department of Physics, Brookhaven National Laboratory,\\ 
Upton, N.Y., 11973, U.S.A.\\[0.2cm]
}
\end{center}

\vspace{0.0cm}
\begin{abstract}
\vskip0.2cm\noindent
We study deep-inelastic scattering (DIS) in the threshold region $x\to 1$ beyond leading power in $1-x$ at leading twist. In the framework of  position-space soft-collinear effective theory, we derive a factorization theorem for the off-diagonal parton-scattering channel, which starts at next-to-leading power (NLP). The had\-ronic tensor separates into an anti-hardcollinear contribution from power-sup\-pressed currents and a soft-collinear contribution from subleading Lagrangian interactions, whose convolution integrals exhibit endpoint divergences.
We show that these divergences are resolved by endpoint factorization relations, implying that the parton distribution function (PDF) near threshold becomes a two-scale object at NLP. We derive the factorization of the NLP gluon distribution at large $x$ in terms of hardcollinear matching coefficients and a subtracted soft-collinear function.
We resum the leading logarithms of $1-x$ both from $d$-dimensional factorization in the $\overline{\rm MS}$ scheme and in an endpoint scheme based on refactorization, and establish the all-order relation between the two definitions of the PDF.
\end{abstract}

\end{titlepage}

\pagenumbering{roman}
{\hypersetup{hidelinks}
\pdfbookmark[1]{Contents}{ToC}
\setcounter{tocdepth}{2}
\tableofcontents}
\vspace{6mm}
\newpage
\pagenumbering{arabic}

\section{Introduction}
\label{sec:introduction}

Deep inelastic scattering (DIS) of a colour-neutral particle on a hadron is a   classical process in the study of quantum chromodynamics (QCD), which originally revealed the partonic structure of hadrons and still offers unparalleled insights into the internal structure of hadrons and the fundamental interactions of quarks and gluons. The precise theoretical description of DIS cross sections is also crucial for extracting the parton distribution functions (PDFs), which are indispensable for predictions of hadronic cross sections at high-energy colliders. 

The kinematic regime in DIS where the Bjorken scaling variable $x$ approaches unity $(x \to 1)$ and one parton carries almost all the momentum of the hadron is special. In this limit, the invariant mass of the hadronic final state becomes parametrically small compared to the scale $Q$ of the hard DIS process. This kinematic constraint leads to the dominance of soft and collinear gluon emissions, which manifest
themselves as large logarithms of $(1-x)$ in fixed-order perturbative calculations.   
An all-order resummation is required to ensure reliable theoretical predictions. 
Systematic diagrammatic factorization methods extending the factorization of the DIS process into short-distance coefficients and the PDF were developed to achieve 
such resummations beyond the exponentiation of leading double 
logarithms \cite{Sterman:1986aj,Catani:1989ne,Catani:1990rp}.  Among the 
early treatments of the problem, Ref.~\cite{Korchemsky:1992xv} is particularly 
noteworthy as it connects it to the renormalization of soft Wilson lines, 
which are key elements of the effective field theory approach to resummation near kinematic boundaries developed about 15 years later. In terms of parameters, 
the above large-$x$ resummations of DIS all refer to leading twist in the 
strong-interaction scale $\LQCD$, i.e. leading power in the expansion in 
$\LQCD/Q$, {\em and} to leading power (LP) in the small kinematic variable $1-x$. 

To go beyond the LP in $1-x$, but still at leading twist, which is the subject of this work, the factorization of scales that precedes resummations is most easily handled by soft-collinear effective theory (SCET) \cite{Bauer:2000yr,Bauer:2001yt}, here in the position-space representation~\cite{Beneke:2002ph,Beneke:2002ni}. At leading power in the $(1-x)$ expansion, the DIS short-distance coefficients factorize into a jet function characterising the energetic final-state jet and 
a hard function describing the production of a single parton, which develops into the jet. Both are convoluted with the PDF, which accounts for the non-perturbative structure of the proton. The rederivation of the results of \cite{Sterman:1986aj,Catani:1989ne,Catani:1990rp} within the SCET framework 
\cite{Manohar:2003vb,Pecjak:2005uh,Chay:2005rz,Becher:2006mr,Idilbi:2006dg,Chen:2006vd,Chay:2012jr,Chay:2017bmy} represented one of the first applications of SCET to the factorization of high-energy collisions. 

Many of the mentioned references fall short of a satisfactory presentation of the SCET derivation, instead discussing subtleties such as mode decomposition, frame (in)de\-pen\-dence, the relevance of zero-bin subtraction and the disentanglement of ultraviolet (UV) and infrared (IR) singularities of the functions that appear in the factorization formula, thereby providing further insights into the foundations of the diagrammatic derivations. Ref.~\cite{Becher:2006mr} is the first to clarify that in mode language, the low-energy modes are in fact not soft, but soft-collinear. However, as we emphasize in the present work, their assumed scaling is not the right one. While this does not affect the leading-power factorization theorem nor any of the results of \cite{Becher:2006mr}, the proper scaling and virtuality of soft-collinear fluctuations are crucial at the next-to-leading power. We shall therefore devote some space in Secs.~\ref{sec:SCETbasics} and~\ref{sec:LPfact} to the rederivation of the leading-power factorization theorem that emphasizes this point.
The mode structure that is developed in these sections coincides 
with the one that appeared in the context of the LP discussion of the variable flavour-number scheme for massive-quark jets in 
large-$x$ DIS \cite{Hoang:2015iva}. 

The NLP contributions arise from power-suppressed currents and Lagrangian interactions within the effective theory, which represent novel physical effects such as the final-state jet being sourced by double-collinear parton emission from the hard vertex and the emission of transversely polarized soft gluons. These effects present substantial theoretical challenges for factorization, related to the so-called endpoint singularities in convolution integrals 
\cite{Beneke:2003pa,Moult:2019uhz,Beneke:2019oqx,Liu:2019oav, Hurth:2023paz,Cornella:2022ubo}, which spoil naive factorization among the various modes. This challenge implies that beyond LP the conventional definitions or treatments of PDFs prove insufficient or require reinterpretation to fully capture the multi-scale  dynamics of the PDFs at the kinematic edge. 

Here we focus on the so-called off-diagonal channel in DIS, where a quark is struck in a process primarily mediated by gluons, because it starts at NLP and displays all significant aspects of NLP factorization at its leading non-vanishing order. The NLP contributions in the off-diagonal channel manifest themselves through two distinct physical mechanisms. One arises from the anti-hardcollinear contribution, involving the emission of two energetic partons from the hard vertex into the final-state jet. The other arises from the soft-collinear contribution, where the struck quark is effectively converted into a gluon by emitting a soft-collinear quark into the hadronic remnant. The former is represented in the EFT by a power-suppressed current, while the latter is reproduced by the time-ordered product of the leading-power current and the power-suppressed Lagrangian. See Sec.~\ref{sec:SCETbasics} for a non-technical overview of the two mechanisms, their relation, and interpretation. 

In~\cite{Beneke:2020ibj}, the leading logarithmic (LL) resummation 
of the above off-diagonal DIS channel at NLP has been achieved using consistency conditions in $d=4-2\eps$ dimensions, which require that singularities are absent in physical observables.  For this purpose, it was sufficient to focus  solely on the anti-hardcollinear term. This approach is similar to the exploitation of $d$-dimensional factorization and infrared-finiteness constraints for DIS and the Drell-Yan process in \cite{Vogt:2010cv,Almasy:2010wn}, and it agrees with these results without relying on an extrapolation of finite (but very high) order expressions to all orders. On the other hand, the assumptions and ansatz made in \cite{Almasy:2010wn} provide powerful statements beyond the LL accuracy and even beyond NLP. 

In the present work, we continue the investigation of threshold production in hadron collisions and focus on the complete structure of the NLP factorization theorem for $x\to 1$ DIS in SCET language, which was only sketched in \cite{Beneke:2020ibj}. In Secs.~\ref{sec:B1term} and \ref{sec:A0term} we derive the factorization of the anti-hardcollinear and Lagrangian interaction parts, respectively. From the perspective of the underlying hard process, the off-diagonal partonic $x\to 1$ DIS amplitudes are related by crossing to ``gluon-thrust" in two-jet kinematics and, indeed, there is a similarity in structure to the treatment of this process \cite{Beneke:2022obx}. However, while thrust is infrared-finite, an important aspect of SCET NLP factorization of $x\to 1$ DIS concerns the definition of the PDF in the effective theory. In Sec.~\ref{sec:A0term} we therefore present the derivation of the threshold factorization of the $\overline{\rm MS}$-scheme PDF defined according to Collins-Soper \cite{Collins:1981uw}. This shows explicitly that the PDF near $x=1$ is itself a two-scale object starting from the NLP and includes the soft-collinear emission from the Lagrangian insertion.

The definition of the PDF that emerges naturally from SCET factorization at the kinematic edge is, however, different from the conventional $\overline{\rm MS}$-scheme PDF. The difference is present even at the level of leading logarithms and arises from the different implicit $d$-dimensional treatment of endpoint divergences in the $\overline{\rm MS}$ definition as compared to the four-dimensional refactorization-based approach in the effective field theory. In Sec.~\ref{sec:resum} we resum both PDFs and off-diagonal DIS with LL accuracy, employing SCET renormalization-group equations for the component functions, and further derive the relation between the two PDF definitions, which must itself be resummed. This relation is a universal ingredient of SCET NLP factorization of hadron collisions near the kinematic edge for the colliding partons.
We summarize in Sec.~\ref{sec:conclusion}.


\section{Threshold factorization with SCET}
\label{sec:SCETbasics}

We review threshold factorization in the SCET framework at leading power in $1-x$ in DIS,  
\begin{equation}
\gamma^*(q)+N(p)\to X(p_X)\,,
\end{equation}
of a virtual photon on a nucleon $N$. 
This serves not only to introduce notation but also to derive some results, which carry over to the next-to-leading power. We also clarify several points that have not been addressed in previous works. We then provide an overview of the anticipated structure of threshold factorization for off-diagonal parton scattering at NLP.

\subsection{Kinematics and modes}
\label{sec:kinematicsmodes}

The hadronic tensor for DIS of a virtual photon on a single quark with electric charge $e_q=1$ is defined as
\begin{eqnarray}
W^{\mu\nu}(p,q) &=& \frac{1}{4\pi}
\sum_{X}\int \!\PS_{X}\left(2\pi\right)^{d}\delta^{(d)}\left(p+q-p_{X}\right)\left\langle N(p)|J^{\dagger\nu}(0)|X\right\rangle \left\langle X|J^{\mu}(0)|N(p)\right\rangle \nonumber\\
&=& -g^{\mu\nu} F_1(Q^2,x) + \ldots\,,
\label{eq:hadtensor}
\end{eqnarray}
where the terms not written out can be ignored since, at leading power, we may assume that $\mu$, $\nu$ are transverse to the nucleon 
momentum and momentum transfer
\begin{equation}
p^\mu=\np p\frac{\nm^\mu}{2}, \qquad 
q^\mu = -Q \frac{\nm^\mu}{2}+Q\frac{\np^\mu}{2}\,,
\end{equation}
respectively. We are allowed to neglect the nucleon mass and set $p^2=0$. 
We recall the standard definitions 
$n_\pm^2=0$, $\np\nm=2$, and 
\begin{equation}
Q^2=-q^2,\qquad x=\frac{Q^2}{2 pq}=\frac{Q}{\np p}\,.
\end{equation}
The nucleon is assumed to be unpolarized, and {\em an average over the nucleon spin is implicitly understood} in the above and later expressions.

The distinctive property of the $x\to 1$ limit is that the total invariant mass 
\begin{equation}
p_X^2 = (p+q)^2 = Q^2\frac{1-x}{x}    
\end{equation}
of the final state is parametrically small. The final state forms a jet in the $\np^\mu$ direction, referred to as 
``anti-collinear'' in the following.  Consequently, the light-cone momentum fraction $\np p_{\rm rem}$ of the target remnant, which is part of the final state $X$, must be much smaller than $Q$, as most of the nucleon momentum is carried by the struck quark. This kinematic restriction identifies the relevant on-shell modes for this process: 
\begin{align}
&\mbox{collinear (struck quark)} \hspace*{1cm}  Q (1,\xi,\xi^2)\,,\nonumber \\[0.2cm]
&\mbox{anti-hardcollinear (jet)} \hspace*{1.15cm} Q(\lambda^2,\lambda,1)\,,
\label{eq:modes}\\[0.2cm]
&\mbox{soft-collinear (remnant)} \hspace*{1cm}  Q(\lambda^2,\xi\beta,\xi^2 (\beta/\lambda)^2)\,,
\nonumber
\end{align}
where the momentum scalings are specified as $p=(\np p, p_\perp, \nm p)$, and the two small power-counting parameters are defined as
\begin{equation}
\lambda=\sqrt{1-x},\qquad  \xi=\frac{\Lambda_{\rm QCD}}{Q}\,.  
\end{equation}
The following remarks are now in order:
\begin{itemize}
\item The only requirements for the soft-collinear mode are that 
$\np p_{sc} \sim Q (1-x)\sim Q \lambda^2$ and that $p_{\perp sc}$ and 
$\nm p_{sc}$ vanish as $\xi\to 0$ for fixed $\lambda$. We therefore introduce a scaling parameter $\beta\sim \lambda^a$, where $a \geq 0$ is 
not specified. The soft-collinear scaling assumed in \cite{Becher:2006mr} amounts to $a=1$, and the soft-collinear mode that appeared for the QED electron PDF in \cite{Beneke:2007zg} corresponds to $a=2$.  However, in QCD, any $a>0$ leads to modes with virtuality $\Lambda_{\rm QCD}^2(1-x)^a$, parametrically below $\Lambda_{\rm QCD}^2$, which cannot exist non-perturbatively due to confinement. Hence, the derivation of threshold factorization must also apply to the case $a=0$.
\item One does not need standard hardcollinear modes with scaling $Q(1,\lambda,\lambda^2)$, since kinematics does not allow such modes in the final state $X$, while virtual hardcollinear loops are scaleless. Furthermore, while soft modes can be radiated from the anti-hardcollinear jet, there is no kinematic constraint on the phase space of the soft modes; hence, after squaring the amplitude, the soft loops and phase-space integrals are effectively scaleless. 
However, to cover the case $\beta\sim 1$, a hardcollinear mode with scaling
\begin{equation}
\mbox{hardcollinear (struck quark) \hspace*{1cm}} Q(1,\xi/\lambda,\xi^2/\lambda^2)
\label{eq:hcmode}
\end{equation}
is needed, as will be seen below. The fact that this mode turns out to be irrelevant at LP, but not at NLP, is one of the important differences between factorization at LP and in higher powers.
\item The collinear and soft-collinear modes have non-perturbative virtualities and must be factorized into hadronic matrix elements. The virtuality $\Lambda_{\rm QCD}^2/(1-x)$ of the hardcollinear mode does not depend on the hard scale $Q$, but is parametrically above 
$\Lambda_{\rm QCD}^2$, allowing it to be treated perturbatively.
\end{itemize}
 In this work, we consider DIS at leading twist, that is, leading power in 
$\xi$, but at subleading power ($\lambda^2$) in the threshold variable.

\subsection{Wilson lines}
\label{sec:wilsonlines}

The factorization of soft-collinear gluons introduces semi-infinite soft-collinear Wilson lines of various kinds, which we summarize here:
\begin{eqnarray}
\mbox{incoming particle}&\quad&
Y_{sc,n_-}(x)=\mathcal{P} \exp\left[i g_s T^C\!\int_{-\infty}^0 \!ds \,n_-A^C_{sc}(x+sn_-) \,e^{s 0^+}\right],\quad
\label{eq:wilsondef}\\
\mbox{outgoing particle}&\quad&
\overline{Y}^\dagger_{sc,n_+}(x)=\mathcal{P} \exp\left[i g_s T^C\!\int_0^{\infty}\! ds \,n_+A^C_{sc}(x+sn_+) \,e^{-s 0^+}\right],\qquad
\label{eq:wilsonbardef}
\end{eqnarray}
where $T^C$ denotes the generator in a given representation of SU($N_c$) and $\mathcal{P}$ denotes the path-ordering of the matrix exponential. The relevant cases are the fundamental and the adjoint representation for the decoupling of soft-collinear gluons from an energetic quark and gluon, respectively. For the adjoint, $[T^C_{\rm ad}]^{AB} = (-i) f^{CAB}$, and we denote the 
corresponding Wilson line as $\mathcal{Y}_{sc,\nm}^{AB}$. The factor $e^{\pm s 0^+}$ in the integrand is a remnant of the $i0^+$ prescription of the Feynman propagator from which 
the Wilson line originates.

For incoming / outgoing anti-particles in the Dirac representation, the Wilson lines are the hermitian conjugates of the above, $Y^\dagger_{sc,n_-}(x)$ and $\overline{Y}_{sc,n_+}(x)$, respectively. The conjugation then implies anti-path-ordering of these Wilson lines. Note that the Wilson line for the outgoing particle is not identical to the hermitian conjugate of that of the incoming particle, i.e.~$\overline{Y}_{sc,n}\not= Y_{sc,n}$, and we use the overbar to distinguish the two.\footnote{See also \cite{Beneke:2019slt} and note that, compared to that reference, the notation convention for the barred and unbarred Wilson lines is interchanged.} We further note the identities
\begin{equation}
       Y_n(x) T^A Y_n^\dagger(x) = T^B\,\mathcal{Y}^{BA}_{n}(x), 
       \qquad
       Y^\dagger_{n}(x) T^A Y_{n}(x) = T^B\,[\mathcal{Y}^{-1}]^{BA}_{n} (x)\,,
       \label{eq:wilsonfadrel}
\end{equation}
which also hold under the substitutions $Y_n\to \overline{Y}_n^\dagger$, $\mathcal{Y}_n\to \overline{\mathcal{Y}}_n^\dagger$.\footnote{The adjoint representation is real, hence $\mathcal{Y}^\dagger =\mathcal{Y}^T$. However, we do not make use of this property in the following since some rapidity regulators (for example, the $\delta$-regulators that correspond to a real off-shellness of the emitting particle) render the adjoint Wilson line complex. For the same reason, in the second equation in \eqref{eq:wilsonfadrel}, we did not use the unitarity of the Wilson line to simplify $\mathcal{Y}^{-1}=\mathcal{Y}^\dagger$, since it can also be violated by the regulator. In this case, making use of unitarity or reality (in the case of the adjoint) results in different definitions of the regularized soft function. However, one expects suitably subtracted soft functions 
(see, for example \cite{Beneke:2024cpq}), which have finite limits when the regulator is removed, to be independent of this ambiguity.}
Finally, the finite-distance Wilson line is defined as 
\begin{equation}
\left[t_1 n,t_2 n\right]\equiv {Y}_n(t_1 n)Y_n^\dagger(t_2 n) = \mathcal{P} \exp\left[i g_s T^C \!\int_{t_2}^{t_1}\! ds \,n\cdot A^C(sn)\right].
\label{eq:finitewilsondef}
\end{equation}
The same definition applies to the adjoint representation by substituting the colour generator $(T^C)_{AB} = -if^{CAB}$.

\subsection{Overview}
\label{sec:overview}

\begin{figure}[t]
\begin{center}
\includegraphics[width=\textwidth]{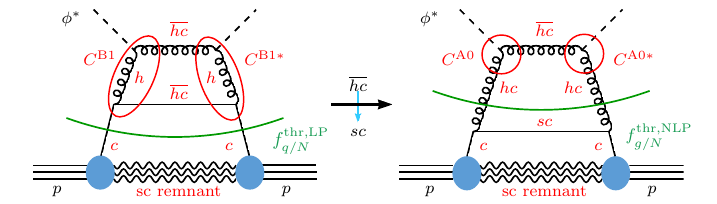}
\end{center}
\vspace*{-0.0cm}
\caption{Sketch of the physics of factorization at next-to-leading power: An anti-hardcollinear quark in the final-state jet moves into the soft-collinear proton remnant as its $\nm\cdot p$ light-cone momentum component and virtuality decrease from $Q$ and $Q^2(1-x)$, respectively, to $Q(1-x)$ and $\Lambda^2_{\rm QCD}$.}
\label{fig:sketch}
\end{figure}

For the LP discussion, we used standard photon-initiated DIS to establish the mode structure and notation. For the NLP off-diagonal channel, however, it is convenient to switch to a scalar probe coupled to $G_{\mu\nu}^A G^{\mu\nu A}$, because this isolates the gluon-initiated hard interaction already at tree level.
Before turning to their formal derivations, we describe the two mechanisms that contribute to $x\to 1$ DIS at NLP in non-technical terms, which also explains the origin of the endpoint divergence. The characteristic feature of DIS as $x\to 1$ is the collimation of the hadronic final state into a jet of invariant mass $Q^2(1-x)$, which is back-to-back with the struck parton in the Breit frame of the proton. In this frame, at leading power, the colour singlet with mass $Q$, assumed to be a virtual Higgs  boson, couples to a gluon pair. Emitting an energetic (anti-hardcollinear) parton into the final state from the hard process costs a factor of $\lambda$. At next-to-leading power ($\lambda^2$ suppression for the square of the amplitude), the hard process for quark scattering of a virtual Higgs boson $\phi^*$ is therefore 
$q+\phi^*\to g+q$, as shown at tree-level in the left diagram of Fig.~\ref{fig:sketch},
where $g$ and $q$ are both energetic and part of the final-state jet. The hard amplitude $C^{B1}(r)$ depends on $r$, which determines the fraction of total large light-cone momentum $\nm\cdot (p_g+p_q)$ carried by the quark. The gluon enclosed by the red ellipse in Fig.~\ref{fig:sketch} is hard, resulting in a local vertex. Contracting it to a point and identifying the part below the green line with the parton distribution function, the process shown on the left is very similar to the leading-power process, and one anticipates that the corresponding hadronic tensor can be factorized into
\begin{equation}
W_\phi^{\overline{hc}} = C^{B1*}(r^\prime)C^{B1}(r) 
\stackrel{\,r,r^\prime}{\otimes}\mathcal{J}^{(qg)}_{\overline{hc}}(\xi,r,r^\prime)\stackrel{\xi}{\otimes}f_{q/N}(\xi)\,,
\label{eq:B1schematic}
\end{equation}
where the convolution variables are indicated above the $\otimes$ symbol. The convolution in $\xi$ is the standard convolution in the large $\np\cdot p\sim Q$ light-cone momentum component of the quark in the proton, described by the PDF $f_{q/N}(\xi)$. The new feature is the convolution in $r,r^\prime$ related to the large opposite light-cone momentum components $\nm\cdot p$ in the final state and a corresponding generalized jet function. The derivation of \eqref{eq:B1schematic} is provided in Sec.~\ref{sec:B1term}.

The second NLP process is depicted in the right diagram in Fig.~\ref{fig:sketch}. In this process, the hard part remains the same as in leading power $g+\phi^*\to g$ scattering, represented by the amplitude $C^{A0}$. The power suppression arises from the splitting $q\to g(hc)+q(sc)$ of the struck quark into a hardcollinear gluon and a soft-collinear quark  through a power-suppressed interaction in the SCET Lagrangian before the gluon enters the hard process. Notice that the struck quark, which itself carries almost all of the  momentum of the proton, must transfer almost all of its momentum to the gluon in order to realize DIS with $x\approx 1$. For this reason, the emitted quark must be {\em soft}-collinear with $\np\cdot p$ light-cone momentum of order $Q(1-x)$, of the same order as the partons in the proton remnant. As indicated in the right figure, this process is contained within the gluon parton distribution since the soft-collinear modes have transverse momentum of order $\Lambda_{\rm QCD}$. With respect to this process, the hadronic tensor factorizes into 
\begin{equation}
W_\phi^{sc} = |C^{A0}|^2\,\mathcal{J}^{(g)}_{\overline{hc}}(\xi)\stackrel{\xi}{\otimes}f^{\rm thr, NLP}_{g/N}(\xi)\,,
\label{eq:A0schematic}
\end{equation}
which has the same form as at LP in the expansion in $(1-x)$, except that the parton distribution refers to the NLP term in {its} expansion, which itself factorizes as 
\begin{equation}
f^{\rm thr, NLP}_{g/N}(\xi) = D^{B1*}(\omega^\prime)D^{B1}(\omega) 
\stackrel{\,\omega,\omega^\prime}{\otimes}S^{\rm NLP}_{sc,\rm sub}(1-\xi^\prime,\omega,\omega^\prime)\stackrel{\,\xi^\prime}{\otimes}f_{q/N}^{\rm thr, LP}(1+\xi-\xi^\prime)\,.
\label{eq:PDFschematic}
\end{equation}
The gluon PDF at NLP near threshold is a convolution of a soft-collinear function with the 
{\em quark} PDF near threshold at LP, and a function $D^{B1}(\omega)$ (not indicated in the figure), which accounts for hardcollinear quantum corrections to the $q\to g(hc)+q(sc)$ splitting vertex. The derivation is presented in detail in Sec.~\ref{sec:A0term}. 

A complication with important consequences now arises because the new convolutions in $r,r^\prime$ and $\omega,\omega^\prime$ at NLP diverge logarithmically when the various hard, collinear, and soft-collinear factors are renormalized, and the limit $\eps\to 0$ of dimensional regularization has been taken, rendering the anti-hardcollinear and soft-collinear contributions to the hadronic tensor ill-defined separately. In the anti-hardcollinear term \eqref{eq:B1schematic}, the divergence arises from the endpoint $r,r^\prime =0$, while in the soft-collinear term, it arises from $\omega,\omega^\prime=\infty$. This can easily be seen from the left diagram in Fig.~\ref{fig:sketch}. When $r\to 0$ the originally hard gluon propagator 
with a virtuality of order $r Q^2$ in the red ellipse goes on-shell and 
produces a $1/r$ singularity. At the same time, the assumption that this propagator is hard is violated.

We now observe that the anti-hardcollinear and soft-collinear modes have 
$\np\cdot p$ light-cone momentum of the same order $Q(1-x)$. Since both are on-shell modes, their virtuality is controlled by the $\nm\cdot p$ light-cone component, which is of order $r Q$ for the $\overline{hc}$ quark above the green line in the left figure, and $\omega \sim \Lambda_{\rm QCD}^2/(Q (1-x))$ on the right for the quark below the line. As $r$ is taken from $\mathcal{O}(1)$ to $\mathcal{O}(\Lambda_{\rm QCD}^2/(Q^2 (1-x)))$ at fixed $\np\cdot p$, the $\overline{hc}$ mode turns smoothly into the $sc$ one. There is an intermediate region where the mode can be considered as either one, implying that both terms \eqref{eq:B1schematic} and \eqref{eq:A0schematic} must be identical, which leads to endpoint factorization consistency or ``refactorization'' relations for the hard, collinear, and soft-collinear  factors. Similar consistency relations related to endpoint divergence have been observed in other cases of NLP factorization \cite{Liu:2019oav, Beneke:2020ibj, Liu:2020wbn, Beneke:2022obx,Hurth:2023paz,Cornella:2022ubo,Liu:2022ajh}. The logarithmic divergence of the convolution integrals means that the overlap region must be factorized, but since renormalization of the component functions has already been performed, the dimensional regulator is no longer available. 
Instead one can adopt an endpoint factorization and rearrangement by introducing an explicit factorization scale $\Lambda$ in $\nm\cdot p$ \cite{Beneke:2022obx},\footnote{See \cite{Beneke:2008pi} for an early application of this idea in the context of exclusive $B$-meson decays.}  formally satisfying $\Lambda_{\rm QCD}^2/(Q^2 (1-x)) < \Lambda <  Q$ to subtract the contribution from $r,r' < \Lambda/Q$ from 
the anti-hardcollinear contribution \eqref{eq:B1schematic} and absorb it into the soft-collinear term \eqref{eq:A0schematic}.   
The PDF defined by \eqref{eq:PDFschematic} then acquires a dependence on $\Lambda$, $f_{q/N}^{\rm thr, LP}(\xi,\Lambda)$, and therefore differs from the usual $\overline{\rm MS}$ definition. The latter would be obtained by 
not renormalizing the hard, collinear and soft functions separately, but instead performing the divergent convolutions in dimensional regularization and subtracting the total poles afterwards. This, however, obscures the physics and factorization as $x\to 1$, and suggests that the $\overline{\rm MS}$ definition of the PDFs is not the most adequate one near the kinematic edge. We discuss this point and the relation between the two PDF definitions in Sec.~\ref{sec:resum}.


\section{Threshold factorization at leading power}
\label{sec:LPfact}

Here we consider the scattering of a nucleon off a virtual photon at leading power as $x\to 1$, i.e. the partonic process $q+\gamma^* \to X$.

\subsection{Threshold factorization of DIS}

We first match the electromagnetic current to the SCET 
current at leading power,
\begin{equation}
\bar\psi\gamma_\perp^\mu\psi(0) = \int dtd\bar{t}\,
\widetilde{C}^{\rm A0}(t,\bar{t})\,
\bar{\chi}_{\overline{hc}}(\bar{t}\nm)\gamma_\perp^\mu\chi_{hc}(t\np) \,.
\label{eq:vectortoA0}
\end{equation}
$\chi_{hc (\overline{hc})}$ 
denotes the collinear gauge-invariant (anti-) hardcollinear SCET quark field, which contains a collinear Wilson line in its definition. The momentum-space hard coefficient is defined as
\begin{equation}
 \label{eq:CA0momentumspace}
C^{\rm A0}(n_+p_c, n_-p_{\bar{c}}) =  \int dtd\bar{t}\,e^{-it n_+ p_c+i \bar{t}\nm p_{\bar c}}\, \widetilde{C}^{\rm A0}(t,\bar{t} \,)\,.
\end{equation}
It depends only on the product $n_+p_c n_-p_{\bar{c}}$  of 
the large collinear and anti-collinear momentum components. 
At leading order (LO), $C^{\rm A0}(n_+p_c, n_-p_{\bar{c}})=1+\mathcal{O}(\alpha_s)$.

Next, we decouple soft-collinear gluons from the anti-hardcollinear fields by the field redefinition
\begin{equation}
\bar{\chi}_{\overline{hc}}(x) \to \bar{\chi}^{(0)}_{\overline{hc}}(x) \overline{Y}^\dagger_{sc, n_+}(x_+), 
\end{equation}
with 
\begin{equation}
x_\mp^\mu = (n_\pm x)\frac{n_\mp^\mu}{2}\,,
\label{eq:xminusdef}
\end{equation}
and $\overline{Y}^\dagger_{sc, n_+}$ as defined in \eqref{eq:wilsonbardef}, 
which represents the soft-collinear analogue of the standard soft decoupling transformation \cite{Bauer:2001yt} for DIS near threshold. The above decoupling transformation is independent of the precise scaling of the soft-collinear mode (determined by $\beta$) since it relies only on the fact that the leading-power coupling of the soft-collinear to the anti-hardcollinear mode is of the eikonal type, which holds for any non-negative $a$.

The following comment needs to be made: The hardcollinear field $\chi_{hc}(x)$ 
in \eqref{eq:vectortoA0} is the collinear-gauge invariant field 
$W^\dagger_{hc}(x) \xi(x)$, where 
\begin{equation}
W^\dagger_{hc}(x) = \mathcal{P} \exp\left[i g_s \int_0^\infty ds \,n_+A_{hc}(x+sn_+)\right]
\end{equation}
denotes a hardcollinear Wilson line, which arises from decoupling {\em generic} collinear modes from the anti-collinear part of the operator. In this general form, $W^\dagger_{hc}$ includes any collinear mode whose coupling to the anti-collinear sector is of the eikonal type at leading power, among them the soft-collinear modes discussed above. To avoid double counting with $\overline{Y}^\dagger_{sc,n_+}$, we now specify that $W^\dagger_{hc}$ contains any collinear mode with $\np p\sim Q$, while the 
soft-collinear ones with $\np p\sim Q(1-x)$ are part of $\overline{Y}^\dagger_{sc,n_+}$.

In the absence of leading-power interactions between the anti-hardcollinear and soft-collinear modes, the final state can now be written as  $|X\rangle = |X_{\overline{hc}}\rangle|X_{sc}\rangle$, 
and hence the matrix element of the electromagnetic current factorizes into
\begin{eqnarray}
\left\langle X|J^{\mu}(0)|N(p)\right\rangle &=&    
\int dtd\bar{t}\,
\widetilde{C}^{\rm A0}(t,\bar{t})\,
\langle X_{\overline{hc}}|\bar{\chi}^{(0)}_{\overline{hc}}(\bar{t}\nm)_{a\alpha }|0\rangle\,[\gamma_\perp^\mu]_{\alpha\beta} 
\nonumber\\
&&\times\,\langle X_{sc}|[\overline{Y}_{sc,n_+}^\dagger(0)\chi_{hc}(t\np)]_{a \beta }|N(p)\rangle\,.
\label{eq:A0amp}
\end{eqnarray}
Field products in an amplitude (its conjugate) are understood to be T-ordered (anti-T-ordered). Lower Latin indices refer to colour in the fundamental representation, Greek indices to Dirac spinor indices.  
When squaring the matrix element and summing over the final state $X$ according to \eqref{eq:hadtensor}, we introduce the anti-hardcollinear ``jet'' function
\begin{eqnarray}
&&\frac{1}{2\pi}\sum_{X_{\overline{hc}}}\int \!\PS_{X_{\overline{hc}}}
\,e^{-i p_{X_{\overline{hc}}}\cdot x}\,\langle 0|\chi_{\overline{hc}}^{(0)}(0)_{a'\alpha'}|X_{\overline{hc}}\rangle 
\langle X_{\overline{hc}}|\bar{\chi}_{\overline{hc}}^{(0)}(0)_{a\alpha}|0\rangle 
\nonumber\\
&& \equiv\,\delta_{a'a}\left(\frac{\slashed n_{+}}{2}\right)_{\!\alpha^\prime\alpha}\,\int\!\frac{d^d k}{(2\pi)^d}\,n_{-}k\,e^{-ikx}
\mathcal{J}^{(q)}_{\overline{hc}}(k^2)\,,
\label{eq:LPjet2}
\end{eqnarray}
which equals  $\mathcal{J}^{(q)}_{\overline{hc}}(k^2)=\delta^{+}(k^2)
\equiv \theta(k^0)\delta(k^2)$ at LO in the strong coupling. This allows us to write \eqref{eq:hadtensor} as
\begin{eqnarray}
W^{\mu\nu} &=& \frac{1}{2}
\left[\gamma_\perp^\nu \frac{\slashed n_{+}}{2}\gamma_\perp^\mu\right]_{\beta'\beta}\,\int d^d z \int \frac{d^d k}{(2\pi)^d}\,n_{-}k\,
\mathcal{J}^{(q)}_{\overline{hc}}(k^2)\,
\sum_{X_{sc}}\int \!\PS_{X_{sc}}\,e^{i (p+q-k-p_{X_{sc}})\cdot z}
\nonumber\\
&& \times\int dtd\bar{t}\,e^{i \bar{t} \nm k}\,
\widetilde{C}^{\rm A0}(t,\bar{t})\,
\int dsd\bar{s}\,e^{-i \bar{s} \nm k}\,
[\widetilde{C}^{\rm A0}(s,\bar{s})]^*\nonumber\\[0.1cm]
&& \times \,
\langle N(p)| [\bar{\chi}_{hc}(s \np) \overline{Y}_{sc,n_+}(0)]_{a\beta'}|X_{sc}\rangle
\langle X_{sc}|[\overline{Y}_{sc,n_+}^\dagger(0)\chi_{hc}(t\np)]_{a \beta }|N(p)\rangle\,.
\label{eq:W1}
\end{eqnarray}

It is now tempting to proceed by using a similar completeness relation for the sum over the soft-collinear final-state particles as done in \cite{Becher:2006mr}. However, this is technically incorrect, as the (hard-) collinear $\chi_{(h)c}$ field operates on a larger Hilbert space containing collinear modes up to a larger virtuality than the soft-collinear one. The matrix element that would follow from the last line of \eqref{eq:W1} by this procedure would contain contributions from collinear emissions into the final state, which are kinematically forbidden and were eliminated ``by hand'' in \cite{Becher:2006mr}. This problem is solved by first integrating out the (hard-) collinear fluctuations of the $\chi_{(h)c}$ field, matching the latter onto a field with a frozen $n_+ p$ component \cite{Beneke:2003xh,Beneke:2004km}.

\begin{figure}[t]
\centering
\includegraphics[width=0.45\textwidth]{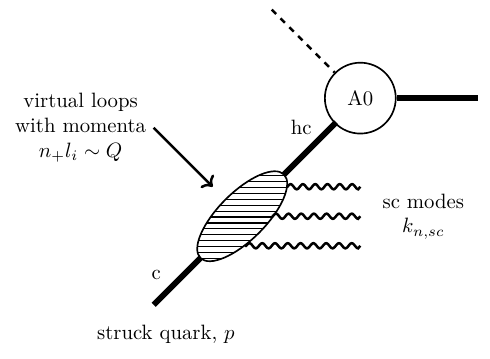}
 \caption{Diagram to be considered for integrating out fluctuations with $\np l\sim Q$.}
\label{fig:scfactorization}
\end{figure}

For this purpose, we consider an $N$-loop diagram of the form shown in Fig.~\ref{fig:scfactorization}, where the shaded blob contains loops with momenta $l_i$, satisfying $\np l\sim Q$. Such modes cannot be part of the final state $X$. The external lines are, therefore, soft-collinear with momenta $k_{n,sc}$, and the collinear struck quark carries momentum $p_c$.
The initial $N$-loop diagram takes the form
\begin{equation}
\label{eq:initialdiagram}
\int\prod_{i=1}^N d^dl_i\,\prod_k \frac{1}{(p_{k,c}+\sum l_i 
 +\sum k_{n,sc})^2}\times {\rm polynomial},
\end{equation}
where the product over $k$ refers to the internal lines, and the
sums denote line-specific linear combinations of loop and external $sc$ momenta with
coefficients $\pm 1, 0$, which we denote by $L_{k}$,
and $K_{k,sc}$, respectively, while $p_{k,c}=p_c$ or 0, if $p_c$ does not flow through the line $k$. We note that the virtuality of the 
momentum $P=p_c-\sum k_{n,sc}$ is $P^2\sim Q^2(\xi\beta/\lambda)^2$, 
that is, $P$ is hardcollinear as defined in \eqref{eq:hcmode} 
when $\beta\sim 1$, and collinear when $\beta\sim \lambda$. 

To integrate out the modes with $\np l\sim Q$, the propagators in \eqref{eq:initialdiagram} are expanded in the small scalar products.
Using the scaling rules for
the various momenta and dropping terms that are always small compared to the dominant term, we find
\be
(L_{k} + p_{k,c} + K_{k,sc})^2 \approx 
(L_{k}+p_{k,c})^2 + \np (L_k+p_{k,c}) (\nm K_{k,sc})\,.
\ee
The integrand of \eqref{eq:initialdiagram} must be Taylor-expanded in all terms that are suppressed by at least one power of $\lambda$, and we obtain a sum of terms of the form
\be
\label{expandeddiagram}
\int\prod_{i=1}^N d^dl_i\,\prod_k
\frac{1}{((L_{k}+p_{k,c})^2 + \np (L_k+p_{k,c}) (\nm K_{k,sc}))^{a_k}}
\times {\rm polynomial}
\ee
with integers $a_k$. For on-shell matching, $p_c^\mu=(\np p_c) \frac{\nm^\mu}{2}$, such that the only invariants are of the form $\np p_c \nm k_{n,sc}$ with scaling $(\xi \beta/\lambda)^2$, hence the integrals are scaleless and vanish in dimensional regularization unless $L_k^2$ has the same 
scaling, which identifies the modes as hardcollinear or collinear. 

However, at leading power, the coupling of the soft-collinear modes to the (hard-) collinear ones with $\np l\sim Q$ is always of the eikonal type, independent of the scaling of the soft-collinear modes. Hence, the coupling can be removed by a field redefinition with the incoming-particle Wilson line $Y_{sc,n_-}(x)$ from \eqref{eq:wilsondef}.
The decoupling transformation detaches the soft-collinear lines from the shaded blob in Fig.~\ref{fig:scfactorization}. But without external soft-collinear momenta, the loops in the shaded blob are scaleless, and the hardcollinear function is trivial for the same reason that there is no collinear function in the threshold factorization of the partonic Drell-Yan process {\em at leading power} \cite{Beneke:2019oqx}. When the fluctuations have collinear virtuality, the corresponding collinear function is 
non-perturbative and we therefore allow for a matching coefficient $C_\Lambda$, 
which depends on the strong interaction scale and, as will be seen, the factorization scale, and accounts for the purely virtual collinear fluctuations on the external parton leg.

This discussion justifies the leading-power matching equation
\begin{equation}
\chi_{hc}(x) =\chi_{c}(x)= e^{-i p x} C_\Lambda \,\chi_{fc}(x) = e^{-i p x}\, C_\Lambda \,Y_{sc,n_-}(x_-)\chi_{fc}^{(0)}\,,
\label{eq:frmatching}
\end{equation}
where, in the first step, the hardcollinear modes are integrated out, which at leading power trivially match onto the collinear ones, as discussed above. In the second step, the collinear virtuality is integrated out, such that after taking out the rapidly varying phase, $\chi_{fc}(x)$ 
has only soft-collinear fluctuations. Here, $p$ is taken to be the external nucleon momentum. Moreover, the soft-collinear fluctuations can be decoupled by a Wilson line field redefinition. 
The redefined  ``frozen-collinear" field $\chi_{fc}^{(0)}$ is sterile, and its 
spatial dependence becomes irrelevant. We emphasize again that the above holds independent of the scaling of $\beta$, covering in particular the case $\beta\sim 1$.
The leading-power Lagrangian for the ``frozen-collinear" field $\chi_{fc}(x)$ is (compare \cite{Beneke:2003xh,Dugan:1990de})
\begin{align}
    \mathcal{L}_{fc} = \overline{\chi}_{fc}(x) \frac{\not\!\np}{2}\, n_- \cdot i D_{sc}(x_-) \chi_{fc}(x),
\end{align}
with the soft-collinear covariant derivative $n_- \cdot iD_{sc}(x) =  n_-\cdot i\partial + g_sn_- \cdot A_{sc}(x_-)$. 

Substituting \eqref{eq:frmatching} into \eqref{eq:W1} and making use of the momentum-space hard functions \eqref{eq:CA0momentumspace}, we simplify \eqref{eq:W1} to
\begin{eqnarray}
W^{\mu\nu} &=& \frac{1}{2}
\left[\gamma_\perp^\nu \frac{\slashed n_{+}}{2}\gamma_\perp^\mu\right]_{\beta'\beta}\,\int d^d z \int \frac{d^d k}{(2\pi)^d}\,n_{-}k\,
\mathcal{J}^{(q)}_{\overline{hc}}(k^2)\,
\sum_{X_{sc}}\int \!\PS_{X_{sc}}\,e^{i (p+q-k-p_{X_{sc}})\cdot z}
\nonumber\\
&& \times \left|C^{\rm A0}(\np p,\nm k)\right|^2 
\left| C_\Lambda\right|^2
\nonumber\\[0.2cm]
&& \times \,
\langle N(p)| [\bar{\chi}^{(0)}_{fc} Y_{sc,n_-}^\dagger \overline{Y}_{sc,n_+}]_{a\beta'}(0)|X_{sc}\rangle
\langle X_{sc}|[\overline{Y}_{sc,n_+}^\dagger Y_{sc,n_-}\chi^{(0)}_{fc}]_{a \beta }(0)|N(p)\rangle\,.\qquad
\label{eq:W2}
\end{eqnarray}
Next, we consider the integration over $z_\perp$ and $\np z$. Approximating $(p+q-k-p_{X_{sc}})_\perp \approx k_\perp$, 
$\nm\cdot (p+q-k-p_{X_{sc}}) \approx Q-\nm k$ with leading-power accuracy, we obtain
\begin{eqnarray}
W^{\mu\nu} &=& \frac{1}{2}
\left[\gamma_\perp^\nu \frac{\slashed n_{+}}{2}\gamma_\perp^\mu\right]_{\beta'\beta}\,
\left|C^{\rm A0}(\np p,Q)\right|^2 
\left| C_\Lambda\right|^2 
 \int \frac{d\np k}{2\pi}\,Q\,
\mathcal{J}^{(q)}_{\overline{hc}}(Q\np k)\nonumber\\
&& \times
\int d\left(\frac{\nm z}{2}\right)\sum_{X_{sc}}\int \!\PS_{X_{sc}}\,e^{i \np\cdot (p+q-k-p_{X_{sc}}) \nm \cdot z/2}\nonumber\\[0.0cm]
&& \times \,
\langle N(p)| [\bar{\chi}^{(0)}_{fc} Y_{sc,n_-}^\dagger \overline{Y}_{sc,n_+}]_{a\beta'}(0)|X_{sc}\rangle
\langle X_{sc}|[\overline{Y}_{sc,n_+}^\dagger Y_{sc,n_-}\chi^{(0)}_{fc}]_{a \beta }(0)|N(p)\rangle\,.\qquad
\label{eq:W3}
\end{eqnarray}
At this point, the sum over the soft-collinear final state can be done by using completeness, resulting in 
\begin{eqnarray}
&&\sum_{X_{sc}}\int \!\PS_{X_{sc}}\,e^{-i \np\cdot p_{X_{sc}} \nm \cdot z/2}
\langle N(p)| [\bar{\chi}^{(0)}_{fc} Y_{sc,n_-}^\dagger \overline{Y}_{sc,n_+}]_{a\beta'}(0)|X_{sc}\rangle\nonumber\\[0.0cm]
&& \hspace*{1cm}\times \,
\langle X_{sc}|[\overline{Y}_{sc,n_+}^\dagger Y_{sc,n_-}\chi^{(0)}_{fc}]_{a \beta }(0)|N(p)\rangle\nonumber\\[0.2cm]
&&=\,\langle N(p)| [\bar{\chi}^{(0)}_{fc} Y_{sc,n_-}^\dagger \overline{Y}_{sc,n_+}]_{a\beta'}(z_+) [\overline{Y}_{sc,n_+}^\dagger Y_{sc,n_-}\chi^{(0)}_{fc}]_{a \beta }(0)|N(p)\rangle\nonumber\\[0.2cm]
&&=\,\langle N(p)| \bar{\chi}^{(0)}_{fc,a'\beta'}\chi^{(0)}_{fc,a\beta}|N(p)\rangle
\times\langle 0|[Y_{sc,n_-}^\dagger \overline{Y}_{sc,n_+}]_{a b}(z_+) [\overline{Y}_{sc,n_+}^\dagger Y_{sc,n_-}]_{ba'}(0)|0\rangle\nonumber\\[0.2cm]
&&\equiv \,
\left(\frac{\slashed n_{-}}{2}\right)_{\!\beta\beta'}\, \frac{1}{2}\,\np p \,K_{q/N}\,
\int d\Omega\,e^{-i\Omega\frac{\nm z}{2}}\,S_{sc}^{\rm LP}(\Omega)\,,
\label{eq:scsum}
\end{eqnarray}
with $z_+^\mu=(n_- z)\frac{\np^\mu}{2}$. 
To arrive at the second-to-last line, we used that the frozen-collinear field is sterile to factorize the matrix element into the (implicitly spin-averaged) nucleon matrix element
\begin{equation}
 \langle N(p)| \bar{\chi}^{(0)}_{fc,a'\beta'}\chi^{(0)}_{fc,a\beta}|N(p)\rangle = \frac{\delta_{a a'}}{2 N_c} \,\left(\frac{\slashed n_{-}}{2}\right)_{\!\beta\beta'} \np p \,K_{q/N}\,,
 \label{eq:KqN}
\end{equation}
where $K_{q/N}$ is a dimensionless constant.  
The leading-power soft-collinear function is defined as 
\begin{equation}
\frac{1}{N_c}\langle 0|\,\mbox{tr}\,\big([Y_{sc,n_-}^\dagger \overline{Y}_{sc,n_+}](z_+) [\overline{Y}_{sc,n_+}^\dagger Y_{sc,n_-}](0)\big)|0\rangle = 
 \int d\Omega\,e^{-i\Omega\frac{\nm z}{2}}\,S_{sc}^{\rm LP}(\Omega)\,. 
\label{eq:LPscfunction}
\end{equation}
The trace is taken over the colour indices of the Wilson lines in the fundamental representation. At leading order in the strong coupling $S_{sc}^{\rm LP}(\Omega)= \delta(\Omega)$. 

Finally, substituting \eqref{eq:scsum} into \eqref{eq:W3}, the integration over $\frac{\nm z}{2}$ can be performed and sets 
$\np k=\frac{Q(1-x)}{x}-\Omega$, upon which \eqref{eq:W3} becomes
\begin{eqnarray}
W^{\mu\nu} &=& -g_\perp^{\mu\nu}\, 
 \left|C^{\rm A0}(\np p,Q)\right|^2
 \nonumber\\
&&\times  \int d\Omega\,Q\,
\mathcal{J}^{(q)}_{\overline{hc}}\left(Q \left(\frac{Q(1-x)}{x}-\Omega\right)\right)\, 
\frac{1}{2}\,\np p\,K_{q/N} \left| C_\Lambda\right|^2 
S_{sc}^{\rm LP}(\Omega)\,. \qquad
\label{eq:W4}
\end{eqnarray}
Since the anti-hardcollinear function has support only for $k^2\geq 0$, and since the invariant mass of the final state $p_X^2$ equals $Q^2(1-x)/x$, $\Omega$ is restricted 
to $0\leq \Omega\leq Q(1-x)/x$. Introducing the variable $\xi$ through 
\begin{equation}
\Omega=\frac{Q(1-\xi)}{x}\,,
\end{equation}
we obtain the final result
\begin{eqnarray}
W^{\mu\nu} &=& -g_\perp^{\mu\nu}\, 
\left|C^{\rm A0}(\np p,Q)\right|^2\,\frac{Q^2}{x}\nonumber\\
&&\times \int_x^1 d\xi\,
\mathcal{J}^{(q)}_{\overline{hc}}\left(Q^2\frac{\xi-x}{x}\right)\, 
\frac{1}{2}\,\np p \,K_{q/N} 
\left| C_\Lambda\right|^2  S_{sc}^{\rm LP}\left(\frac{Q(1-\xi)}{x}\right)\,.
\label{eq:LPfact}
\end{eqnarray}

\subsection{LP threshold factorization of the parton distribution}
\label{sec:quarkPDFfactLP}

We next show that the leading-power soft-collinear function is directly related to the ordinary parton distribution function (PDF) in the $\xi\to 1$ limit. The PDF is defined in SCET as 
\begin{equation}
f_{q/N}(\xi) =\frac{1}{2\pi} \int dt \,e^{-i \xi t  \np p}\, \langle N(p)| \bar{\chi}_{c}(t n_+)\frac{\slashed n_{+}}{2}
\chi_c(0)|N(p)\rangle\,.
\label{eq:PDFdef}
\end{equation}
We examine the leading-power term in the $\xi\to 1$ limit by first writing: 
\begin{eqnarray}
f_{q/N}(\xi) &\stackrel{\xi\to 1}{=}&
\sum_{X_{sc}}\int \!\PS_{X_{sc}}\!\int \frac{dt}{2\pi}\, e^{i (1-\xi) t  \np p-i t \np p_{X_{sc}}}\,\langle N(p)| \bar{\chi}_{c}(0)|X_{sc}\rangle\,\frac{\slashed n_{+}}{2} \nonumber\\
&&\times\,
\langle X_{sc}|
\chi_c(0)|N(p)\rangle\,,\quad
\label{eq:PDF2}
\end{eqnarray}
where we accounted for the fact that the intermediate states must be soft-collinear 
for kinematic reasons. The collinear field in the ordinary PDF definition is a generic collinear field that contains all fluctuations with $p_{c\perp}\ll Q$ and 
$\np p_c\gg p_{c\perp}\gg \nm p_c$. It therefore includes both
$W^\dagger_{c}$ and $\overline{Y}^\dagger_{sc,n_+}$. The collinear fluctuations with $\np p_c\sim Q$ cannot be real modes in the final state and must be purely virtual. Integrating them out in the presence of the soft-collinear modes yields an equation analogous to \eqref{eq:frmatching} with matching coefficient $C_\Lambda$ for the virtual collinear modes. Hence, integrating out the collinear modes  in the presence of the soft-collinear fluctuations yields 
\begin{equation}
\chi_{c}(z) \to  \overline{Y}^\dagger_{sc,n_+}(z) \times C_\Lambda \, e^{-i p z} \chi_{fc}(z) = \overline{Y}^\dagger_{sc,n_+}(z) \times C_\Lambda\, e^{-i p z} \, 
Y_{sc,n_-}(z_-)\chi_{fc}^{(0)}\,,
\label{eq:frmatchingPDF}
\end{equation}
resulting in 
\begin{eqnarray}
f_{q/N}(\xi) &\stackrel{\xi\to 1}{=}& \int \frac{dt}{2\pi}\, e^{i (1-\xi) t  \np p}\, |C_\Lambda|^2 \, \langle N(p)| \bar{\chi}^{(0)}_{fc}[Y^\dagger_{sc,n_-}\overline{Y}_{sc,n_+}](t \np)\,\frac{\slashed n_{+}}{2}\nonumber\\
&&\times\,[\overline{Y}^\dagger_{sc,n_+} Y_{sc,n_-}](0) 
\chi_{fc}^{(0)}|N(p)\rangle\nonumber\\
&=& \np p \,K_{q/N}\, |C_\Lambda|^2\,S^{\rm LP}_{sc}\left(\np p(1-\xi)\right) \equiv f^{\rm thr,LP}_{q/N}(\xi) \,,
\label{eq:LPPDFfact}
\end{eqnarray}
where $\np p=Q/x$ has been used. This shows explicitly that, at leading power, the ordinary PDF in the $\xi \to 1$ limit coincides with the soft-collinear function times the virtual collinear function without another non-trivial matching coefficient. Substituting the previous expression into \eqref{eq:LPfact}, the DIS hadronic tensor at leading power in the $x \to 1$ limit can be expressed in terms of the ordinary PDF as
\begin{equation}
W^{\mu\nu} = -g_\perp^{\mu\nu}\, \frac{1}{2x}
\left|C^{\rm A0}(\np p,Q)\right|^2\,Q^2
 \int_x^1 d\xi\,
\mathcal{J}^{(q)}_{\overline{hc}}\left(Q^2\frac{\xi-x}{x}\right) 
f_{q/N}(\xi)\,.
\label{eq:LPfact2}
\end{equation}
This represents the final result at leading power and formally agrees with \cite{Sterman:1986aj,Catani:1989ne} 
and \cite{Becher:2006mr}, although the derivation given here provides a more precise analysis of the implications of the phase-space restrictions on the mode structure and does not require excluding contributions ``by hand''. 

Eq.~\eqref{eq:LPPDFfact} represents the factorization formula for the unrenormalized PDF. 
In order to complete the proof that the threshold PDF corresponds to the $\xi\to 1$ limit of the usual $\overline{\rm MS}$-renormalized PDF, it needs to be shown that $\overline{\rm MS}$ renormalization commutes with the $\xi\to 1$ limit and does not alter the relation between the left- and right-hand sides of that equation. In other words, the kernel renormalizing the product of the soft function defined in~\eqref{eq:LPscfunction} and the matching coefficient $C_\Lambda$ must coincide with the $\xi\to 1$ limit of the DGLAP kernel, which is obtained by computing the right-hand side of \eqref{eq:PDFdef} in $d$ dimensions and taking the $\xi\to 1$  limit  of the pole part afterwards. 

To demonstrate this, we evaluate these quantities at next-to-leading order (NLO); see also App.~B of~\cite{Beneke:2025ufd}. Since renormalization concerns ultraviolet physics, we can replace the nucleon $N$ with a quark $q$ for this purpose. With dimensional regularization as the only regulator, the loop integrals are scaleless, and the unrenormalized $f_{q/q}$, $C_\Lambda$ and $S_{sc}^{\rm LP}$ receive no corrections. The corresponding $\overline{\rm MS}$-renormalized quantities are equal to the counterterm contributions. To compute the counterterm, we introduce an off-shellness $p^2$ of the external quark to regulate the IR divergences and extract the UV poles in dimensional regularization with $d=4-2\eps$. 

For the quark state, we find 
$K_{q/q}=1$ (recall the implicit spin-average in taking the matrix element), 
while the collinear factor can be obtained to NLO from the corresponding QED expression  \cite{Beneke:2019slt} by replacing the charge factor $Q_\ell^2\to C_F$:
\begin{equation}
C_\Lambda = 1 + \frac{\alpha_s C_F}{4\pi} \left[\frac{2}{\epsilon^2} +\frac{2}{\epsilon} \ln \frac{\mu^2}{-p^2}+ \frac{3}{2\epsilon}\right]+ \mathcal{O}(\epsilon^0)\,.
\label{eq:CLam1loop}
\end{equation}
The soft-collinear function is given by 
\begin{eqnarray}
    S_{sc}^{\rm LP} (\Omega) &=& \delta(\Omega) + g_s^{2}C_F\!\int\!\frac{ d^{d}k}{\left(2\pi\right)^{d}}\,\Bigg\{\,\frac{n_{-}\cdot n_{+}}{n_{+}k\left(n_{-}k+\delta+i0^+\right)}\nonumber\\
    &&\times\,\left[2\pi\theta(k_0)\delta(k^2)\delta\left(\Omega-n_{+}k\right) +\frac{i}{k^2}\delta(\Omega)\right] + {\rm c.c.}\,\Bigg\}
    \nonumber\\
    &=&\delta(\Omega )+ \frac{\alpha_s C_F}{2\pi} \Theta(\Omega)\,\frac{1}{\Omega  } \Gamma(\epsilon)\Bigg\{
    \left(\frac{\mu^2 e^{\gamma_E}}{(-\delta-i 0^+) \Omega}\right)^{\!\epsilon}
     + {\rm c.c.}\,\Bigg\}
    \,,
    \label{eq:softLPOmega}
\end{eqnarray}
where $\delta=\frac{p^2}{n_+p}$. 
Setting $\Omega=\np p (1-\xi)$ and expanding in the standard plus-distribution gives 
\begin{eqnarray}
S_{sc}^{\rm LP} (\Omega) &=&\frac{1}{\np p}\,\bigg(\delta(1-\xi)+
\frac{\alpha_s C_F}{4\pi} \bigg[\,\left(-\frac{4}{\eps^2}-\frac{4}{\eps} \ln\frac{\mu^2}{\np p |\delta|} \right)\delta(1-\xi)\nonumber\\
&&+\,\frac{1}{\eps}\,\frac{4}{[1-\xi]_+}+\mathcal{O}(\epsilon^0)\bigg]\,\bigg)\,.
\label{eq:SscLP1loop}
\end{eqnarray}
Multiplying \eqref{eq:CLam1loop} and \eqref{eq:SscLP1loop}, the double poles and $\delta$-dependent terms cancel and we obtain  
\begin{equation}
f^{\rm thr, LP}_{q/q}(\xi)=\;\delta(1-\xi)+
\frac{\alpha_s C_F}{2\pi} \,\frac{1}{\eps}\left(\frac{2}{[1-\xi]_+}+\frac{3}{2}\delta(1-\xi)\right)+\mathcal{O}(\epsilon^0)\,
\label{eq:PDFrhs}
\end{equation}
for the UV pole part of all factors on the right-hand side of \eqref{eq:LPPDFfact}. 

The UV pole part of the standard dimensionally regulated bare PDF at one-loop is obtained from 
\begin{equation}
f_{q/q}(\xi)|_{\mathcal{O}(\alpha_s)}  =  \frac{1}{\epsilon}
\int_\xi^1 \frac{dx}{x} \,P_{qq}\!\left(\frac{\xi}{x}\right)\,f_{q/q,\rm LO}(x) + \mathcal{O}(\epsilon^0)\,,
  \end{equation}
where $f_{q/q,\rm LO}(y)=\delta(1-y)+\mathcal{O}(\alpha_s)$ and 
\begin{align}
 P_{qq}(x)  = \frac{\alpha_s C_F}{2\pi}\left[\frac{1+x^2}{1-x}\right]_+
\end{align}
denotes the one-loop DGLAP kernel. Including the lowest-order term and taking the $\xi\to 1$ limit, we find 
\begin{align}
   f_{q/q}(\xi) &= \delta(1-\xi) + \frac{\alpha_s C_F}{2\pi}
   \frac{1}{\epsilon} \left[\frac{1+\xi^2}{1-\xi}\right]_+\nonumber\\
   &\mathop{ = }\limits^{\xi \to 1} \delta(1-\xi) + \frac{\alpha_s C_F}{2\pi}
   \frac{1}{\epsilon}\left(2\left[\frac{1}{1-\xi}\right]_+ + \frac{3}{2}\delta(1-\xi) \right) 
   \label{eq:fqqMSbar}
\end{align}
 for the standard dimensionally regulated quark-in-quark PDF, in agreement with \eqref{eq:PDFrhs}. Since $f^{\rm thr, LP}_{q/q}(\xi)$ and $f_{q/q}(\xi)$ are rendered finite by the same counterterm, we conclude that at LP, the renormalized threshold PDFs defined through SCET factorization coincide with the threshold limit of the standard $\overline{\rm MS}$ renormalized collinear PDFs. 

\subsection{Threshold resummation of the parton distribution}
\label{sec:LPPDFresum}

For later purposes, we rederive known results \cite{Beneke:2020ibj,Vogt:2010cv} for the $d$-dimensional threshold resummation of the unrenormalized parton distribution function $f_{q/N}(\xi)$ and its $\msbar$ counter\-term at leading power. We restrict ourselves to the double-logarithmic approximation (DLA); that is, we focus on leading $\eps$-pole terms.\footnote{Note that individual component functions $C_\Lambda$ and $S_{sc}^{\rm LP}$ feature a double-pole expansion, while the PDF $f_{q/N}(\xi)$ has only single poles.} 

The UV renormalization of the PDFs is conventionally done in Mellin space, where it becomes multiplicative. We denote by $N$ the Mellin variable conjugate to $\xi$, and define the Mellin transform by
\begin{align}
  f(N) \equiv \int_0^1\! d\xi\, \xi^{N-1} f(\xi)\,,
  \label{eq:mellin2}
\end{align}
In Mellin space, the $\xi \to 1$ limit corresponds to $N \to \infty$. Up to power-suppressed terms in these limits, it suffices to replace $\bar{\xi}\equiv 1-\xi \to 1/N$. We will suppress the dependence of component functions on the Mellin variable $N$ for brevity in the following.

The PDF at leading power is defined in terms of component functions in \eqref{eq:LPPDFfact}. For the nucleon, both $C_\Lambda$ and $S_{\rm sc}^{\rm LP}$ are non-perturbative. However, their UV poles can be obtained in perturbation theory and are given by \eqref{eq:CLam1loop} and \eqref{eq:SscLP1loop} at one-loop order, respectively. For the purpose of resummation, we take $\mu_0\equiv \sqrt{-p^2}$ to be the initial scale of the renormalization-group (RG) evolution and assume that $\mu_0$ is a low-energy scale, sufficiently but not parametrically large compared to $\Lambda_{\rm QCD}$, such that $\alpha_s(\mu_0)$ is small enough. Note that $\as\equiv\as(\mu)$ is the $\msbar$ running coupling, and we write the argument explicitly only when the coupling is to be evaluated at a different scale.

First, we demonstrate how to resum the leading poles of a general bare Wilson coefficient $\mathcal{C}$ and the corresponding operator matrix element $\mathcal{O}$, before applying it to the component functions entering the LP PDF. We renormalize operator matrix elements and coefficients according to 
\begin{equation}
\label{eq:OCZfactors}
    \mathcal{O}(\mu)=Z(\mu)\mathcal{O},\qquad \mathcal{C}(\mu)=\mathcal{C}Z^{-1}(\mu)\,.
\end{equation}
The renormalization group equations (RGEs) of these objects then read:
\begin{equation}
\frac{d\mathcal{O}(\mu)}{d\ln\mu}=-\Gamma(\as,\mu) \mathcal{O}(\mu)\,,
\qquad \frac{d\mathcal{C}(\mu)}{d\ln\mu}=\Gamma(\as,\mu)^T \mathcal{C}(\mu)\,.
\end{equation}
Here, $\Gamma(\as,\mu)$ denotes the anomalous dimension matrix, and $\Gamma(\as,\mu)^T$ its transpose. Plugging \eqref{eq:OCZfactors} into these RGEs, we find the relation
\begin{equation}\label{eq:Gh}
    \Gamma(\as,\mu)=-\frac{d Z(\mu)}{d\ln\mu}Z^{-1}(\mu)\,.
\end{equation}
In general, the renormalization equation \eqref{eq:OCZfactors} is a matrix equation, and different operators mix under RG evolution. Consequently, the anomalous dimension becomes a matrix, and the order of objects in \eqref{eq:OCZfactors} to \eqref{eq:Gh} is relevant. In the double-logarithmic approximation (DLA), the anomalous dimension is approximated by 
\begin{equation}
    \Gamma(\as,\mu)=\gamma_\text{cusp}(\alpha_s(\mu))\,C_\mathcal{O}\ln\frac{\mu^2}{\tilde\mu^2}\,,
\end{equation}
where $C_\mathcal{O}$ denotes the corresponding colour factor, and 
\begin{equation}
\label{eq:cusp}
    \gamma_\text{cusp}(\as)=\frac\as\pi+\mathcal{O}(\as^2)
\end{equation}
is the cusp anomalous dimension. The scale $\tilde\mu$ is the natural scale of the operator, such that when $\mu\sim\tilde\mu$ no large logarithmic corrections are left in $\mathcal{O}(\tilde\mu)$. With the help of the $d$-dimensional relation
\begin{equation}
    \frac{d\as}{d\ln\mu}=-2\eps\as+\beta(\as)
\end{equation}
we find 
\begin{equation}\label{eq:Zres}
    \ln Z(\mu)=-\int_0^{\as}\frac{d\alpha}{\alpha}\frac{1}{2\eps-\beta(\alpha)/\alpha}\left(\Gamma(\alpha,\mu)+\int_0^\alpha\frac{d\alpha'}{\alpha'}\frac{2C_\mathcal{O}\gamma_\text{cusp}(\alpha')}{2\eps-\beta(\alpha')/\alpha'}\right).
\end{equation}
In the DLA, contributions from the $\beta$-function can be neglected, as well as contributions from non-cusp parts in the anomalous dimension $\Gamma$. We therefore find 
\begin{equation}
\label{eq:Zhresum}
    \ln Z(\tilde\mu)\overset{\text{DLA}}{=}
    \frac{\as(\tilde\mu)C_\mathcal{O}}{2\pi\eps^2}=\frac{\as C_\mathcal{O}}{2\pi\eps^2}\left(\frac{\mu^2}{\tilde\mu^2}\right)^{\!\eps}\,,
\end{equation}
where the $d$-dimensional evolution of the coupling in the DLA ($\beta(\alpha_s)=0$) has been used. As a result, the above $Z$ factor, upon expansion, contains positive powers of $\epsilon$. In the following, we shall denote such $d$-dimensional $Z$ factors by $\mathcal{Z}$ to distinguish them from the $\msbar$ $Z$ factors, which consist of the pole part of $\mathcal{Z}$ only.

From the one-loop expressions \eqref{eq:CLam1loop} and \eqref{eq:SscLP1loop} we deduce the one-loop renormalization factors as
\begin{equation}
\begin{aligned}
    \left(Z_{C_\Lambda}^{(1\text{L})}(\mu)\right)^{-1}=&\,1-\frac{\as C_F}{4\pi}\left(\frac{2}{\eps^2}+\frac2\eps\ln\frac{\mu^2}{\mu_0^2}\right)\,,\\
    Z_\text{sc}^{(1\text{L})}(\mu)=&\,1+\frac{\as C_F}{4\pi}\left(\frac{4}{\eps^2}+\frac4\eps\ln\frac{\mu^2N}{\mu_0^2}\right)\,,
    \end{aligned}
\end{equation}
and the superscript $1$L refers to terms up to one-loop order.
Computing the anomalous dimension with \eqref{eq:Gh} and using \eqref{eq:Zhresum}, we derive 
\begin{equation}
\ln \mathcal{Z}_{C_\Lambda}^{-1}(\tilde\mu)=-\frac{\as C_F}{2\pi\eps^2}\left(\frac{\mu^2}{\mu_0^2}\right)^{\!\eps},\qquad
    \ln \mathcal{Z}_\text{sc}(\tilde\mu)=\frac{\as C_F}{\pi\eps^2}\left(\frac{\mu^2N}{\mu_0^2}\right)^{\!\eps}\,.
\end{equation}
Here we made the identification $\tilde\mu=\mu_0$ for $C_\Lambda$ and $\tilde\mu=\mu_0/\sqrt{N}$ for the LP soft-collinear function, which follows from \eqref{eq:softLPOmega}. Inverting the relations \eqref{eq:OCZfactors} and using the fact that large logarithmic corrections are absent when the initial conditions are evaluated at their respective natural scale $\tilde\mu$, the $d$-dimensional resummed expressions for the bare component functions read
\begin{align}
    C_\Lambda^\text{DLA}=\,&\exp\left[\frac{\as C_F}{2\pi\eps^2}\left(\frac{\mu^2}{\mu_0^2}\right)^{\! \eps}\,\right]C_\Lambda(\mu_0)\,,\\
   S_\text{sc}^\text{LP,DLA}=\,&\exp\left[-\frac{\as C_F}{\pi\eps^2}\left(\frac{\mu^2N}{\mu_0^2}\right)^{\!\eps}\,\right]S_\text{sc}^\text{LP}(\mu_0).
\end{align}
Two remarks are now in order: First, $\alpha_s\times(\mu^2)^\eps$ equals the bare coupling in the DLA, and hence the right-hand sides of the above equations are $\mu$-independent. Second, $C_\Lambda(\mu_0)$ and $S_{\rm sc}^{\rm LP}(\mu_0)$ represent renormalized, i.e. finite, but $d$-dimensional objects at their natural scales, which contain no large logarithms.

The bare LP PDF then follows as
\begin{align}
    f_{q/N}^\text{LP}=\,&\exp\left[-\frac{\as C_F}{\pi\eps^2}\left(\frac{\mu^2}{\mu_0^2}\right)^{\!\eps}(N^\eps-1)\right]n_+pK_{q/N}|C_\Lambda(\mu_0)|^2S_\text{sc}^\text{LP}(\mu_0)\nonumber\\
    \equiv\,&\,\mathcal{Z}^\text{LP}_{qq}(\mu_0)f_{q/N}^\text{LP}(\mu_0)\,,
\label{eq:fqnLPLL}
\end{align}
where we introduced the $d$-dimensional inverse-$\mathcal{Z}$ factor (denoted by $U_{qq}^{\rm LP}$ in \cite{Beneke:2020ibj})
\begin{equation}
    \mathcal{Z}_{qq}^\text{LP}(\mu)=\exp\left[-\frac{\as C_F}{\pi\eps^2}\left(\frac{\mu^2}{\mu_0^2}\right)^{\!\eps}(N^\eps-1)\right]\,,
    \label{eq:ULPqqbare}
\end{equation}
for the PDF and the renormalized $d$-dimensional initial condition $f_{q/N}^\text{LP}(\mu_0)$.
We emphasize that, in general, we do not control the dependence of the initial condition $f_{q/N}^\text{LP}(\mu_0)$ on the Mellin variable $N$, because the dependence of the PDF on the momentum fraction $\xi$ at the initial scale is non-perturbative and not amenable to power counting in $1-\xi \sim 1/N$. 

The renormalized $\msbar$ PDF is defined by $Z_{qq}^{\rm LP}f_{q/N}^\text{LP}$, where the $\msbar$ renormalization factor $Z_{qq}^{\rm LP}$ subtracts precisely the pole part. Since all UV poles are contained in $\mathcal{Z}^\text{LP}_{qq}$, the resummed expression for $Z_{qq}^\text{LP}$ can be immediately read off as
\begin{equation}
Z^\text{LP}_{qq}=\exp\left[\frac{\as C_F}\pi\frac{\ln N}{\eps}\right],
\label{eq:ZLPqq}
\end{equation}
in agreement with \cite{Vogt:2010cv,Beneke:2020ibj}.


\section{Next-to-leading power off-diagonal DIS: The anti-hardcollinear term}
\label{sec:B1term}

To analyze NLP factorization in the off-diagonal parton channel, we consider DIS of a nucleon induced by a scalar ``Higgs'' boson,
\begin{equation}
\phi^*(q) +N(p) \to X\,,
\end{equation}
coupled to gluons through the effective interaction
\begin{equation}
\label{LHiggsEff}
\mathcal{L}_{\rm eff} = 
\frac{\as C_t\left(m_t,\mu\right)}{12\pi} \frac{\phi}{v} G^A_{\mu\nu}G^{\mu\nu A} \,.
\end{equation}

\subsection{Hadronic tensor and hard matching}

The dimensionless DIS hadronic tensor is defined as
\begin{eqnarray}
\label{hadrtensdefPartonicPhiQ}
W_{\phi} = \frac{1}{8\pi Q^2} \int d^4 x \, e^{i q \cdot x} \,
\big\langle N(p) \big| \big[G_{\mu\nu}^A G^{\mu\nu A}\big](x)
\big[G_{\rho\sigma}^B G^{\rho\sigma B} \big](0) \big| N(p) 
\big\rangle\,.
\end{eqnarray}
We are interested in the ``off-diagonal'' contribution, which arises 
when the struck parton in the nucleon is a quark (or antiquark\footnote{However, since the treatment of the antiquark contribution is identical to that of the quark one with obvious substitutions, we will not discuss it explicitly in the following.}). 
In the $x\to 1$ limit, the off-diagonal contribution starts at next-to-leading power since either two energetic partons must be emitted into the final-state jet from the hard vertex (the ``anti-hardcollinear 
contribution"), or the struck quark must be turned into a gluon by 
emitting a soft-collinear quark into the remnant (the ``soft-collinear contribution"). We begin in this section with the anti-hardcollinear term.

The two relevant terms 
in the hard matching of the $G^2$ operator onto SCET are \cite{Beneke:2020ibj}
\begin{eqnarray}
\big[G_{\mu\nu}^A G^{\mu\nu A}\big](0) &=&\frac{1}{g_s^2}  \int dtd\bar{t}\,
\widetilde{C}_g^{\rm A0}(t,\bar{t})\,J^{A0}(t,\bar{t}\,)
\nonumber\\
&&
+\,\int dt d\bar{t}_1d\bar{t}_2\,
\widetilde{C}_g^{\rm B1}(t,\bar{t}_1,\bar{t}_2)\,
J^{B1}(t,\bar{t}_1,\bar{t}_2)\,,
\label{eq:GGhardmatching}
\end{eqnarray}
where 
\begin{eqnarray} 
&& \label{eq:A0operator} 
J^{A0}(t,\bar{t}\,) =  
2 g_{\mu\nu} \, n_- \partial {\cal A}^{\nu A}_{\,\overline{hc} \perp}(\bar t n_-)
\, n_+ \partial {\cal A}^{\mu A}_{\,hc \perp}(t n_+)\,, \\[0.1cm]
&& \label{eq:B1operator} 
J^{B1}(t,\bar{t}_1,\bar{t}_2) =  
\bar \chi_{\overline{hc}}(\bar t_1 n_-){\cal A}^{\mu 
A}_{\,\overline{hc} \perp}(\bar t_2 n_-)T^{A} \gamma_{\perp\mu} \, \chi_{hc}(t n_+)\,.
\qquad\,\,\,
\end{eqnarray}
Here $\mathcal{A}^\mu_{hc\perp} = W^\dagger_{hc} [i D_{\perp hc}^\mu W_{hc}]$ denotes the 
gauge-invariant gluon field-operator dressed with Wilson lines, which in light-cone gauge is related to the gluon field by $\mathcal{A}^\mu_{hc\perp}=
g_s A^\mu_{hc\perp}$. 
The momentum-space A0-coefficient is defined as in \eqref{eq:CA0momentumspace}, and the overall factor $1/g_s^2$ in \eqref{eq:GGhardmatching} implies the normalization $C_g^{\rm A0}(\np p_c,\nm p_{\bar c}) = 1+\mathcal{O}(\alpha_s)$. 
For the B1-coefficient, we define 
\begin{equation}
C_g^{\rm B1}(n_+p_c, r n_-p_{\bar{c}},\bar{r} n_-p_{\bar{c}}) = 
\int dtd\bar{t}_1d\bar{t}_2\,e^{-it n_+ p_c
+i (\bar{t}_1 r + \bar{t}_2 \bar{r})\nm p_{\bar c}}\,
\widetilde{C}_g^{\rm B1}(t,\bar{t}_1,\bar{t}_2)\,.
\label{eq:CB1momentumspace}
\end{equation}
The B1 operator is $\mathcal{O}(\lambda)$ suppressed relative to the A0 operator due to the additional $\mathcal{A}_{\overline{hc} \perp}$ field. For the hadronic tensor, this leads to $\mathcal{O}(\lambda^2)\sim \mathcal{O}(1-x)$ power suppression relative to the leading-power diagonal $\phi^*+g\to X$ process. The operator creates an energetic anti-hardcollinear quark and gluon, which 
 source the final-state jet. The variable $r$ ($\bar r\equiv 1-r$) 
denotes the fraction of the total anti-hardcollinear momentum $\nm p_{\overline{hc}}\equiv Q$ carried by the quark (gluon). The A0 operator cannot contribute directly to the off-diagonal channel but will become relevant for the soft-collinear term in the next section.

\begin{figure}[t]
\begin{center}
\vspace*{-0.3cm}
\includegraphics[width=0.8\textwidth]{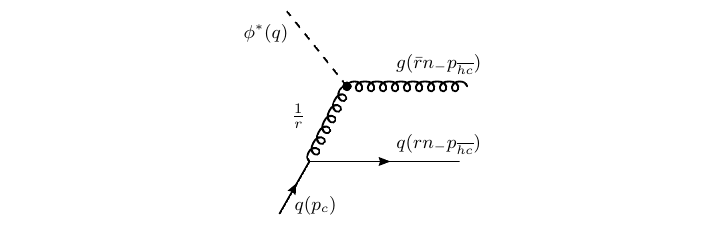}
\vspace*{-0.3cm}
\end{center}
\caption{Hard scattering of a quark off a virtual Higgs boson at
tree level results in the matching coefficient $C_g^{\rm B1}$.}
\label{fig:treeCB1}
\end{figure}


Continuing with the B1 operator, we find its tree-level matching coefficient from the diagram shown in Figure~\ref{fig:treeCB1}, 
\begin{equation}
    C_g^{\rm B1}(n_+p_c, r Q,\bar{r} Q) = -\frac{2\bar{r}}{r}  \,.
    \label{eq:CB1tree}
\end{equation}
The factor $1/r$ arises from the virtuality of the internal gluon line, as indicated in the figure, and will be important later.

The B1 matching coefficient obeys the renormalization-group equation \cite{NLPDIS2} 
\begin{align}
\frac{ d}{ d\ln\mu}C_g^{\rm B1}\klammer{n_+p_c, r Q,\bar{r} Q,\mu^2}=
\int_0^1d\hat{r} \,\Gamma(r,\hat{r}) C_g^{\rm B1}\klammer{n_+p_c, \hat{r} Q,\hat{\bar{r}} Q,\mu^2},
\label{eq:CRGEB1}
\end{align}
with the kernel 
\begin{equation}
        \Gamma(r,\hat{r})=\gamma_\text{cusp}\,\delta(r-\hat{r})\left[\frac{C_A}{2}\ln\left(\bar{r}\frac{-\tilde Q^2}{\mu^2}\right)+\left(C_F-\frac{C_A}{2}\right)\ln\left(r\frac{-\tilde Q^2}{\mu^2}\right)\right]+\dots \,,
\end{equation}
and we introduced the shorthand notation $\tilde Q^2=Qn_+p_c$. Here  $\gamma_{\rm cusp}=\alpha_s/\pi$ is the one-loop cusp anomalous dimension 
\eqref{eq:cusp}, 
and the dots refer to non-cusp terms that can be neglected when resummation is performed in the leading-logarithmic limit. 

\subsection{Factorization of the SCET matrix element}

Proceeding as at leading power by decoupling the soft-collinear modes from the anti-hardcollinear final-state jet, the NLP analogue of~\eqref{eq:A0amp} now reads
\begin{eqnarray}
\left\langle X|\big[G_{\mu\nu}^A G^{\mu\nu A}\big](0)|N(p)\right\rangle &=&    
\int dtd\bar{t}_1 d\bar{t}_2\,
\widetilde{C}_g^{\rm B1}(t,\bar{t}_1,\bar{t}_2)\,
\langle X_{\overline{hc}}|[\bar{\chi}^{(0)}_{\overline{hc}}(\bar{t}_1\nm)\slashed{\mathcal{A}}_{\overline{hc}\perp}^{(0)}(\bar{t}_2\nm)]_{a\alpha }|0\rangle
\nonumber\\
&&\hspace*{-2cm}\times\,\langle X_{sc}|[\overline{Y}_{sc,n_+}^\dagger(0)\chi_{hc}(t\np)]_{a \alpha }|N(p)\rangle\,.
\label{eq:B1amp}
\end{eqnarray}
We note that the (soft-) collinear sector is the same as in~\eqref{eq:A0amp}. This suggests introducing a generalized anti-hardcollinear composite-operator jet 
function. Let 
\begin{equation}
\mathcal{Q}_{a\alpha}(r) = 
\frac{\nm p_{\overline{hc}}}{2\pi}\int d\bar{t}\,
e^{-i r \bar{t} \nm\cdot p_{\overline{hc}}}\,
[\bar{\chi}^{(0)}_{\overline{hc}}(\bar{t}\nm)\slashed{\mathcal{A}}_{\overline{hc}\perp}^{(0)}(0)]_{a\alpha}\,,
\label{eq:Qdef}
\end{equation}
and then define, in analogy with \eqref{eq:LPjet2}, 
\begin{eqnarray}
&&\frac{1}{2\pi}
\sum_{X_{\overline{hc}}}\int \!\PS_{X_{\overline{hc}}}\,e^{-i p_{X_{\overline{hc}}}\cdot x}\,
\langle 0|\mathcal{Q}_{b\beta}^{\dagger}(r')
|X_{\overline{hc}}\rangle 
\langle X_{\overline{hc}}|\mathcal{Q}_{a\alpha}(r)|0\rangle 
\nonumber\\
&&\hspace*{0cm}=\,
\delta_{ba}\left(\frac{\slashed n_{+}}{2}\right)_{\!\beta\alpha}\,\int\!\frac{d^d k}{(2\pi)^d}\,e^{-ikx}\,\nm k
\,\mathcal{J}_{\overline{hc}}^{(qg)}(k^2,r,r')\,,
\label{eq:NLPqgjet}
\end{eqnarray}
which is a NLP object. 
Inverting \eqref{eq:Qdef}, we express \eqref{eq:B1amp} as 
\begin{eqnarray}
\left\langle X|\big[G_{\mu\nu}^A G^{\mu\nu A}\big](0)|N(p)\right\rangle &=&    
 \int \frac{d\np p_c}{2\pi} \int_0^1 dr\,C_g^{\rm B1}(\np p_c,r \nm p_{\overline{hc}},\bar{r} \nm p_{\overline{hc}})\,
\nonumber\\
&&\hspace*{-3cm}\times\,\langle X_{\overline{hc}}|\mathcal{Q}_{a\alpha}(r)|0\rangle
\int dt \,e^{i t \np p_c}\langle X_{sc}|[\overline{Y}_{sc,n_+}^\dagger(0)\chi_{hc}(t\np)]_{a \alpha }|N(p)\rangle\,.\qquad
\label{eq:B1amp2}
\end{eqnarray}

At this point, we square the matrix element and perform the sum over the anti-hardcollinear final state by introducing the composite-operator jet function 
\eqref{eq:NLPqgjet}. The hadronic tensor then reads
\begin{eqnarray}
W_\phi &=& \frac{1}{4 Q^2}
\left[\frac{\slashed n_{+}}{2}\right]_{\beta\alpha}\,\int d^d z \int \frac{d^d k}{(2\pi)^d}\,n_{-}k\,
\sum_{X_{sc}}\int \!\PS_{X_{sc}}\,e^{i (p+q-k-p_{X_{sc}})\cdot z}
\nonumber\\
&&  \hspace*{-0.7cm}\times \int_0^1 dr dr'\,C_g^{\rm B1*}(\np p,r' \nm k,\bar{r}'\nm k)
C_g^{\rm B1}(\np p,r \nm k,\bar{r}\nm k)\,
\mathcal{J}^{(qg)}_{\overline{hc}}(k^2,r,r')\,\nonumber\\[0.2cm]
&& \hspace*{-0.7cm} \times \,|C_\Lambda|^2\,
\langle N(p)| [\bar{\chi}^{(0)}_{fc} Y_{sc,n_-}^\dagger \overline{Y}_{sc,n_+}]_{a\beta}(0)|X_{sc}\rangle
\langle X_{sc}|[\overline{Y}_{sc,n_+}^\dagger Y_{sc,n_-}\chi^{(0)}_{fc}]_{a \alpha }(0)|N(p)\rangle\,.\qquad
\label{eq:Wphi2}
\end{eqnarray}
The nucleon matrix element with the soft-collinear final state in the last line of this equation is identical to the one in \eqref{eq:W2}. We therefore 
match to the field $\chi^{(0)}_{fc}$ with frozen large momentum $\np p$ using  \eqref{eq:frmatching} and introduce the soft-collinear function \eqref{eq:LPscfunction}  to obtain 
\begin{eqnarray}
W_{\phi}^{\overline{hc}} &=& \frac{1}{2 Q^2}\,
 \int_0^1 dr dr'\,C_g^{\rm B1*}(\np p,r' Q,\bar{r}' Q)
C_g^{\rm B1}(\np p,r Q,\bar{r}Q)\nonumber\\
&&\times\,
 \int d\Omega\,Q\,
\mathcal{J}^{(qg)}_{\overline{hc}}\left(Q \left(\frac{Q(1-x)}{x}-\Omega\right),r,r'\right)\, 
\frac{1}{2}\,\np p\,K_{q/N} \,|C_\Lambda|^2\,S_{sc}^{\rm LP}(\Omega)\,,\quad
\label{eq:Wphi3}\\
&=&  \frac{1}{4x}\,
 \int_0^1 dr dr'\,C_g^{\rm B1*}(\np p,r' Q,\bar{r}' Q)
C_g^{\rm B1}(\np p,r Q,\bar{r}Q)\nonumber\\
&&\times\,\int_x^1 d\xi\,
\mathcal{J}^{(qg)}_{\overline{hc}}\left(\frac{Q^2(\xi-x)}{x},r,r'\right)\, 
f_{q/N}^{\rm thr,LP}(\xi)\,,
\label{eq:Wphi3b}
\end{eqnarray}
which is the final form for the anti-hardcollinear term, given in schematic form in \eqref{eq:B1schematic}. We note the similarity to the leading-power results \eqref{eq:W4} and \eqref{eq:LPfact2}. The essential difference is that the hard interaction produces two anti-hardcollinear particles. Consequently, the final-state jet is described by a generalized jet function, weighted with the primary momentum distribution of the two energetic particles from the hard interaction,~$C_g^{\rm B1}$.

\subsection{Calculation of the composite-operator jet function}

To obtain $\mathcal{J}_{\overline{hc}}^{(qg)}(k^2,r,r')$ at the 
lowest non-vanishing order $\mathcal{O}(\alpha_s)$, we 
consider the $X_{\overline{hc}} = qg$ final-state in \eqref{eq:NLPqgjet}, resulting in 
\begin{equation}
\langle q(k_1,b)g(k_2,A)|\mathcal{Q}_{a\alpha}(r)|0\rangle = 
\nm p_{\overline{hc}}\,\delta(r \nm p_{\overline{hc}}-\nm k_1) \,g_s(\bar{\xi}_{n_-}\gamma_\perp^\mu)_{\alpha}T^A_{ba}\,\epsilon_\mu(k_2)\,,
\end{equation}
where $\xi_{\nm}$ denotes the anti-hardcollinear quark spinor and 
$p_{\overline{hc}}=k_1+k_2$. 
Squaring this expression, performing the two-particle phase-space integral, and carrying out a Fourier transformation, we obtain
\begin{equation}
\mathcal{J}_{\overline{hc}}^{(qg)}(k^2,r,r')=\frac{\alpha_{s}C_F}{4\pi}
\,\frac{d-2}{\Gamma\left(1-\epsilon\right)}\,\delta(r-r')\,r\left(\frac{\mu^{2}e^{\gamma_E}}{k^{2}r \bar{r}}\right)^{\!\epsilon}\,.
\end{equation}

Evaluating \eqref{eq:Wphi3b} with the leading-order expressions 
for $C_g^{\rm B1}$ and $\mathcal{J}_{\overline{hc}}^{(qg)}$, we encounter 
\begin{equation}
\label{eq:NLPjetLO}
\int_0^1 dr dr'\,C_g^{\rm B1*}(\np p,r' Q,\bar{r}' Q)
C_g^{\rm B1}(\np p,r Q,\bar{r}Q)\,
\mathcal{J}^{(qg)}_{\overline{hc}}(k^2,r,r') 
\propto \int_0^1 dr\, \frac{\bar{r}^{2-\epsilon}}{r^{1+\epsilon}}\,, 
\end{equation}
which diverges at $r=0$ in four dimensions. While the divergence is regulated dimensionally, its existence implies that one cannot renormalize the hard and jet functions separately, resum large logarithms by solving the renormalization-group equations of the renormalized functions, and then integrate over $r$ to obtain the resummed hadronic tensor free from large logarithms. The endpoint divergence in the above expression is precisely analogous to the corresponding divergence in the crossed process, the ``gluon thrust'' distribution in $e^+ e^-$ annihilation \cite{Beneke:2022obx}, and indicates a missing piece, which we discuss next.


\section{Next-to-leading power off-diagonal DIS: The soft-collinear term}
\label{sec:A0term}

\subsection{Hadronic tensor and hard matching}

The missing piece originates from combining the leading-power A0 operator~\eqref{eq:A0operator} in a time-ordered product with the subleading-power $\mathcal{O}(\lambda)$ SCET Lagrangian interaction that converts the struck quark into an energetic gluon and a soft-collinear quark. We start with the amplitude
\begin{eqnarray}
&&\big\langle X \big| \big[G_{\mu\nu}^A G^{\mu\nu A}\big](0)
\big| N(p) \big\rangle
= \frac{2}{g_s}\int dt d\bar{t} \,\widetilde{C}_g^{A0}(t,\bar t)\,
g^\perp_{\mu\nu}\,
\langle X_{\overline{hc}}|
\nm\partial \mathcal{A}^{(0)\nu B}_{\overline{hc}\perp}
(\bar t\nm) |0\rangle
\nonumber \\ &&\hskip1cm \times \,
\frac{i}{g_s}\int d^4 y\,\langle X_{sc}| \mbox{T}\left(
\np\partial \mathcal{A}^{\mu A}_{hc\perp}(t \np) 
\overline{\mathcal{Y}}_{sc,\np}^{\dagger BA}(0),
\mathcal{L}^{(1)}(y)\right)|N(p)\rangle\,,
\label{eq:A0termME}
\end{eqnarray}
where $\overline{\mathcal{Y}}_{sc, \np}^{\dagger BA}$ denotes the adjoint Wilson line, defined in \eqref{eq:wilsonbardef}, associated with the outgoing energetic gluon, which arises from the decoupling 
of soft-collinear gluons from the energetic gluon in the anti-hardcollinear direction. The subleading-power Lagrangian contains a hardcollinear gluon, a collinear quark, and a soft-collinear quark, and its precise form will be discussed below. 
All essential aspects 
of NLP factorization now concern the collinear sector 
and arise from the matrix element 
\begin{eqnarray}
\mathcal{M}_{sc}^{\mu_\perp B}(t)\equiv 
\frac{i}{g_s}\int d^4 y\,\langle X_{sc}| \,\mbox{T}\left(
\np\partial \mathcal{A}^{\mu A}_{hc\perp}(t \np) 
\overline{\mathcal{Y}}_{sc,\np}^{\dagger BA}(0), 
\mathcal{L}^{(1)}(y)\right)|N(p)\rangle
\label{eq:msc}
\end{eqnarray}
of the time-ordered product in the second line.

After squaring the matrix element \eqref{eq:A0termME} and summing over the anti-hardcollinear final state, 
the anti-hardcollinear part of the hadronic tensor is expressed in terms of the leading-power gluon jet function
\begin{eqnarray}
&&\frac{1}{2\pi}\frac{1}{g_{s}^2}
\sum_{X_{\overline{hc}}}
\int \!\PS_{X_{\overline{hc}}}\,
\langle 0|\mathcal{A}_{\overline{hc}\perp}^{(0)\nu B}(x)|X_{\overline{hc}}\rangle 
\langle X_{\overline{hc}}|\mathcal{A}_{\overline{hc}\perp}^{(0)\nu' C}(0)|0\rangle 
\nonumber\\
&& \equiv\,\delta^{BC}\,(-g^{\nu\nu'}_\perp)
\int\!\frac{d^{d}k}{(2\pi)^d}\,e^{-ikx}
\mathcal{J}^{(g)}_{\overline{hc}}(k^2).
\label{eq:LPgluonjet}
\end{eqnarray}
At LO, we have $\mathcal{J}^{(g)}_{\overline{hc}}(k^2)=\delta^+(k^2)$. 
The hadronic tensor then reads 
\begin{eqnarray}
W_\phi &=& \frac{4\cdot 2\pi}{8\pi Q^2}\int \!d^d z \int \!\frac{d^d k}{(2\pi)^d}\,(n_{-}k)^2\,
\mathcal{J}^{(g)}_{\overline{hc}}(k^2)\,
\sum_{X_{sc}}\int \!\PS_{X_{sc}}\,e^{i (p+q-k-p_{X_{sc}})\cdot z}
\nonumber\\
&& \times\int dtd\bar{t}\,e^{i \bar{t} \nm k}\,
\widetilde{C}_g^{\rm A0}(t,\bar{t})\,
\int dsd\bar{s}\,e^{-i \bar{s} \nm k}\,
[\widetilde{C}_g^{\rm A0}(s,\bar{s})]^*
\nonumber\\[0.2cm]
&& \times
\,(-1)\, [\mathcal{M}_{sc,\mu_\perp}^{B }(s)]^*\,\mathcal{M}_{sc}^{\mu_\perp B}(t)\,.
\label{eq:WphiA1}
\end{eqnarray}
The analysis in the following subsection will demonstrate that the typical virtualities in the object $\mathcal{M}_{sc}^{\mu_\perp A}(t)$ are at most of order $\Lambda_{\rm QCD}^2/(1-x)$, and that the $t$-dependence is 
$e^{-it\np p} \mathcal{M}_{sc}^{\mu_\perp A}(0)$. Assuming this, we can proceed with the remaining integrations as for \eqref{eq:W1}, and by performing similar steps that lead to \eqref{eq:W4}, we obtain 
\begin{eqnarray}
W_\phi^{sc} &=& 
\left|C_g^{\rm A0}(\np p,Q)\right|^2
\int d\Omega\,
\mathcal{J}^{(g)}_{\overline{hc}}\left(Q \left(\frac{Q(1-x)}{x}-\Omega\right)\right)\, 
\mathcal{S}_{sc}^{\rm NLP}(\Omega)\,,
\label{eq:WphiA2}
\end{eqnarray}
where 
\begin{eqnarray}
\mathcal{S}_{sc}^{\rm NLP}(\Omega)&=&
\int \frac{du}{2\pi} \,e^{i u\Omega}\,\sum_{X_{sc}}\int \!\PS_{X_{sc}}\,e^{-i \np\cdot p_{X_{sc}} u}\,
(-1)\, [\mathcal{M}_{sc, \mu_\perp}^{B}(0)]^*\,\mathcal{M}_{sc}^{\mu_\perp B}(0)\,.\quad
\label{eq:NLPscfunction}
\end{eqnarray}

\subsection{Analysis of the soft-collinear matrix element}
\label{sec:scME}

\subsubsection{Subleading-power Lagrangian and power counting}

\begin{figure}[t]
\begin{center}
  \includegraphics[width=0.8\textwidth]{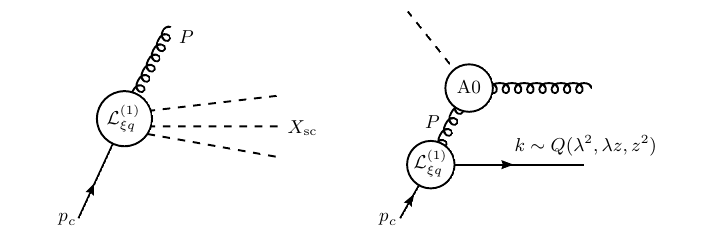}
\end{center}
\caption{(left) Graphical representation of the soft-collinear amplitude $\mathcal{M}_{sc}^{\mu_\perp B}(t)$ \eqref{eq:msc}, and (right) soft-collinear momentum scaling at the $\mathcal{L}^{(1)}$ vertex.}
\label{fig:NLPscamplitude}
\end{figure}

The leading interaction involving a {\em soft} quark and (hard-) collinear modes 
is 
$\mathcal{O}(\lambda)$ suppressed and reads \cite{Beneke:2002ph,Beneke:2002ni}
\begin{equation}
\mathcal{L}^{(1)}_{\xi q}(y)=\bar{q}_s(y_{-})\slashed{\mathcal{A}}_{hc\perp}(y)\chi_{hc}(y)+{\rm h.c.}
=[\bar{q}_s Y_{s,\nm}](y_{-})
\frac{\nms\nps}{4}\,\slashed{\mathcal{A}}^{(0)}_{hc\perp}(y)\chi^{(0)}_{hc}(y)+\rm{h.c.}\,,
\label{eq:Lxiq}
\end{equation} 
where the second equality follows from $\nms \chi_{hc}=0$.\footnote{In this general expression, the $hc$ field still contains the collinear mode. In a tree-level interaction such as the one shown in Fig.~\ref{fig:NLPscamplitude}, the external $\chi$ field is collinear, and the vertex is \eqref{eq:Lxiq} with $\chi_{hc}(y)\to \chi_{c}(y_+)$. However, when the soft-collinear interaction vertex is dressed by hardcollinear loops, the $\chi$-field at the interaction vertex can also be hardcollinear.} 
This power-suppressed interaction arises in SCET$_{\rm II}$ problems with soft modes and collinear modes of different virtuality
\begin{equation}
\mbox{(c)} \hspace*{0.4cm} Q'(1,\lambda^2,\lambda^4),\qquad
\mbox{(s)} \hspace*{0.4cm} Q'(\lambda^2,\lambda^2,\lambda^2),\qquad
\mbox{(hc)} \hspace*{0.4cm} Q'(1,\lambda,\lambda^2)\,.
\label{eq:SCETIIvertex}
\end{equation}

To obtain the corresponding interaction for the kinematics of $x\to 1$ DIS, we boost the frame 
by $\np l\to \kappa \np l$, $\nm l\to \kappa^{-1}\nm l$, leaving the transverse components unchanged. Choosing $\kappa=\lambda^2/(\xi\beta)$ and setting $Q=Q' \lambda^2/(\xi\beta)$, the modes above are transformed into 
\begin{equation}
\mbox{(c)} \hspace*{0.4cm} Q(1,\xi\beta,(\xi\beta)^2),\qquad
\mbox{(sc)} \hspace*{0.4cm} Q(\lambda^2,\xi\beta,(\xi\beta)^2/\lambda^2),\qquad
\mbox{(hc)} \hspace*{0.4cm} Q(1,\xi\beta/\lambda,(\xi\beta/\lambda)^2)\,,
\label{eq:L1scvertex}
\end{equation}
which for $\beta\sim 1$ conforms with the scaling of the collinear, soft-collinear and 
hardcollinear modes in \eqref{eq:modes}, \eqref{eq:hcmode}, respectively. Note that collinear and soft-collinear modes have the same virtuality. After the boost, ${q}_s$ becomes a soft-collinear field, and $\frac{\nms\nps}{4}$ projects onto the 
small component \cite{Beneke:2002ph}
\begin{equation}
\eta_{sc} = -\frac{\nps}{2}\frac{1}{i\np D_{sc}} i\slashed{D}_{{\perp, sc}} \xi_{sc}
\label{eq:etacomponent}
\end{equation}
of the soft-collinear spinor. The Lagrangian that describes the conversion of the struck collinear quark into a soft-collinear quark and a hardcollinear gluon is therefore obtained from \eqref{eq:Lxiq} as 
\begin{equation}
\mathcal{L}^{(1)}(y)=[\bar{\eta}_{sc} Y_{sc,\nm}](y_{-})
\,\slashed{\mathcal{A}}^{(0)}_{hc\perp}(y)\chi^{(0)}_{c}(y_+)+\rm{h.c.}\,.
\label{eq:Lxietasc}
\end{equation} 
We note that $[\bar{\eta}_{sc} Y_{sc,\nm}](y_{-})$ remains  invariant under soft-collinear gauge trans\-form\-ations $U_{sc}(y)$, since $\eta_{sc}(y_-)$ and $Y_{sc,\nm}(y_-)$  both transform with the same $U_{sc}(y_-)$.

As an aside, we note that by performing a boost with $\kappa\ll 1$ in the opposite direction, one obtains instead
\begin{equation}
\mathcal{L}^{(1)}_{\xi q}(y)\quad\to\quad 
\bar{\chi}_{\overline{hc}}(y_-) \gamma^\mu_\perp T^A \chi_{hc}(y_+)\,\mathcal{A}_{\perp h}^{\mu A}(y)\,,
\label{eq:hardLagrangian}
\end{equation}
with
\begin{equation}
\mbox{(c)} \hspace*{0.4cm} Q(1,\lambda,\lambda^2),\qquad
\mbox{($\overline{\rm hc}$)} \hspace*{0.4cm} Q(\lambda^2,\lambda,1),\qquad
\mbox{(h)} \hspace*{0.4cm} Q(1,1,1)\,,   
\end{equation}
where in this case $Q=Q'\lambda$. This is nothing but the back-to-back two-jet operator 
\eqref{eq:vectortoA0}, generated by a hard colour-octet source represented by 
$\mathcal{A}_{\perp h}^{\mu A}(y)$.

Let us parameterize the momentum of the outgoing quark as 
\begin{equation}
    k\sim Q (\lambda^2,\lambda z,z^2)\,,
    \label{eq:kz}
\end{equation}
as shown in Fig.~\ref{fig:NLPscamplitude}. Then, as $z$ increases from 
$\xi/\lambda$ through $\lambda$ to 1, $k$ transitions from soft-collinear to soft to anti-hardcollinear. Thus, varying $z$ smoothly interpolates between the remnant scaling and that of the anti-hardcollinear jet. At the same time, the hardcollinear gluon with momentum $P=p_c-k$, which connects the $\mathcal{L}^{(1)}$ vertex to the hard A0-vertex, is off-shell by an 
amount $\np p_c \nm k\sim Q^2 z^2$ increasing from $Q^2 (\xi/\lambda)^2 \sim \Lambda_{\rm QCD}^2/(1-x)$ 
to $Q^2(1-x)$ to $Q^2$. This corresponds to the three cases represented by 
the Lagrangians \eqref{eq:Lxietasc}, \eqref{eq:Lxiq},  \eqref{eq:hardLagrangian} above. In the latter case, the hard gluon should be integrated out in matching to SCET in the first place, resulting in the B1 operator that sources the anti-hardcollinear jet at NLP.

Before proceeding with the calculation of the time-ordered product of 
$\mathcal{L}^{(1)}_{\xi q}$, we discuss the power counting for the soft-collinear contribution and demonstrate that it is of the same NLP order 
$\mathcal{O}(\lambda^2)$ as the anti-hardcollinear B1 term. All three fields 
in the version \eqref{eq:Lxietasc} of $\mathcal{L}^{(1)}_{\xi q}$ relevant to DIS are of the collinear type. For such modes, the large component of the on-shell spinor scales as $\chi\sim \xi\sim p_\perp \sqrt{\np p} \sim \np p \sqrt{\nm p}$, where $p$ is the typical mode momentum. For the small spinor component, 
$\eta \sim p_\perp^2/\sqrt{\np p}\sim \nm p\sqrt{\np p}$ from \eqref{eq:etacomponent}. 
The hardcollinear transverse gluon field with virtuality of order $P^2$ counts as  
$\sqrt{P^2}\sim P_\perp$, and since the space-time measure $d^4y$ is dominated by the shortest fluctuations, we have $d^4y\sim 1/P^2$. Adopting the scaling 
\eqref{eq:L1scvertex}, we obtain
\begin{equation}
\int d^4y \,\mathcal{L}^{(1)}_{\xi q} \sim \left(\frac{\lambda}{\xi \beta}\right)^4 \times \frac{(\xi\beta)^2}{\lambda} \times \frac{\xi\beta}{\lambda}\times \xi\beta =  \lambda^2\,,
\end{equation}
with the four factors referring to $d^4y$ and the three (sc, hc, c) fields in 
\eqref{eq:Lxietasc}. This scaling corresponds to $z= \xi\beta/\lambda$ in \eqref{eq:kz}. For general $z$, 
\begin{equation}
\int d^4y \,\mathcal{L}^{(1)}_{\xi q} \sim \left(\frac{1}{z}\right)^4 \times \lambda z^2 \times {z}\times \lambda z =  \lambda^2    
\end{equation}
gives the same power suppression as it should. 

\subsubsection{Computation of the matrix element}

We now return to the computation of \eqref{eq:msc}.
Dressing the $\mathcal{L}^{(1)}$ vertex with loops of hardcollinear virtuality $P^2 = \np p_c \nm k \sim Q^2 z^2$ and integrating out the hardcollinear modes gives rise to amplitude-level ``collinear functions'' similar to those in the Drell-Yan process near threshold \cite{Beneke:2018gvs,Beneke:2019oqx}. The operator definition of the matching coefficient $\tilde J$ in position-space reads \cite{Beneke:2020ibj}
\begin{eqnarray}
&& \frac{i}{g_s}\int d^4 y\,\mbox{T}\left(
\np\partial\mathcal{A}^{(0)\mu C}_{hc\perp}(t \np), 
\mathcal{L}^{(1)}(y)\right) = 
2\pi g_s\int du\int d\left(\frac{\np y}{2}\right)\,
\tilde{J}\left(t,u;\frac{\np y}{2}\right) \,
\nonumber \\ &&\hskip1cm \times \,
\left[\bar{\eta}_{sc} Y_{sc,\nm}\right]\!(\np y\frac{\nm}{2})\,T^C \gamma^{\mu_\perp}
\chi^{(0)}_c(u\np)\,.
\label{eq:tproduct}
\end{eqnarray}

The equation is formulated in terms of the decoupled (incoming) hardcollinear gluon field rather than $\mathcal{A}^{\mu A}_{hc\perp}(t \np) = \mathcal{Y}_{sc,\nm}^{AC}(0)\mathcal{A}^{(0)\mu C}_{hc\perp}(t \np) $, which appears in \eqref{eq:msc}. Since 
$\left[\bar{\eta}_{sc} Y_{sc,\nm}\right]$ and $\chi^{(0)}_c(u\np)$ are themselves gauge-invariant, this defines $\tilde J$ in a manifestly gauge-invariant way. Denoting
\begin{equation} 
v\equiv \frac{\np y}{2}\,
\end{equation}
we find 
\begin{eqnarray}
\mathcal{M}^{\mu_\perp B}_{sc}(t) &=& 
2\pi g_s\int du\int dv\,\tilde{J}(t,u;v) \,
\nonumber \\ && 
\hspace*{-1.5cm}\times \,\langle X_{sc}| \,\mbox{T}\left(
 \overline{\mathcal{Y}}_{sc,\np}^{\dagger BA}(0)\mathcal{Y}_{sc,\nm}^{AC}(0)
\left[\bar{\eta}_{sc} Y_{sc,\nm}\right]\!(v\nm)T^C \gamma^{\mu_\perp}
\chi_c^{(0)}(u\np)\right)|N(p)\rangle\,.\quad
\label{eq:msc2}
\end{eqnarray}
The momentum-space hardcollinear function $J$ is defined in terms of $\tilde{J}$ as 
\begin{eqnarray}
\int dt\int du \,  \tilde{J}(t,u; v)
\, e^{i (\np p_{hc} )\,t}\,e^{-i(\np p_c)\,u}=
\int \frac{d\omega}{2\pi}\,  e^{-i \omega v} \, 
J(\np p_{hc},\np p_c; \omega)\,,
\label{eq:JFT}
\end{eqnarray}
where, in fact, 
\begin{equation}
J(\np p_{hc},\np p_c; \omega) = \delta(\np p_{hc}-\np p_c) \,
\frac{-i \np p_c}{p^2}\, D_g^{\rm B1}(p^2)\,,\quad p^2=-\np p_c\omega+i0^+
\label{eq:DB1definition}
\end{equation}
by momentum conservation, since the soft-collinear line does not change 
the $\mathcal{O}(1)$ $\np p$-components of the (hard) collinear momenta.
The function $D_g^{\rm B1}(p^2)$ as well as its cusp anomalous dimension were already introduced in the context of DIS in 
\cite{Beneke:2020ibj}. The corresponding coefficient $D_q^{\rm B1}(p^2)$, obtained by interchanging the external quark and off-shell gluon with an external gluon and off-shell quark, respectively,
appears as a hardcollinear function in the factorization theorem for Higgs production in 
gluon fusion through light-quark loops \cite{Liu:2021mac,Liu:2022ajh}, as well as in the refactorization of the gluon contribution to the thrust 
event shape in $e^+ e^-$ annihilation \cite{Beneke:2022obx}.\footnote{In the literature, these collinear matching coefficients are sometimes called ``radiative jet functions". A misnomer, since the coefficients encapsulate purely virtual effects contrary to true inclusive final-state jet functions such as in \eqref{eq:LPjet2}, \eqref{eq:NLPqgjet}.}

The momentum-space matching coefficient is determined by sandwiching the operator definition \eqref{eq:tproduct} between an on-shell outgoing soft-collinear and an ingoing collinear quark state, $\langle q_{sc}(l)|\ldots |q(p_c)\rangle$, identifying $\omega=\nm l$. 
At leading order in the strong coupling, one finds $D_g^{\rm B1}(p^2)=1+\mathcal{O}(\as)$. Inverting the Fourier transform results in  
\begin{equation}
 \tilde{J}(t,u; v) = \int\frac{d \np p_c}{2\pi}  \,e^{-i \np p_c (t-u)}  \int \frac{d\omega}{2\pi}\,  e^{-i \omega v} \, \frac{i}{2\pi \omega} D_g^{\rm B1}(-\np p_c\omega)\,.
 \label{eq:JintermsofDB1}
\end{equation}

We match the collinear field in the nucleon matrix element \eqref{eq:msc2} to the frozen-collinear mode $\chi_c^{(0)}(u\np) = e^{-i (\np p) u} \,C_\Lambda \,\chi_{fc}^{(0)} = e^{-i (\np p) u} \chi_c^{(0)}(0)$ equivalently to 
\eqref{eq:frmatching}, which allows us to extract the $u$-dependence and perform 
the corresponding integration to obtain 
\begin{eqnarray}
\mathcal{M}^{\mu_\perp B}_{sc}(t) &=& 
 i g_s\,e^{-i \np p t} \int \frac{d\omega}{2\pi}\ \frac{1}{\omega}\,D_g^{\rm B1}(-\np p\omega)
\int dv\,e^{-i \omega v} \,
\nonumber \\ && 
\hspace*{-1.5cm}\times \,\langle X_{sc}| \,\mbox{T}\left(\overline{\mathcal{Y}}_{sc, \np}^{\dagger BA}(0)\mathcal{Y}_{sc,\nm}^{AC}(0)
\left[\bar{\eta}_{sc} Y_{sc,\nm}\right]\!(v\nm)T^C \gamma^{\mu_\perp}
\chi_c^{(0)}(0)\right)|N(p)\rangle\,.\quad
\label{eq:msc4}
\end{eqnarray}
This demonstrates the $t$-dependence of the form $e^{-i \np p t}$ stated below \eqref{eq:WphiA1}, and identifies 
$\mathcal{M}^{\mu_\perp B}_{sc}(0)$, which enters \eqref{eq:NLPscfunction}.

To rewrite the Wilson lines in the $\nm$ direction in a more 
intuitive form, we insert the identity $1=[Y^\dagger _{sc,\nm}Y_{sc,\nm}](v\nm)$ immediately to the right of $T^C$ in the second line of \eqref{eq:msc4}. We then employ the hermitian conjugate 
of the first identity of \eqref{eq:wilsonfadrel}
and the definition 
\begin{equation}
\left[t_1 n,t_2 n\right]^{AB}\equiv \mathcal{Y}^{AD}_n(t_1 n)\mathcal{Y}^{\dagger DB}_n(t_2 n) = \mathcal{P} \exp\left[i g_s T^C_{\rm ad} \int_{t_2}^{t_1} ds \,n\cdot A^C(sn)\right]
\end{equation}
of the (adjoint) finite-distance Wilson line to rearrange
\begin{equation}
\mathcal{Y}_{\nm}^{AC}(0)
\left[\bar{\eta}_{sc} Y_{sc,\nm}\right]\!(v\nm)T^C 
= \left[0,v\nm\right]^{AC}_{sc,\nm} \bar{\eta}_{sc}(v\nm) T^C Y_{sc,\nm}(v\nm) \,.
\label{eq:wilsonrewrite}
\end{equation}
Reading this expression from right to left, the Wilson lines now describe the emission of soft-collinear gluons 
from an energetic colour triplet (quark) travelling from $-\infty$ to $v\nm$. At this point the quark converts into an outgoing soft-collinear quark and an energetic colour-octet (gluon) that moves from $v\nm$ to the hard interaction point at 0, emitting soft-collinear gluons.

\begin{figure}
\centering
\includegraphics[width=0.7\textwidth]{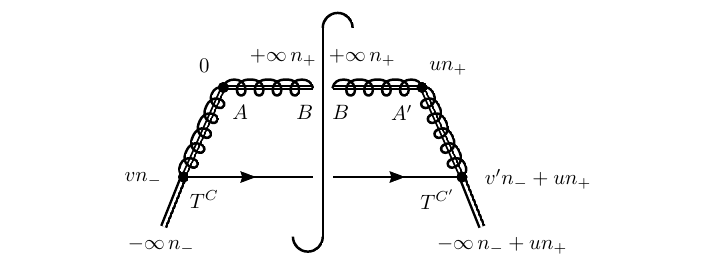}
\centering
\caption{Pictorial representation of the soft-collinear operator \eqref{eq:scWilsonlineamp} and its hermitian conjugate entering the soft-collinear function \eqref{eq:diffNLPscfunction}, showing the Wilson lines together with their direction, position and colour dependence. Double lines indicate Wilson lines in the fundamental representation, curly double lines indicate Wilson lines in the adjoint representation. The solid line stands for the soft-collinear quark field.}
\label{fig:softfunc}
\end{figure}

Evaluating \eqref{eq:NLPscfunction} to obtain the NLP soft-collinear function now follows the same steps as at leading power. The sum over the soft-collinear final state is performed by translating the position of the conjugate amplitude $\mathcal{M}^*$, and the matrix element factorizes into the nucleon matrix element of the frozen-collinear field and a  vacuum correlator of soft-collinear Wilson lines and a soft-collinear quark field. For the latter, it is convenient to abbreviate 
\begin{eqnarray}
\mathcal{O}^B_{sc,\alpha a}(v\nm,0) &=& 
\mbox{T}\left(\overline{\mathcal{Y}}_{sc,\np}^{\dagger BA}(0)
[0,v \nm]_{sc,\nm}^{AC}\,
\left[\bar{\eta}_\alpha(v\nm) T^C Y_{sc,\nm}(v\nm)\right]_a\,\right),
\label{eq:scWilsonlineamp}
\end{eqnarray}
which carries a (transposed) Dirac spinor and a fundamental as well as adjoint colour index. We define the momentum-space NLP soft-collinear two-point function of this operator as 
\begin{eqnarray}
S_{sc}^{\rm NLP}(\Omega,\omega,\omega')&=& (d-2)\,\frac{g_s^2}{N_c} 
\int \frac{du}{2\pi} \,e^{i u\Omega}\int \frac{dv}{2\pi}\,e^{-i\omega v}\int \frac{dv'}{2\pi}\,e^{i \omega' v'}
\nonumber\\
&&\times\,\langle 0 | \,\mbox{tr}\left(\frac{\slashed n_{-}}{2}\,\mathcal{O}^{\dagger B}_{sc}(u\np+v'\nm,u\np)\mathcal{O}^{B}_{sc}(v\nm,0)\right)|0\rangle\,.\quad
\label{eq:diffNLPscfunction}
\end{eqnarray}
Here, in the hermitian conjugate of $\mathcal{O}^B$ all field positions are translated by $u\np$ relative to \eqref{eq:scWilsonlineamp}. Also, it  
is understood that in the hermitian conjugate, the 
time-ordering is replaced by anti-time ordering, and the trace is over the Dirac and fundamental colour indices. The field content of the soft-collinear  operator is shown in Fig.~\ref{fig:softfunc}.

Proceeding as described, we arrive at the expression
\begin{eqnarray}
\mathcal{S}_{sc}^{\rm NLP}(\Omega)&=& 
|C_\Lambda|^2\, \frac{1}{2}\,\np p\, K_{q/N} \nonumber\\
&&\times\,
\int_0^\infty \frac{d\omega}{\omega} \frac{d\omega'}{\omega'}\,D_g^{\rm B1}(-\np p\omega) \,D_g^{{\rm B1}*}(-\np p\omega')\,S_{sc}^{\rm NLP}(\Omega,\omega,\omega')\,\quad
\label{eq:NLPscfunctionfinal}
\end{eqnarray}
for \eqref{eq:NLPscfunction}. Note that while $-\np p\omega$ has the $+i0^+$ prescription, $-\np p\omega^\prime$ should be interpreted as $-\np p \omega^\prime -i0^+$ due to complex conjugation. 
We emphasize that the left-hand side is a multi-scale 
object. It has been factorized above into a convolution of perturbative collinear functions at amplitude level associated with hardcollinear virtuality $\Lambda^2_{\rm QCD}/(1-x)$, an inclusive non-perturbative soft-collinear function 
(virtuality $\Lambda_{\rm QCD}^2$), and a non-perturbative collinear function at amplitude level, $C_{\Lambda}$, related to the struck quark.

\subsubsection{Soft rearrangement}

The definitions introduced so far are suitable for computing the bare, dimensionally regulated factorized hadronic tensors at fixed order. However, when evaluated with partonic states, the pole parts of the bare objects mix UV and IR divergences and therefore do not allow us to read off the anomalous dimension of the NLP functions and resum large logarithmic corrections. Worse, the anomalous dimensions of the collinear and soft-collinear functions are not separately well-defined: when one introduces a suitable non-dimensional IR regulator, 
the anomalous dimensions are not separately IR finite. The problem is remedied by performing a soft subtraction / rearrangement between soft and collinear functions (reminiscent of the eikonal subtraction in  \cite{Akhoury:1998gs}). This procedure was first applied in \cite{Beneke:2019slt,Beneke:2022msp} to the QED anomalous dimension of the B-meson decay soft function and subsequently to various NLP soft function renormalization kernels \cite{Beneke:2024cpq,Beneke:2025ufd}. 

In the present case, the soft(-collinear) rearrangement of the NLP soft-collinear function 
(in position space with respect to the first argument) amounts to dividing this function by the leading power soft-collinear function \eqref{eq:LPscfunction}, which defines the subtracted soft-collinear function 
\begin{align}
    S_{\rm sc, sub}^{\rm NLP}(n_-z,\omega,\omega') =  \frac{ S_{ sc}^{\rm NLP}(n_-z,\omega,\omega')}{S_{sc}^{\rm LP}(n_-z)} \,.
    \label{eq:SCsub}
\end{align}
At LO in $\alpha_s$,  $S_{sc}^{\rm LP}(n_-z) =1$, hence $S_{\rm sc, sub}^{\rm NLP,\, LO}(n_-z,\omega,\omega')=S_{\rm sc}^{\rm NLP, \, LO}(n_-z,\omega,\omega')$.
Combining \eqref{eq:NLPscfunctionfinal} with \eqref{eq:LPPDFfact} for the leading-power quark-PDF allows us to express $\mathcal{S}_{sc}^{\rm NLP}(n_-z)$
in the form of the convolution 
\begin{eqnarray}
\label{eq:factorizationSSC}
\mathcal{S}_{sc}^{\rm NLP}(n_-z)&=&\frac{1}{2}\int_0^\infty \frac{d\omega}{\omega} \frac{d\omega'}{\omega'}\,D_g^{\rm B1}(-n_+ p\omega) \,D_g^{B1\ast}(-n_+ p\omega') 
\notag
\\
&\times&\,S_{\rm sc, sub}^{\rm NLP}(n_-z,\omega,\omega')
\, f^{\rm LP}_{q/N}(n_-z)
\end{eqnarray}
of NLP functions with the leading-power PDF. 
For the expression \eqref{eq:NLPscfunctionfinal} in terms of the momentum-space variable $\Omega$ conjugate to $n_- z$, the above factorization formula reads
\begin{align}
\mathcal{S}_{\rm sc}^{\rm NLP}(\Omega) 
&= \frac{1}{2} \int_0^\infty \frac{d\omega}{\omega} \frac{d\omega'}{\omega'} \, 
D_g^{\rm B1}(-n_+ p\omega) \, D_g^{\rm B1\ast}(-n_+ p\omega') \notag \\
&\quad \times \int d\Omega' \, S_{\rm sc, sub}^{\rm NLP}(\Omega',\omega,\omega') \, 
f^{\rm LP}_{q/N}(1-\frac{\Omega - \Omega'}{\np p}) \,,
\label{eq:NLPscfunctionfinalsub}
\end{align}
which makes explicit that the NLP soft-collinear function is given by a convolution 
of the subtracted soft function with the leading-power PDF in the variable $\Omega$.

Unlike the original factors 
$|C_\Lambda|^2$ and $S_{\rm sc}^{\rm NLP}$ appearing in \eqref{eq:NLPscfunctionfinal}, after this rearrangement, the new soft-collinear and collinear factors $S_{\rm sc, sub}^{\rm NLP}$ and $f^{\rm LP}_{q/N}$, respectively, have well-defined anomalous dimensions separately.  The $\msbar$ UV renormalization of the LP PDF has been discussed before  in Sec.~\ref{sec:quarkPDFfactLP}, and its anomalous dimension can be read off from 
\eqref{eq:PDFrhs}. To obtain the renormalization of the subtracted NLP soft-collinear function, we follow the approach of \cite{Beneke:2025ufd}, using the $\delta$ IR regulator \cite{Beneke:2019slt} for the soft-collinear Wilson lines. The details of the computation will be given in~\cite{NLPDIS2}. 
The subtracted NLP soft-collinear function obeys the RGE
\begin{equation}
    \frac{d}{d\ln\mu}S^\text{NLP}_\text{sc,sub}(\Omega,\omega,\omega')=-\int d\hat{\Omega}\int d\hat{\omega}\int d\hat{\omega}' \gamma^\text{NLP}_\text{sc}(\Omega,\omega,\omega';\hat{\Omega},\hat{\omega},\hat{\omega}^\prime)\,S^\text{NLP}_\text{sc,sub}(\hat{\Omega},\hat{\omega},\hat{\omega}')\,.
    \label{eq:SNLPscRGE}
\end{equation}
For the purpose of this work, only the cusp part of the one-loop anomalous dimension kernel is needed, which is given by
\begin{eqnarray}
    \gamma_\text{sc}^\text{NLP}(\Omega,\omega,\omega';\hat{\Omega},\hat{\omega},\hat{\omega}^\prime)&=&(C_F-C_A)\,\gamma_\text{cusp}(\alpha_s)\left[\ln\frac{-\omega\Omega+i0^+}{\mu^2}+\ln\frac{-\omega'\Omega-i0^+}{\mu^2}\right]\nonumber\\
    &&\times\,\delta(\Omega-\hat{\Omega})\delta(\omega-\hat{\omega})\delta(\omega'-\hat{\omega}')\,
    \label{eq:gammasc}
\end{eqnarray}
with $\gamma_\text{cusp}(\alpha_s)$ defined in \eqref{eq:cusp}.

\subsubsection{NLP threshold factorization of the bare gluon PDF}

We tentatively identify $\mathcal{S}_{sc}^{\rm NLP}(\Omega)$ with the NLP term $f^{\rm thr, NLP}_{g/N}(x)$ in the $x\to 1$ asymptotic behaviour of the {\em gluon} PDF arising from off-diagonal parton-scattering in the nucleon. 
By adapting the leading-power factorization formula \eqref{eq:LPfact2} to the gluon PDF and to the process $\phi^*+g\to X$, we can write
\begin{equation}
W_\phi =
\left|C_g^{\rm A0}(\np p,Q)\right|^2\,\frac{Q^2}{x^2}
 \int_x^1 d\xi\,
\mathcal{J}^{(g)}_{\overline{hc}}\left(Q^2\frac{\xi-x}{x}\right)\, 
f_{g/N}(\xi)\,.
\label{eq:WLPgluon}
\end{equation}
We then expand the PDF in a power series
\begin{equation}
f_{g/N}(\xi)\stackrel{\xi\to 1}{=} f^{\rm thr, LP}_{g/N}(\xi) + 
f^{\rm thr, NLP}_{g/N}(\xi) +\ldots
\label{eq:fthrexpansion}
\end{equation}
near the threshold. Comparing to \eqref{eq:WphiA2}, we deduce that 
\begin{equation}
\label{eq:536}
\np p\,\mathcal{S}_{sc}^{\rm NLP}(\Omega) = \frac{Q^2}{x^2}\,f_{g/N}^{\rm thr, NLP}(\xi)
\end{equation}
with $\xi=1-\frac{\Omega}{\np p}$ for the off-diagonal contribution to the NLP gluon PDF.

In the following, we derive this relation directly from the operator definition of the gluon PDF. The nucleon-spin averaged gluon distribution function in the nucleon is defined in SCET as\footnote{And agrees with the standard definition in QCD\cite{Collins:1981uw} as shown 
in Appendix~A of \cite{Beneke:2010da}. 
} 
\begin{equation}
f_{g/N}(\xi) = -\frac{\xi \np p}{2 \pi g_s^2}
  \int dt \,e^{-i \xi t \np p} \,\langle N(p)|
  \mathcal{A}_{c\perp}^{\mu A}(t n_+)
  \mathcal{A}^{A}_{c\perp,\mu}(0)|N(p)\rangle\,.
\end{equation}
At leading power, the factorization of the gluon PDF can be derived in a manner identical to that of the quark PDF in Sec.~\ref{sec:quarkPDFfactLP}. In the resulting expression, the quantities $K_{q/N}$, $C_\Lambda$, and $S_{\rm sc}^{LP}$ shall be replaced by their corresponding matrix elements relevant to gluons. 

Here, we are interested in the next term in the expansion near $\xi=1$, which involves time-ordered products of subleading-power soft-collinear interactions in the SCET Lagrangian, similar to the soft-collinear contribution to DIS factorization discussed above. However, there is no analogue of the anti-hardcollinear term. We distinguish 
two types of terms: those with soft-collinear gluon fields, which contribute to the diagonal gluon-scattering channel, and those with soft-collinear quark fields, which are related to the off-diagonal quark-scattering channel. We focus on the latter, which is relevant to NLP factorization of the quark contribution to the $\phi^*+N\to X$ DIS process.

As before, we examine the $\xi\to 1$ limit. In analogy with \eqref{eq:PDF2}, we find 
\begin{eqnarray}
f_{g/N}(\xi) &\stackrel{\xi\to 1}{=}&
-\frac{\np p}{g_s^2} \sum_{X_{sc}}\int \!\PS_{X_{sc}}\!\int \frac{dt}{2\pi}\, e^{i (1-\xi) t  \np p-i t \np p_{X_{sc}}}\,\langle N(p)|  \mathcal{A}_{c\perp}^{\mu A}(0)|X_{sc}\rangle\, \nonumber\\
&&\times\,
\langle X_{sc}|
\mathcal{A}^{A}_{c\perp,\mu}(0)|N(p)\rangle\,.
\end{eqnarray}
To identify the off-diagonal contribution, we match the collinear gluon field entering the PDF operator onto the frozen-collinear quark field. The matching equation at NLP accuracy reads
\begin{eqnarray}
&&\frac{1}{g_s} \langle X_{sc}|
\mathcal{A}^{\mu, A}_{c\perp}(t n_+)|N(p)\rangle\ =
 i g_s\,e^{-i \np p t} \int \frac{d\omega}{2\pi}\ \frac{1}{n_+ p \omega}\,D_g^{\rm B1}(-\np p\omega)
\int dv\,e^{-i \omega v} \,
\nonumber \\ && 
\hspace*{.0cm}\times \,\langle X_{sc}| \,\mbox{T}\left(\overline{\mathcal{Y}}_{sc, \np}^{\dagger BA}(0)\mathcal{Y}_{sc,\nm}^{AC}(0)
\left[\bar{\eta}_{sc} Y_{sc,\nm}\right]\!(v\nm)T^C \gamma^{\mu_\perp}
\chi_c^{(0)}(0)\right)|N(p)\rangle\, + \ldots\,,\qquad
\label{eq:PDFA0term}
\end{eqnarray}
where the dots refer to terms that do not contribute to the off-diagonal matrix element. Eq.~\eqref{eq:PDFA0term} is derived by generalizing~\eqref{eq:frmatchingPDF} and following the steps outlined in Sec.~\ref{sec:scME}. 
Inserting the matching equation into the definition of the gluon PDF, we obtain the relation anticipated in \eqref{eq:536},
\begin{equation}
f_{g/N}^{\rm thr, \, NLP}(\xi) =\frac{1}{n_+p}\mathcal{S}_{sc}^{\rm NLP}((1-\xi)n_+p) \,,
\label{eq:NLPthrpdf}
\end{equation}
which holds for the dimensionally regulated bare PDF.

We emphasize again that this identification holds for the off-diagonal quark-parton scattering channel. There is a similar contribution to the left-hand side of \eqref{eq:PDFA0term} from scattering off an antiquark and another from the diagonal channel that involves the analogue of \eqref{eq:NLPscfunctionfinalsub} containing the leading-power gluon threshold PDF in the nucleon and appropriately modified soft- and hardcollinear functions. The diagonal channel is algebraically much more involved and will be considered separately. 
Writing the relationship between bare PDFs in the form 
\begin{equation}
f_{i/N}^{\rm thr, \, NLP}(\xi) = \sum_{j=g,q,\bar{q}}\, \int d\xi^\prime\, V_{ij}(\xi^\prime)\,f_{j/N}^{\rm LP}(1-(\bar{\xi}-\bar{\xi}^\prime))\,, 
\label{eq:Uijdef}
\end{equation}
which defines the NLP conversion factor $V_{ij}(\xi)$, one can relate the NLP threshold PDFs to the leading-power PDF. Eqs.~\eqref{eq:NLPthrpdf}, \eqref{eq:NLPscfunctionfinalsub} imply
\begin{eqnarray}
V_{gq}(\xi) 
&=& \frac{1}{2} \int_0^\infty \frac{d\omega}{\omega} \frac{d\omega'}{\omega'} \, 
D_g^{\rm B1}(-n_+ p\omega) \, D_g^{\rm B1\ast}(-n_+ p\omega') 
\, S_{\rm sc, sub}^{\rm NLP}(\np p\,(1-\xi),\omega,\omega') \,.
\qquad
\label{eq:Ugqfact}
\end{eqnarray}
Together with the $d$-dimensional $\mathcal{Z}$ factors \eqref{eq:ULPqqbare} for the diagonal channels, this defines the corresponding off-diagonal NLP inverse-$\mathcal{Z}$ factors as\footnote{No summation over $j$ is implied. Only the diagonal LP $\mathcal{Z}$ factors are non-zero.} 
\begin{equation}\label{eq:UVrel}
    \mathcal{Z}(\mu_0)_{ij}^\text{NLP}=V_{ij}\mathcal{Z}_{jj}^\text{LP}(\mu_0)
\end{equation}
for $i\neq j$, which relate the bare NLP off-diagonal PDF to the $d$-dimensional renormalized $f_{j/N}^{\rm LP}(\mu_0)$ defined in \eqref{eq:fqnLPLL}. The objects $\mathcal{Z}_{ij}^{\rm (N)LP}$  agree with the ones denoted by $U_{ij}^{\rm (N)LP}$ in \cite{Beneke:2020ibj}.
Eq.~\eqref{eq:Ugqfact} provides an operator definition and factorization formula for them. The dependence on the nucleon state is now factored into the LP PDF. 
Nevertheless, the $V_{ij}$ cannot be computed in perturbation theory since the soft-collinear function lives at virtualities of order $\Lambda^2_{\rm QCD}$, and \eqref{eq:diffNLPscfunction} must be evaluated in the non-perturbative QCD vacuum state.

These results also show that, unlike at LP, the NLP term in the threshold limit of the PDF is intrinsically a two-scale object. The soft-collinear emission of modes with momenta $Q(\lambda^2,\xi,\xi^2/\lambda^2)$ into the remnant generates modes with virtuality of order $\Lambda_{\rm QCD}^2/(1-x)$, which is large compared to 
$\Lambda_{\rm QCD}^2$. These modes are factored into the $D_g^{\rm B1}$ coefficients, which can be perturbatively calculated, since $\Lambda_{\rm QCD}^2/(1-x)\gg\Lambda_{\rm QCD}^2$. However, they are contained in the standard PDF at NLP accuracy, since their transverse momenta $\Lambda_{\rm QCD}/\sqrt{1-x}$ are below the factorization scale $\mu\ll Q$ of collinear factorization away from the endpoint. The existence of two scales allows for the presence of large logarithms $\ln(1-x)$ within the PDF, which are not controlled by the DGLAP kernels but must be summed by other means; see Sec.~\ref{sec:resum}. 

Even though these logarithms are contained in the hadronic PDF, they can be sum\-med systematically in perturbation theory due to the following two observations: 1) Eq.~\eqref{eq:NLPscfunctionfinalsub} involves the leading-power PDF near threshold and the subtracted soft-collinear function, both of which are non-perturbative, because they capture virtualities of order $\Lambda_{\rm QCD}^2$, when $\Omega$ and $\omega^{(\prime)}$ are of the order of their natural size $Q(1-x)$ and $\Lambda_{\rm QCD}^2/(Q (1-x))$, respectively. However, although $\Omega$ and $\omega^{(\prime)}$ are very different, the soft-collinear function cannot depend on the large ratios $\Omega/\omega^{(\prime)}$ due to boost invariance under rescaling $\np\to \kappa\np$, $\nm\to \kappa^{-1}\nm$ (while the ratio $\omega^\prime/\omega$ is allowed). Thus, when the renormalization scale for the RG evolution is taken at a few times $\Lambda_{\rm QCD}$, the soft-collinear function and LP threshold PDF may depend on $1-x$, but do not exhibit large logarithms in this quantity at this scale. 2) Large logarithms of $1-x$ can arise from two sources. First, from RG evolution between the scales $\Lambda_{\rm QCD}$ and $\Lambda_{\rm QCD}/\sqrt{1-x}$, which can be controlled perturbatively by RGE evolution of the hardcollinear $D_g^{\rm B1}$ function from the initial scale to $\Lambda_{\rm QCD}/\sqrt{1-x}$. Second, from an endpoint divergence at large $\omega^{(\prime)}$ in the convolution integrals, which signals that there is a relevant contribution when $\omega^{(\prime)}$ is not of its natural size. However, in this case, the virtuality-invariants $\Omega\omega^{(\prime)}$ in the soft-collinear function are much larger than $\Lambda_{\rm QCD}^2$, and the effect can again be calculated perturbatively and is accounted for by refactorization as discussed in Sec.~\ref{sec:resum}.

It should be possible to formalize the above discussion in terms of a short-distance expansion of the NLP soft-collinear function, which, by boost invariance, must be organized as an expansion in powers of $\Lambda_{\rm QCD}^2/(\Omega \omega^{(\prime)})$.  The large-$\omega^{(\prime)}$ expansion corresponds to the small-$v^{(\prime)}$ expansion of the part $
[0,v \nm]_{sc,\nm}^{AC}\,
\left[\bar{\eta}_\alpha(v\nm) T^C Y_{sc,\nm}(v\nm)\right]_a$ of the operator $\mathcal{O}^B_{sc,\alpha a}(v\nm,0)$ and its conjugate in \eqref{eq:diffNLPscfunction}, when some soft-collinear particles in the final state have  large $\nm p$ momenta of order $\omega^{(\prime)}$, while those with $\nm p$ of natural soft-collinear magnitude $\Lambda_{\rm QCD}^2/(Q (1-x))$ are treated as background fields. All terms except the first perturbative term will, however, be suppressed by powers of $\Lambda_{\rm QCD}^2$, and hence count as ``higher-twist'' effects, and are therefore beyond the accuracy of the present work, which concerns {\em leading}-twist DIS and PDFs at next-to-leading power in $1-x$.

\subsubsection{Tree-level soft-collinear function and endpoint divergence}

The large-$\omega^{(\prime)}$ limit of the soft-collinear function corresponds to fluctuations with large virtuality of order $\omega^{(\prime)}Q(1-x)$ and can be computed in the perturbative vacuum state. One then finds at lowest order in the strong coupling
\begin{equation}
S_{sc}^{\rm NLP, \,LO}(\Omega,\omega,\omega')= \frac{\alpha_s C_F}{4\pi}\,\frac{2(d-2)}{\Gamma(1-\epsilon)}\,\omega\, \delta(\omega-\omega') \left(\frac{\mu^2 e^{\gamma_E}}{\Omega \omega}\right)^{\epsilon} \,.
\label{eq:diffNLPsctree}
\end{equation}
Inserting this expression into \eqref{eq:NLPscfunctionfinal} and 
employing $K_{q/q}=1$ and $C_\Lambda=1+\mathcal{O}(\alpha_s)$ results in 
\begin{equation}
\label{eq:SscLO}
   \mathcal{S}_{sc}^{\rm NLP,\,LO}(\Omega) =
   \frac{1}{2}\,\np p\, 
 \frac{\alpha_s C_F}{4\pi}\,\frac{2(d-2)}{\Gamma(1-\epsilon)}  \int_{0}^\infty \frac{d\omega}{\omega}  \left(\frac{\mu^2 e^{\gamma_E}}{\Omega \omega}\right)^{\!\epsilon} \,,
\end{equation}
which, in four dimensions ($\epsilon=0$), is logarithmically divergent at both $\omega\to0$ and $\omega \to\infty$. The divergence at $\omega=0$ is spurious since the right-hand side is a good approximation only at large $\omega$. The UV singularity at $\omega=\infty$, on the other hand, is not an artefact of the perturbative approximation and is of crucial importance for the following discussion.

Continuing with $\epsilon\not= 0$, the UV pole at large $\omega$ translates to 
\begin{align}
    f_{g/q}^{\rm thr, \, NLP}(\xi) =  \frac{\alpha_s C_F}{2\pi}\frac{1}{\epsilon}  + \mathcal{O}(\epsilon^0)
    \label{eq:LOfgqpole}
\end{align}
for the bare PDF, where \eqref{eq:NLPthrpdf} has been used.
That is, the pole part of the bare PDF indeed coincides at leading order $x\to 1$ with the UV pole of the standard definition of the dimensionally regulated bare off-diagonal PDF, given by 
\begin{align}
f_{g/q}^{\rm thr, \, NLP}(\xi)  = \int\! dx dy \,\delta(\xi -x y)
  \,\frac{1}{\epsilon} \left[P_{gq}(x)\right]_{x \to 1 } f_{q/q}^{\rm thr, \, LP}(y) + \mathcal{O}(\epsilon^0)
  \label{eq:qgDGLAPconvolution}
\end{align}
with $f_{q/q}^{\rm thr,\, LP}(y)=\delta(1-y)$ and 
\begin{align}
 P_{gq}(x)  = \frac{\alpha_s C_F}{2\pi} \frac{1+(1-x)^2}{x}\,.
\end{align}
Interestingly, the $\mathcal{O}(\alpha_s)$ off-diagonal splitting kernel in the $x\to 1$ limit naturally arises from an endpoint divergence in the longitudinal momentum integration in the opposite light-cone momentum component  $\nm p$, rather than from a divergence in the transverse momentum integral.

Nonetheless, the divergence of \eqref{eq:SscLO} signals the failure of \emph{na\"ive} NLP factorization, since the integral represents the product of the $D_g^{\rm B1}$ hardcollinear and $S_{sc}^{\rm NLP}$ soft-collinear functions, both of which should  be {\em first} renormalized separately and only then convoluted {\em after} taking the $d\to 4$ limit. 
Since the partonic expression \eqref{eq:SscLO} captures the UV behaviour of the nucleon PDF, this implies that $f^{\rm thr, NLP}_{g/N}(x)$ defined 
in this way is ill-defined because of the endpoint divergence at $\omega\to \infty$. 
The fact that the limit $d\to 4$ does not commute with the convolution integral spoils the interpretation of $\mathcal{S}_{sc}^{\rm NLP}(\Omega)$ 
as part of the standard $\msbar$-scheme PDF, since the $\msbar$ definition implies that the convolution is performed in $d$ dimensions with poles subtracted afterwards.

This divergence from large $\omega$ should, in fact, be expected and is related to the divergence manifested in \eqref{eq:NLPjetLO} for the anti-hardcollinear term in the convolution over the momentum fraction of the anti-hardcollinear quark at {\em small} momentum fraction. In both cases, the momentum fraction ($\omega$ resp.~$r$) refers to the $\nm k$ component, that is, the component of the opposite light-cone direction relative to the large momentum components of the initial state relevant to standard collinear factorization. The endpoint divergences as $r^{(\prime)}\to 0$ and $\omega^{(\prime)}\to \infty$ arise instead from the need for factorization in $\nm k$, which distinguishes whether a mode belongs to the soft-collinear remnant or the anti-hardcollinear jet. This factorization is a new feature that appears 
only in the threshold limit and starting at NLP, and it differs from standard collinear factorization in transverse momentum. Conceptually and technically, these endpoint divergences are the exact analogues of the endpoint divergences for the crossed, time-like ``gluon-thrust'' process \cite{Beneke:2022obx}. This suggests applying the same endpoint rearrangement, here between the anti-hardcollinear and soft-collinear contributions to $W^\phi$. As will be seen below, this leads to a PDF definition that differs from but can be related to the conventional $\msbar$ PDF definition.

\subsection{Factorization formula for the hadronic tensor}

To summarize this section, we derived the factorization for the NLP soft-collinear matrix element of DIS in the threshold region $x\to1$ and showed how to match it onto the threshold PDF at NLP. After soft subtraction, the factorization formula has been written in terms of multiple single-scale objects with well-defined anomalous dimensions, a prerequisite for resumming large logarithms with RGEs for the separate factors. Additionally, this soft subtraction demonstrates how perturbative and non-perturbative physics can be split in a consistent way. As a result, the power-corrections to the gluon PDF can be formulated in terms of a  hardcollinear function with virtuality $\Lambda_{\rm QCD}^2/(1-x)$ and the LP quark PDF, which reflects the two-scale nature of the PDF near threshold. 

The factorization allows us to write the full hadronic tensor in the already anticipated form \eqref{eq:B1schematic}, \eqref{eq:A0schematic}
\begin{eqnarray}
W_\phi^{\rm NLP} &=&\frac{1}{4x}\,
 \int_0^1 dr dr'\,C_g^{\rm B1*}(\np p,r' Q,\bar{r}' Q)
C_g^{\rm B1}(\np p,r Q,\bar{r}Q)\nonumber\\
&&\quad\times\,\int_x^1 d\xi\,
\mathcal{J}^{(qg)}_{\overline{hc}}\left(\frac{Q^2(\xi-x)}{x},r,r'\right)\, 
f_{q/N}^\text{thr, LP}(\xi)\nonumber\\
&&+\left|C_g^{\rm A0}(\np p,Q)\right|^2\,\frac{Q^2}{x^2}
 \int_x^1 d\xi\,
\mathcal{J}^{(g)}_{\overline{hc}}\left(Q^2\frac{\xi-x}{x}\right)\, 
f_{g/N}^\text{thr, NLP}(\xi)\\
&=&\frac{1}{4x}\,
 \int_0^1 dr dr'\,C_g^{\rm B1*}(\np p,r' Q,\bar{r}' Q)
C_g^{\rm B1}(\np p,r Q,\bar{r}Q)\nonumber\\
&&\quad\times\,\int_x^1 d\xi\,
\mathcal{J}^{(qg)}_{\overline{hc}}\left(\frac{Q^2(\xi-x)}{x},r,r'\right)f_{q/N}^\text{thr, LP}(\xi)\nonumber\\
&&+\left|C_g^{\rm A0}(\np p,Q)\right|^2\,\frac{Q^2}{2x^2}
 \int_x^1 d\xi\,
\mathcal{J}^{(g)}_{\overline{hc}}\left(Q^2\frac{\xi-x}{x}\right)\int_0^\infty \frac{d\omega}{\omega} \frac{d\omega'}{\omega'}\,D_g^{\rm B1}(-n_+ p\omega) \quad
\notag
\\
&&\quad\times D_g^{{\rm B1}\ast}(-n_+ p\omega')\int d\xi^\prime\,S_{\rm sc, sub}^{\rm NLP}(\np p(1-\xi^\prime),\omega,\omega')f_{q/N}^\text{thr, LP}(1+\xi-\xi^\prime)\,,\qquad
\label{eq:Wtogether}
\end{eqnarray}
where $f_{g/N}^\text{thr,NLP}(\xi)$ is given by \eqref{eq:NLPthrpdf} and factorizes as in~\eqref{eq:factorizationSSC}. 

We emphasize again that this formula refers only to the off-diagonal parton-scattering contributions and is strictly defined only in $d$ dimensions with $d\neq4$ due to the endpoint divergences of the $r^{(\prime)}$ and $\omega^{(\prime)}$ convolutions (the latter being contained implicitly in $f_{g/N}^\text{thr,NLP}(\xi)$). 
We also recall that when the antiquark contribution is explicitly included, there is an identical term to \eqref{eq:Wtogether} with $f_{q/N}^\text{thr,LP}$ replaced by $f_{\bar{q}/N}^\text{thr,LP}$. In particular, $f_{g/N}^\text{thr,NLP}(\xi)$ absorbs a term from soft-collinear antiquark emission into the remnant.  
In the next section, we explain how two different methods can be used to resum logarithms $\ln(1-x)$ to all orders in perturbation theory at NLP. 


\section{Resummation of large logarithms}
\label{sec:resum}

The occurrence of endpoint divergences spoils the standard resummation of large logarithms since four-dimensional (hard, anti-hardcollinear, soft-collinear, etc.) component functions cannot be used. In this section, we present two approaches to solve this problem. First, we follow the approach presented in \cite{Beneke:2020ibj}, in which the leading $1/\eps$ poles of the bare component functions of the factorization formula are resummed rather than the large logarithms themselves.
Since the full cross section is finite, after adding all contributions, all pole terms, UV and endpoint, cancel. The $d$-dimensional treatment is possible because the endpoint divergences do not arise from rapidity factorization between modes with the same virtuality.
They can therefore be regulated dimensionally and subtracted minimally in the $\overline{\rm MS}$ scheme. In the second approach, we propose a new non-minimal ``endpoint scheme'' to define the PDF, which differs from the standard $\overline{\rm MS}$ scheme through the implementation of refactorization-based subtractions to regularize the endpoint divergences. This method has been developed in other contexts \cite{Beneke:2008pi,Liu:2020tzd,Liu:2020wbn,Liu:2022ajh,Beneke:2020ibj,Beneke:2022obx,Cornella:2022ubo,Hurth:2023paz,Cornella:2026lkp} and exploits relations between the integrands in the problematic endpoint regions. The second approach is advantageous because it avoids the need to trace the dimensional regulator, which can only be removed after all UV and endpoint divergences are added up in the $\msbar$ scheme.  This approach also allows for the use of standard methods to sum large logarithms for renormalized component functions.

Because the $d$-dimensional approach first resums the leading poles of bare functions before deriving renormalized expressions, from this point onward, we make bare functions explicit by adding a subscript.

\subsection{Resummation of the $\overline{\rm MS}$ PDF in $d$ dimensions} 
\label{sec:MSbar}

To demonstrate that the factorization equations \eqref{eq:Uijdef}, \eqref{eq:Ugqfact} for the off-diagonal contribution to the gluon PDF near threshold  reproduce the result obtained previously in \cite{Beneke:2020ibj} through consistency conditions, we  resum the leading double-logarithmic corrections in $d$ dimensions and perform $\overline{\rm MS}$ subtraction of the divergences. We refrain from attempting to rectify the endpoint divergence in this approach, as the endpoint divergence is inherently regularized dimensionally.  

In the double-logarithmic (DLA) approximation, the hardcollinear matching function $D_g^{\rm B1}$ obeys the RGE \cite{Beneke:2020ibj}
\begin{align}
    \frac{d}{d\ln\mu} D_g^{\rm B1}(-n_+p \omega,\mu)  =& \,\gamma_{\rm cusp}(\alpha_s) (C_A-C_F) \ln\frac{\mu^2}{-n_+ p \omega}  D_g^{\rm B1}(-n_+p \omega,\mu)\,.
\end{align}
To derive its $d$-dimensional solutions, we follow the same steps as laid out at LP in Sec.~\ref{sec:LPPDFresum}. Given the LO initial condition $ D_g^{\rm B1}(-n_+p \omega,-n_+p \omega) = 1$, the $d$-dimensional solution for the bare function reads:
\begin{equation}\label{eq:resumDB1}
D^{\rm B1}_{g,\rm bare}(-\np p\omega) =  \exp\left[
\frac{\alpha_s(C_A-C_F)}{2\pi\eps^2}\left(\frac{\mu^2}{-\np p\omega}\right)^{\!\eps}\,\right].
\end{equation}

Similarly, since the cusp part of the anomalous dimension \eqref{eq:gammasc} of the NLP soft-collinear function is diagonal in all convolution arguments, the $d$-dimensional solution to the RGE \eqref{eq:SNLPscRGE} is given in the DLA by 
\begin{equation}
\begin{aligned}
\label{eq:resummSsc}
S_{\rm sc, sub , bare}^{\,\rm NLP}(\Omega,\omega,\omega') =&  \exp\left[
-\frac{\alpha_s(C_A-C_F)}{2\pi\eps^2}\left(\left(\frac{\mu^2 e^{-i\pi}}{\omega\Omega}\right)^{\!\eps }+\left(\frac{\mu^2 e^{+i\pi}}{\omega' \Omega}\right)^{\!\eps }\,\right)\,\right]\\
&\times S_{\rm sc}^{\rm NLP, LO}(\Omega,\omega,\omega') \,,
\end{aligned}
\end{equation}
where $S_{\rm sc}^{\rm NLP, LO}(\Omega,\omega,\omega')$ is  the renormalized, $d$-dimensional, non-perturbative initial condition evaluated at a few times its natural scale $\mu^2\sim \Omega \omega^{(\prime)}\sim \Lambda_{\rm QCD}^2$.  The $e^{\pm i \pi}$ factors resolve the $\pm i0^+$ prescriptions in 
\eqref{eq:SNLPscRGE} using the fact that $\Omega, \omega,\omega^\prime$ are positive. 
Substituting this expression into \eqref{eq:Ugqfact} and accounting for the implicit $+i0^+$ prescription on $-\np p\omega$ in \eqref{eq:resumDB1}, we obtain  
\begin{eqnarray}
V_{gq,\rm bare}(\xi) 
&=& \frac{1}{2} \int_0^\infty \frac{d\omega}{\omega} \frac{d\omega'}{\omega'} \, 
\exp\left[
\frac{\alpha_s(C_A-C_F)}{2\pi\eps^2}\left(\left(\frac{\mu^2 e^{-i\pi}}{\np p\omega}\right)^{\!\eps}+\left(\frac{\mu^2 e^{+i\pi}}{\np p\omega'}\right)^{\!\eps}\;\right)\,\right]
\qquad
\nonumber \\[0.2cm]
&&
\times \exp\left[
-\frac{\alpha_s(C_A-C_F)}{2\pi\eps^2}\left(\left(\frac{\mu^2 e^{-i\pi}}{\omega\Omega}\right)^{\!\eps }+\left(\frac{\mu^2 e^{+i\pi}}{\omega' \Omega}\right)^{\!\eps }\;\right)\,\right]
\nonumber\\[0.2cm]
&& \times 
\,S_{\rm sc}^{\rm NLP, LO}(\Omega,\omega,\omega')\,,
\label{eq:UgqDLA}
\end{eqnarray}
where $\Omega$ should be set to $\np p(1-\xi)$. If the 
non-perturbative soft-collinear function on the right-hand side were known at a scale of several $\Lambda_{\rm QCD}$, such that $\alpha_s$ at this scale is sufficiently small, the above convolution would allow one to compute the leading logarithms of the off-diagonal contribution to the nucleon gluon PDF in terms of the LP nucleon PDF according to \eqref{eq:Uijdef}.

In the absence of non-perturbative data, we instead proceed perturbatively. As discussed above, this is justified when
$\omega^{(\prime)} \gg \Lambda_{\rm QCD}^2/(1-x)$. In the DLA, we may employ  the leading-order perturbative expression \eqref{eq:diffNLPsctree} to obtain 
\begin{eqnarray}
\hat V_{gq,\rm bare}(\xi)&=& 
\frac{1}{2}\,\frac{\alpha_s C_F}{4\pi}\,\frac{2(d-2)}{\Gamma(1-\epsilon)}\int_{\mu_0^2/(\np p)}^\infty \frac{d\omega}{\omega} \left(\frac{\mu^2 e^{\gamma_E}}{\omega\Omega}\right)^{\!\epsilon}\nonumber\\
&&\times\,
\exp\left[
\frac{\alpha_s(C_A-C_F)}{\pi\eps^2}\left(\left(\frac{\mu^2}{\np p\omega}\right)^{\!\eps}-\left(\frac{\mu^2}{\omega \Omega}\right)^{\!\eps}\,\right) \cos\pi\eps\,\right].
\label{eq:Ugqpertmomspace}
\end{eqnarray}
The lower limit on the $\omega$ integral corresponds to the scale $\mu_0$ introduced in Sec.~\ref{sec:LPPDFresum} and separates the perturbative part from the non-perturbative initial condition.  In the following, the perturbative part of the conversion factors $V$ will be indicated by a hat,~$\hat V$. 
The integral over $\omega $ can be performed with the help of 
\begin{equation}\int_{\Lambda_{\rm IR}}^\infty \frac{d\omega}{\omega^{1+\eps}}\,
e^{Y \omega^{-\eps}} = \frac{e^{Y (\Lambda_{\rm IR})^{-\eps}}-1}{\eps Y}\,.
\label{eq:omY}
\end{equation}
 Using the standard bar notation $\bar{\xi}\equiv 1-\xi$ gives 
\begin{eqnarray}
\hat V_{gq,\rm bare}(\xi)&=& \,  \frac{1}{2}\,
\frac{C_F}{C_A-C_F}\,\frac{(1-\epsilon)e^{\epsilon \gamma_E}}{\cos(\pi\eps)\Gamma(1-\epsilon)}\frac{\epsilon \bar{\xi}^{-\epsilon}}{1 -\bar{\xi}^{-\epsilon}}\nonumber\\
&&\times\,
\left(\exp\left[
\frac{\alpha_s(C_A-C_F)}{\pi\eps^2}\left(\frac{\mu^2}{\mu_0^2}\right)^{\!\epsilon}\left(1 -\bar{\xi}^{-\epsilon}\right) \cos\pi\eps\,\right] - 1\right).\quad
\label{eq:Ugqbare}
\end{eqnarray}
Since in the DLA one keeps only the leading poles and their associated explicit $\mu$-dependence, the $\cos \pi\eps$ arising from the phases can be dropped. 
We can further drop the prefactor $\frac{(1-\epsilon)e^{\epsilon \gamma_E}}{\cos(\pi\eps)\Gamma(1-\epsilon)}$ in the DLA, as it only contributes at a higher logarithmic order. Passing to Mellin-moment space by replacing $\bar{\xi}\to 1/N$ (see Sec.~\ref{sec:LPPDFresum}), the $d$-dimensional off-diagonal NLP $\mathcal{Z}_{gq}$ factor is then found by 
\begin{eqnarray}
    \mathcal{Z}_{gq}^\text{NLP}(\mu_0)&=&\hat V_{gq,\text{bare}}\,\mathcal{Z}^\text{LP}_{qq}(\mu_0)\nonumber\\
    &=&\frac1{2N}\frac{C_F}{C_F-C_A}\frac{\eps N^\eps}{N^\eps-1}\,\Bigg(\exp\bigg[-\frac{\as C_A}{\pi\eps^2}\left(\frac{\mu^2}{\mu_0^2}\right)^\eps(N^\eps-1)\bigg]\nonumber\\
    &&-\, \exp\bigg[-\frac{\as C_F}{\pi\eps^2}\left(\frac{\mu^2}{\mu_0^2}\right)^\eps(N^\eps-1)\bigg]  \Bigg)\,,\quad
\label{eq:Ugqexpr}
\end{eqnarray}
and agrees with eq.~(3.51) in \cite{Beneke:2020ibj} upon identifying $\mu_0^2=\Lambda^2$. In this work, however, we argue that $\mu_0$ should be chosen as a factor several times larger than the QCD scale, such that $\as(\mu_0)$ remains perturbative. 

Under renormalization, the quark and gluon $\msbar$ PDFs mix. Expanding to NLP and using that $Z_{ij}^{\rm LP}$ is diagonal, 
the following NLP combination must be UV finite: 
\begin{align}
  Z^{\text{NLP}}_{gq}\, f^{\text{thr,LP}}_{q/N,\text{bare}} \;+\; Z^{\text{LP}}_{gg}\,  f^{\text{thr,NLP}}_{g/N,\text{bare}}
  \;\stackrel{!}{=}\; \text{finite.}
  \label{eq:rencond_chapter}
\end{align}
As shown in Sec.~\ref{sec:LPPDFresum} (and before in \cite{Vogt:2010cv}), in the DLA at large $N$ the LP renormalization constants exponentiate, 
\begin{align}
Z^{\text{LP}}_{qq}=\exp\!\Big[\frac{\alpha_s C_F}{\pi}\,
  \frac{\ln N}{\epsilon}\Big]\,,
  \qquad
  Z^{\text{LP}}_{gg}=\exp\!\Big[\frac{\alpha_s C_A}{\pi}\,
  \frac{\ln N}{\epsilon}\Big]\,.
  \label{eq:diagZLP}
\end{align}
We now express both $f^{\text{thr,LP}}_{q/N,\text{bare}}$ and $f^{\text{thr,NLP}}_{g/N,\text{bare}}$ in terms of the initial condition $f^{\text{thr,LP}}_{q/N}(\mu_0)$ through the $\mathcal{Z}$ factors $\mathcal{Z}_{qq}^\text{LP}(\mu_0)$, $\mathcal{Z}_{gq}^\text{NLP}(\mu_0)$. The quantity $f^{\text{thr,LP}}_{q/N}(\mu_0)$ is free of UV poles and can then be divided out in \eqref{eq:rencond_chapter}. 
By adapting the reasoning of Sec.~3.2.3 of \cite{Beneke:2020ibj}, we observe that the finiteness condition \eqref{eq:rencond_chapter} and the requirement that the $\overline{\rm MS}$-scheme $Z$-factors consist of pure poles allow us to recast \eqref{eq:rencond_chapter}  into the form 
\begin{align}
\,e^{-\frac{\alpha_s C_F}{\pi\eps}\ln N}\,
Z^{\rm NLP}_{gq} +
\frac{1}{2N}\frac{C_F}{C_A-C_F}\frac{\epsilon N^\epsilon}{N^\epsilon-1}\exp\left[-\frac{\alpha_s}{\pi}(C_F-C_A)\frac{\ln N}{\epsilon}\right]
\stackrel{!}{=} \mbox{finite}\,,
\label{eq:cond1_chapter}
\end{align}
which isolates the UV poles. Picking up the pole part of the second term on the left-hand side and solving for the $\msbar$ 
 $Z^{\rm NLP}_{gq}$ gives the final result for the off-diagonal counterterm $Z^{\rm NLP}_{gq}$, and hence, the anomalous dimension $\gamma^{\rm NLP}_{gq}$ in agreement with \cite{Beneke:2020ibj}, where, however, the anomalous dimension was deduced from the hard-scattering cross section rather than from PDF renormalization \eqref{eq:rencond_chapter} directly.  
 
 We provide here an explicit derivation of the pole part. Starting from \eqref{eq:cond1_chapter}, it is useful to rewrite
\begin{align}
  \frac{\epsilon\,N^\epsilon}{N^\epsilon-1} \;=\; \frac{\epsilon}{1-e^{-\epsilon L_N}}
  \;=\; \frac{1}{L_N}\sum_{m=0}^{\infty}\frac{B_m}{m!}\,(-\epsilon L_N)^m\,,
  \label{eq:BernoulliGF}
\end{align}
where $B_m$ denote the Bernoulli numbers with $B_0=1$, $B_1=-\tfrac12, \ldots$. We also abbreviate 
\begin{equation}
L_N =\ln N 
\end{equation}
here and in the following. Minimal subtraction instructs us to choose $Z^{\text{NLP}}_{gq}$ such that the pole part in $\epsilon$ of \eqref{eq:cond1_chapter} is cancelled, while $Z^{\text{NLP}}_{gq}$ is equal to its principal part of the Laurent expansion in $\epsilon$ around zero. That is,
\begin{align}
Z^{\rm NLP}_{gq}
  \,=\,
  -\frac{1}{2N}\frac{C_F}{C_A-C_F} \exp\left[{\frac{\alpha_s C_F}{\pi\eps}L_N}\right]\,\left[
  \frac{\epsilon}{1-e^{-\epsilon L_N}} 
  \exp\!\Big(-\frac{\alpha_s}{\pi}(C_F-C_A)\frac{L_N}{\epsilon}\Big)
  \right]_{\text{pole part}}\!,
  \label{eq:Zsolve_PP}
\end{align}
where we retain the pole part of the bracketed expression in  the $\epsilon\to 0$ limit before multiplying by the exponential.
Inserting \eqref{eq:BernoulliGF} and expanding the exponent inside the bracket, we obtain the double series
\begin{align}
  Z^{\text{NLP}}_{gq}
  =& -\exp\left[{\frac{\alpha_s C_F}{\pi\eps}L_N}\right]\,\frac{1}{2NL_N}\frac{C_F}{C_A-C_F}\nonumber \\
  &\times\sum_{n=0}^\infty\sum_{m=0}^\infty
  \frac{B_m}{m!\,n!}\,
  (-1)^m\,\left(\frac{\alpha_s}{\pi} (C_A-C_F)\right)^n\,L_N^{n+m}\,\epsilon^{\,m-n}\Big|_{\text{pole part}}\,.
\end{align}
Only terms with $n>m$ contribute to the pole part. Therefore, we introduce $k=n-m\ge 1$ and after some algebraic manipulations obtain 
\begin{align}
  Z^{\text{NLP}}_{gq}
  &= -\exp\left[{\frac{\alpha_s C_F}{\pi\eps}L_N}\right]\,\frac{1}{2N L_N}\frac{C_F}{C_A-C_F}\nonumber\\
  &\times \sum_{k=1}^\infty \frac{1}{(-\epsilon L_N)^{\,k}}
  \sum_{m=0}^\infty
  \frac{B_m}{m!\,(m+k)!}\,
  \,\left(\frac{\alpha_s}{\pi} (C_F-C_A)L_N^2\right)^{m+k}\,.
  \label{eq:double_series}
\end{align}
It is convenient to package the series into \cite{Beneke:2020ibj} 
\begin{align}
  F_{\rm pole}[w,a] \;\equiv\; \sum_{k\geq 1} \frac{1}{w^k} \sum_{n\geq 0} \frac{B_n \, a^{n+k}}{n!\,(n+k)!}
  \label{eq:defF}
\end{align}
with
\begin{equation}
w=-\epsilon L_N,\quad a=\frac{\alpha_s}{\pi}\,(C_F-C_A)\,L_N^2\,,
\label{eq:defwa}
\end{equation}
such that it  yields the compact closed form for the off-diagonal $\overline{\text{MS}}$ counterterm 
\begin{align}
Z^{\rm NLP}_{gq} = \frac{1}{2N L_N} \frac{C_F}{C_F-C_A}\exp\left[\frac{\alpha_s C_F}{\pi}\frac{L_N}{\epsilon}\right] F_{\rm pole}\!\left[-\epsilon L_N, \frac{\alpha_s}{\pi}(C_F-C_A)L_N^2 \right]
  \label{eq:Zqg_final}
\end{align}
as given in \cite{Beneke:2020ibj}. 
The Mellin-space DGLAP kernels $P_{ij} = -\gamma_{ij}$ follow from the PDF renormalization matrix as  
\begin{align}
  \gamma_{ij}(N)
  \;=\; -\Big[\left(\mu\frac{d}{d\mu}Z\right)\,Z^{-1}\Big]_{ij}\,.
\end{align}
In the DLA for the off-diagonal channel, running-coupling effects in the derivative can be neglected, so the $\mu$-dependence is governed by the $1/\epsilon$ pole of $Z$. Extracting the coefficient of the single $1/\epsilon$ pole in \eqref{eq:Zqg_final}, one finds the closed all-order result
\begin{equation}
  \gamma^{\text{LL,NLP}}_{gq}(N)
  \;=\; -\frac{\alpha_s\,C_F}{\pi}\frac{1}{N}\,\mathcal{B}_0\Big(\frac{\alpha_s}{\pi}\,(C_F-C_A)\,L_N^2\Big)\,,
  \label{eq:Pll_B0}
\end{equation}
with 
\begin{equation}
 \mathcal{B}_0(z)\equiv
 \sum_{n=0}^\infty \frac{B_n}{(n!)^2}\,z^n\,.
 \label{eq:B0def}
\end{equation}
This equation reproduces the standard large-$x$ DLA resummation for the \emph{off-diagonal} splitting functions, which was originally obtained in \cite{Vogt:2010cv}. It includes the characteristic colour structure $C_F(C_F-C_A)^n$ and the Bernoulli series $\mathcal{B}_0(z)$, and it has been rederived in closed form via $d$-dimensional refactorization and soft-quark Sudakov exponentiation in \cite{Beneke:2020ibj}. 
The above expression for $\gamma^{\text{LL,NLP}}_{gq}(N)$ sums the leading logarithmic (LL) terms in the DGLAP splitting function $P_{gq}(N)$ of the form $\alpha_s^{n+1} \ln^{2 n}N$ to all orders in $\alpha_s$.

With the $\msbar$ renormalization factors at hand, it is also of interest to extract the finite part of the $d$-dimensional inverse-$\mathcal{Z}$ factors, which we denote by $\mathcal{U}_{ij}^{\rm (N)LP}$.  This follows closely the derivation of the renormalized short-distance coefficients in \cite{Beneke:2020ibj}. Plugging the expressions for the bare PDFs \eqref{eq:fqnLPLL}, \eqref{eq:Uijdef}, and \eqref{eq:UVrel} into \eqref{eq:rencond_chapter}, we find the renormalization condition
\begin{eqnarray}\label{eq:Ugqrendef}
    \mathcal{U}_{gq}^\text{NLP}(\mu_0)=&Z_{gq}^\text{NLP}\,\mathcal{Z}_{qq}^\text{LP}(\mu_0)+Z_{gg}^\text{LP}\mathcal{Z}_{gq}^\text{NLP}(\mu_0)\equiv \hat V_{gq}(\mu)\, \mathcal{U}^{\rm LP}_{qq}(\mu_0)\,,
\end{eqnarray}
which defines the renormalized conversion factor $\hat V_{gq}$.
The finite part of the LP factor 
\begin{equation}
    \mathcal{U}_{qq}^\text{LP}(\mu_0)\equiv Z_{qq}^{\rm LP} \mathcal{Z}^{\rm LP}_{qq}(\mu_0)=\exp\left[-\frac{\as C_F}{\pi}\left(L_N\ln\frac{\mu^2}{\mu_0^2}+\frac{L_N^2}2\right)\right]
\end{equation}
follows from \eqref{eq:ULPqqbare} and \eqref{eq:ZLPqq}. 
The second summand in \eqref{eq:Ugqrendef} is given by
\begin{equation}
\begin{aligned}
    Z_{gg}^\text{LP}\,\mathcal{Z}_{gq}^\text{NLP}(\mu_0)=\,&\frac{1}{2N}\frac{C_F}{C_F-C_A}\frac{\eps N^\eps}{N^\eps-1}\,\Bigg\{\exp\left[-\frac{\as C_A}\pi\left(L_N\ln\frac{\mu^2}{\mu^2_0}+\frac{L_N^2}2\right)\right]\\
    &-\exp\left[-\frac{\as(C_F-C_A)}\pi\frac{L_N}\eps-\frac{\as C_F}{\pi}\left(L_N\ln\frac{\mu^2}{\mu^2_0}+\frac{L_N^2}2\right)\right]\Bigg\},
    \end{aligned}
    \label{eq:ZggUqgbare}
\end{equation}
as can be seen straightforwardly from \eqref{eq:Ugqexpr}. The first summand in this expression is finite for $\eps\to0$, and the second term can be rewritten as
\begin{eqnarray}
  &&  \frac{\eps N^\eps}{N^\eps-1}\exp\left[-\frac{\as(C_F-C_A)}\pi\frac{L_N}\eps-\frac{\as C_F}{\pi}\left(L_N\ln\frac{\mu^2}{\mu^2_0}+\frac{L_N^2}2\right)\right]\nonumber \\
   && = F\left[-\eps L_N,\frac\as\pi(C_F-C_A)L_N^2\right]\exp\left[-\frac{\as C_F}{\pi}\left(L_N\ln\frac{\mu^2}{\mu^2_0}+\frac{L_N^2}2\right)\right]\,,
\end{eqnarray}
where $F[w,a]$ was defined in \cite{Beneke:2020ibj} as
\begin{equation}
    F[w,a]=\frac{we^{a/w}}{e^w-1}\,,
\end{equation}
and $w$ and $a$ are defined as in \eqref{eq:defwa}. It was shown that $F$ splits into a pole part $F_\text{pole}[w,a]$ \eqref{eq:defF} and a finite part containing the Bernoulli function $\mathcal{B}_0(z)$ \eqref{eq:Pll_B0}. The pole part cancels exactly against the pole part of $Z^{\rm NLP}_{gq}$ in \eqref{eq:Ugqrendef}, and consequently, after extracting the leading-power diagonal $\mathcal{U}_{qq}^\text{LP}$, we obtain for the (renormalized) NLP conversion function 
\begin{eqnarray}
\label{eq:Vgqmsbar}
\hat V_{gq}(\mu)&=&\frac{1}{2NL_N}\frac{C_F}{C_F-C_A}\,\Bigg\{\exp\left[\frac\as\pi(C_F-C_A)\left(L_N\ln\frac{\mu^2}{\mu_0^2}+\frac{L_N^2}2\right)\right]\nonumber\\
    &&-\mathcal{B}_0\left(\frac\as\pi(C_F-C_A)L_N^2\right)\Bigg\}\,.
\end{eqnarray}

\subsection{Factorization theorem in the refactorization scheme}

Up to this point, we performed the resummation of the large-$N$ logarithms of the $\msbar$ PDF in $d$ dimensions, where endpoint divergences are regularized by the dimensional regulator $\epsilon$, which eventually cancels out of the full computation. It is typical of SCET-I NLP problems that no additional rapidity regulator is needed for endpoint divergences.
While consistent, this approach has conceptual and practical limitations. When endpoint singularities are regulated dimensionally and cancel only after combining the different contributions to the factorization theorem, intermediate expressions do not admit a straightforward four-dimensional interpretation. As a result, the structure of subleading-power contributions and their associated 
logarithms is obscured at the level of individual component functions, since they are intertwined with dimensional regulator-dependent terms. In \cite{Liu:2019oav, Liu:2020tzd,Liu:2020wbn,Liu:2022ajh,Beneke:2020ibj,Beneke:2022obx,Cornella:2022ubo,Hurth:2023paz},  an alternative approach to regulating endpoint divergences was developed. This method, known as ``refactorization-based subtraction'' or ``endpoint rearrangement'', results in a version of the factorization theorem in which the convolution integrals converge. Hence, one can use renormalized, four-dimensional component functions in the integrands.
Additionally, the resummation of large logarithms in the component functions can be done by solving standard RG equations. Moreover, through specific choices of the scales where refactorization conditions are implemented, large logarithmic corrections can be moved from one term of the factorization theorem to another. 

In this subsection, we demonstrate how to apply refactorization and endpoint rearrangements to the endpoint divergences of the DIS hadronic tensor and the parton distribution functions. The general idea is the following: The DIS hadronic tensor is a physical observable and therefore free of divergences. When deriving a factorization theorem using the method of regions \cite{Beneke:1997zp,Jantzen:2011nz}, one decomposes this finite quantity into contributions associated with different momentum regions. This decomposition can be performed at the level of bare quantities as well as after renormalization, and must reproduce the same finite result in either case. 
The appearance of endpoint divergences in the factorized expressions, here the hardcollinear and soft-collinear terms, indicates that the separation into individual contributions is not yet complete after renormalizing them separately. 

\begin{figure}[t]
\hspace*{-1cm}
\includegraphics[width=0.9\textwidth]{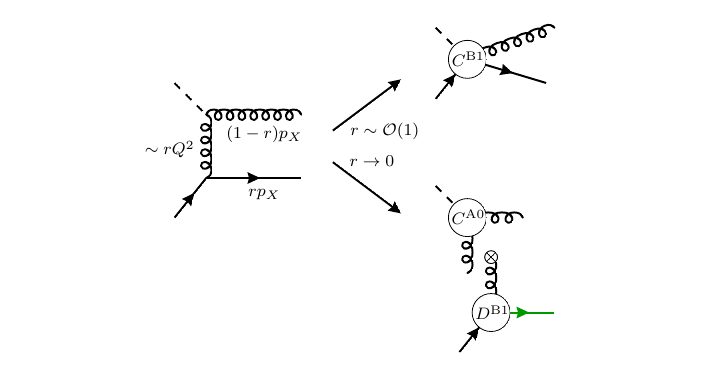}
\centering
\vskip-0.2cm
\caption{Refactorization of $C_g^\text{B1}$ in the endpoint region $r\to0$. The intermediate gluon propagator on the left-hand side ceases to be hard as $r\to0$, while the emitted quark in the final state becomes soft-collinear (indicated by the green line). Hence, matching onto the B1 operator is only valid away from the endpoint.}\label{fig:refac}
\end{figure}

A prime example in this context is the hard matching coefficient 
 $C_g^\text{B1}(n_+p_c, r Q,\bar{r} Q)$, which depends on two scales, $\tilde{Q}^2=\np p_c Q$ and  $r \tilde{Q}^2$ \cite{Beneke:2022obx}, where $r$ is the fraction of the total anti-hardcollinear momentum carried by the quark in the final state. In the derivation of the factorization theorem, the assumption $r\sim\mathcal{O}(1)$ is made, and the two scales need not be distinguished. However, $r$ is subsequently integrated over the full interval $[0,1]$. The endpoint region $r\to0$ that gives rise to the endpoint divergence is exactly where the assumption $r\sim\mathcal{O}(1)$ is violated. In the matching relation for $C_g^\text{B1}$, the intermediate propagator with virtuality $r\tilde{Q}^2$ is contracted to a point. As $r\to0$, however, this propagator does not have hard virtuality and should not be contracted, while the quark that is emitted into the final state becomes soft-collinear. Consequently, the $r\to 0$ endpoint region should instead be matched onto $C^\text{A0}_g(\np p_c,Q)D_g^\text{B1}(r\tilde{Q}^2)/r$. We show this endpoint matching graphically in Fig.~\ref{fig:refac}. Hence, we obtain the \textit{refactorization relation}
\begin{eqnarray}
    \enp{C_g^\text{B1}(\np p_c,rQ,\bar{r}Q)}=-2\,C_g^\text{A0}(\np p_c,Q)\,\frac{D_g^\text{B1}( r \tilde{Q}^2)}{r}\,
    \label{eq:refacto1}
\end{eqnarray}
for the $C_g^{\rm B1}$ coefficient, analogous to the corresponding relations in \cite{Beneke:2020ibj,Beneke:2022obx}, where the sign and normalization follow from the precise definitions of the matching coefficients. Here, the double bracket $\enp{\dotp}$ denotes the leading endpoint behaviour of the enclosed quantity in the limit $r^{(\prime)}\to0$, with relative corrections of order $r^{(\prime)}$ omitted. On the other hand, whenever the argument of the function in $\enp{\dotp}$ is $\omega^{(\prime)}$, the \textit{large} $\omega^{(\prime)}$ limit is meant instead.

Similarly, in the endpoint region, the anti-hardcollinear NLP jet function convoluted with the LP soft-collinear function  can be written as 
\begin{align}
   \enp{\mathcal{J}^{(qg)}_{
   \overline{hc}}(Q\Omega,r,r')}  = \frac{Q}{2}\int d \Omega' \mathcal{J}^{(g)}_{\overline{hc}}(Q\Omega')\enp{ S_\text{sc, sub}^\text{NLP}(\Omega-\Omega',\omega,\omega')}_{\omega^{(\prime)}=r^{(\prime)}Q} \,.
\end{align}

 Recall that in the large $\omega$, $\omega'$ limit, $\enp{ S_\text{sc, sub}^\text{NLP}(\Omega,\omega,\omega')}$ is, by definition, a purely perturbative object. Consequently, the endpoint approximation $\enp{ S_\text{sc, sub}^\text{NLP}}=S_\text{sc, sub}^\text{NLP}$ holds up to non-perturbative power corrections, as long as $\omega$ and $\omega'$ approach infinity at the same rate. 
To derive the factorization formula for the DIS hadronic tensor that is free of endpoint divergences, we also make the important observation that the integrals
\begin{eqnarray}
        0&=&\left|C_g^{\rm A0}(\np p,Q)\right|^2\,\frac{Q^2}{x^2}
 \int_x^1 d\xi\,
\mathcal{J}^{(g)}_{\overline{\text{hc}}}\left(Q^2\frac{\xi-x}{x}\right)\nonumber\\
&&\times\,\frac1{2}\int_0^\infty\frac{d\omega}{\omega}\frac{d\omega'}{\omega'}D^\text{B1}_g(-n_+p\omega)D^{\text{B1}\ast}_g(-n_+p\omega')\nonumber\\
&&\times\int d\xi'\enp{ S_\text{sc, sub}^\text{NLP}(\np p(1-\xi'),\omega,\omega')}f_{q/N}^\text{thr,LP}(1+\xi-\xi')
\label{eq:scalelessA0}\\
&=&\frac{1}{4x}\,
 \int_0^\infty dr dr'\int_x^1 d\xi\enp{C_g^{\rm B1*}(\np p,r' Q,\bar{r}' Q)}\enp{
C_g^{\rm B1}(\np p,r Q,\bar{r}Q)}\nonumber\\
&&\times\,\enp{\mathcal{J}^{(qg)}_{\overline{\text{hc}}}\left(\frac{Q^2(\xi-x)}{x},r,r'\right)}f_{q/N}^\text{thr,LP}(\xi)\,,
\label{eq:scaleless}
\end{eqnarray}
expressed in terms of the dimensionally regulated component functions, are scaleless. Both expressions are also identical when the above two refactorization relations are inserted into the second expression. To derive the factorization theorem in the refactorization scheme, we start by subtracting \eqref{eq:scaleless} from the original factorization theorem \eqref{eq:Wtogether}. The scaleless integral is then split into two regions, $R_1=\omega<\Lambda\land\omega'<\Lambda$ and $R_2=\omega>\Lambda\lor\omega'>\Lambda = [0,\infty)\times [0,\infty)\,\backslash\, R_1$. The scale $\Lambda$ at which the integral is split can be chosen arbitrarily within the constraint $\Lambda_\text{QCD}^2/(Q(1-x))\ll\Lambda\leq Q$. 
Region $R_1$ is then combined with the B1 contribution of the factorization theorem.\footnote{In the terminology of \cite{Beneke:2022obx}, this corresponds to ``version~(1)'' of the subtraction, as detailed in App.~B of that reference.} By construction, the integrand of the B1 term is equal to the integrand of $R_1$ in the small-$r$ limit, and hence the endpoint divergences as $r, r^{\prime} \to 0$ are regulated by this subtraction. Similarly, we combine region $R_2$ in the form of \eqref{eq:scalelessA0} with the A0 contribution. The subtractions are effective in the region where $\omega^{(\prime)}>\Lambda$, and hence the $\omega^{(\prime)}\to\infty$ divergences are regularized. The endpoint-rearranged and divergence-free factorization formula for the hadronic tensor reads 
\begin{eqnarray}
W^\text{NLP}_\phi&=&\frac{1}{4x}\,
 \int_0^1 dr dr'\int_x^1 d\xi\,\Bigg\{C_g^{\rm B1*}(\np p,r' Q,\bar{r}' Q)
C_g^{\rm B1}(\np p,r Q,\bar{r}Q)\nonumber\\
&&\qquad\times\mathcal{J}^{(qg)}_{\overline{hc}}\left(\frac{Q^2(\xi-x)}{x},r,r'\right)\nonumber\\
&&\quad-\theta\left(\frac\Lambda Q-r\right)\theta\left(\frac\Lambda Q-r'\right)\enp{C_g^{\rm B1*}(\np p,r' Q,\bar{r}' Q)}\enp{
C_g^{\rm B1}(\np p,r Q,\bar{r}Q)}\nonumber\\
&&\qquad\times\enp{\mathcal{J}^{(qg)}_{\overline{hc}}\left(\frac{Q^2(\xi-x)}{x},r,r'\right)}\Bigg\} \,f^\text{thr,LP}_{q/N}(\xi)\nonumber\\
&&+\left|C_g^{\rm A0}(\np p,Q)\right|^2\,\frac{Q^2}{x^2}
 \int_x^1 d\xi\,
\mathcal{J}^{(g)}_{\overline{hc}}\left(Q^2\frac{\xi-x}{x}\right)\rf_{g/N}(\Lambda,\xi)\,,
\label{eq:Wtogetherref}
\end{eqnarray}
where the $d\to 4$ limit can now be taken. Since the scale $\Lambda$ arises from splitting a scaleless integral, the dependence on it cancels once the A0 and B1 terms are added. We also introduced the (bare) NLP gluon PDF in the refactorization scheme as
\begin{eqnarray}
    \rf_{g/N}(\Lambda,\xi)&=&\frac{1}{2}\int d\xi'\,\Bigg\{\int_0^\infty \frac{d\omega}{\omega} \frac{d\omega'}{\omega'}D_g^{\rm B1}(-n_+ p\omega) D_g^{{\rm B1}\ast}(-n_+ p\omega')S_{\rm sc, sub}^{\rm NLP}(\np p\bar{\xi}^\prime,\omega,\omega')\nonumber\\
    &&-\int_{R_2} \frac{d\omega}{\omega} \frac{d\omega'}{\omega'}\,D_g^{\rm B1}(-n_+ p\omega) \,D_g^{{\rm B1}\ast}(-n_+ p\omega')\enp{S_{\rm sc, sub}^{\rm NLP}(\np p\bar{\xi}^\prime,\omega,\omega')}\Bigg\}\nonumber\\[0.1cm]
    &&\times f^{\rm thr, LP}_{q/N}(1+\xi-\xi') \,.
\label{eq:refNLPpdfdef}
\end{eqnarray}
This expression can be transformed so that the endpoint finiteness is clearly visible:
\begin{eqnarray}
    \rf_{g/N}(\Lambda,\xi)&=&\frac{1}{2}\int d\xi'\,\Bigg\{\int_0^\Lambda\frac{d\omega}{\omega} \frac{d\omega'}{\omega'}\,D_g^{\rm B1}(-n_+ p\omega) D_g^{{\rm B1}\ast}(-n_+ p\omega')S_{\rm sc, sub}^{\rm NLP}(\np p\bar{\xi}^\prime,\omega,\omega')\nonumber\\
    &&+\int_{R_3} \frac{d\omega}{\omega} \frac{d\omega'}{\omega'}\,D_g^{\rm B1}(-n_+ p\omega) \,D_g^{{\rm B1}\ast}(-n_+ p\omega')\,\Big[S_{\rm sc, sub}^{\rm NLP}(\np p\bar{\xi}^\prime,\omega,\omega')\nonumber\\
    &&-\enp{S_{\rm sc, sub}^{\rm NLP}(\np p\bar{\xi}^\prime,\omega,\omega')}\Big]\Bigg\} \,f^{\rm thr, LP}_{q/N}(1+\xi-\xi')
    \,,
\label{eq:refPDFexpl}
\end{eqnarray}
with $R_3$ denoting the region where $\omega>\Lambda \,\land\, \omega'<\Lambda$ and vice versa. While the integrals in  the first line of this equation are cut off at $\Lambda$, and therefore no $\omega^{(\prime)}\to\infty$ divergence can arise, the endpoint region in the second line where only one of $\omega$, $\omega'$ becomes large is regularized by subtraction. We furthermore made use of the fact that $\Lambda\gg\Lambda_\text{QCD}^2/(Q(1-x))$ 
and therefore when both $\omega^{(\prime)}>\Lambda$, $S_{\rm sc, sub}^{\rm NLP}$ and $\enp{S_{\rm sc, sub}^{\rm NLP}}$ are equal up to non-perturbative power corrections, which can be dropped to leading-twist accuracy. 
We stress that \eqref{eq:refNLPpdfdef} defines a new scheme for the NLP gluon PDF, which differs from the $\overline{\text{MS}}$ scheme at NLP in the $\xi\to 1$ limit. 

Conventional four-dimensional component functions can now be employed, and no endpoint divergences are generated. This also means that standard RGEs for resummation can be used. We have already discussed the freedom in choosing $\Lambda$. Here, we would like to point out a particularly interesting limit: $\Lambda\to Q$. In this limit, all leading logarithmic contributions are assigned to the A0 contribution, which takes the form of the LP factorization equation \eqref{eq:LPfact2}. After the transformation to Mellin space, one solves the RGEs for the component functions in the A0 term to the desired logarithmic accuracy using known methods; hence, $\Lambda\to Q$ is a convenient choice. 

\subsection{Leading-logarithmic resummation in the refactorization\\ scheme}

In this section, we compute the leading-logarithmic (LL) contributions to the hadronic tensor. For resummation at LL accuracy, it is sufficient to solve the RGE with only the one-loop cusp anomalous dimension. Furthermore, there are no contributions from the component functions at the matching scales beyond the first non-vanishing order. In this approximation, all RGEs follow the same structural form 
\begin{equation}
    \frac{d}{d\ln\mu}h(Q_X,\mu)=C_h\gamma_\text{cusp}(\alpha_s(\mu))\ln\frac{-Q_X^2}{\mu^2}\,h(Q_X,\mu)\,,
    \label{eq:genericRGE}
\end{equation}
where $Q_X$ is the scale of the respective function, $C_h$ a corresponding colour factor, and the one-loop cusp anomalous dimension given in \eqref{eq:cusp} is to be used. The general solution to such an RGE is
\begin{equation}
    h(Q_X,\mu)=\exp\left[2C_hS(\mu_X,\mu)\right]\left(\frac{-Q_X^2}{\mu_X^2}\right)^{\!\!-C_h a_\text{cusp}(\mu_X,\mu)}h(Q_X,\mu_X)
\end{equation}
with $\mu_X$ of order $Q_X$ the matching scale chosen such that there are no large logarithms in the initial value $h(Q_X,\mu_X)$. The functions  $S(\nu,\mu)$ and $a_\text{cusp}(\nu,\mu)$ are given by
\begin{equation}
    S(\nu,\mu)=-\int_{\as(\nu)}^{\as(\mu)}d\alpha\,\frac{\gamma_\text{cusp}(\alpha)}{\beta(\alpha)}\int_{\as(\nu)}^\alpha\frac{d\alpha'}{\beta(\alpha')}\,,\qquad a_\text{cusp}(\nu,\mu)=-\int_{\as(\nu)}^{\as(\mu)}d\alpha\frac{\gamma_\text{cusp}(\alpha)}{\beta(\alpha)}\,.
\end{equation}
At LL accuracy, these reduce to
\begin{equation}
S(\nu,\mu)=\frac{4\pi}{\beta_0^2\as(\nu)}\left(1-\frac1r-\ln r\right)\,,\qquad 
a_\text{cusp}(\nu,\mu)=\frac{2}{\beta_0}\ln r,
\end{equation}
with $r=\as(\mu)/\as(\nu)$ and $\beta_0=11-\frac{2 n_f}{3}$ as the one-loop beta-function coefficient.

The scale for the hardcollinear $D_g^{\rm B1}$ coefficient and the NLP soft-collinear function is dependent on the convolution variable $\omega$. Although $\omega$ technically scales as $\Lambda_{\rm QCD}^2/(Q(1-x))$, 
the presence of endpoint divergences implies relevant contributions from $\omega$ values larger than its natural one $\Lambda_{\rm QCD}^2/(Q(1-x))$.  
To avoid large logarithms in the initial conditions, we adopt dynamical scale setting \cite{Beneke:2022obx,Liu:2020eqe,Liu:2022ajh}. This means that at every value of the convolution variable $\omega$, the matching scale is set such that no large logarithms remain in the initial conditions. This ensures that all LL contributions are accounted for by the evolution kernels. 
The matching scales are parametrically of order 
\begin{equation}
    \mu_Q^2\sim Q^2,\quad \mu_{\overline{\rm hc}}^2\sim \frac{Q^2}{N}\,,\quad\mu_{D^{\text{B1}}}^2\sim n_+p\omega\,,\quad\mu_{sc}^2\sim\frac{n_+p\omega}{N}\,,
\label{eq:initialscales}
\end{equation}
and with corresponding $\mu_{D^{\text{B1}}}^{\prime \,2}$, $\mu_{sc}^{\prime\,2}$ defined by replacing $\omega$ with $\omega^\prime$.

As mentioned above, it is convenient to choose $\Lambda=Q$, in which case the anti-hardcollinear B1 term \eqref{eq:Wtogetherref} can be neglected at LL accuracy.\footnote{It would be sufficient at LL accuracy to choose $\Lambda\sim Q$.} The solutions to the RGEs for the component functions needed for the resummation of the remaining soft-collinear A0 term at LL accuracy are then given by
\begin{eqnarray}
C_g^\text{A0}(Q^2,\mu^2)&=&\exp[2C_AS(\mu_Q,\mu)]\left(\frac{-Q^2}{\mu_Q^2}\right)^{\!\!-C_A a_\text{cusp}(\mu_Q,\mu)}
\hspace*{-2.2cm}\,,\\
\mathcal{J}_{\overline{\text{hc}}}^{(g)}(\mu^2)&=&\exp[-4C_AS(\mu_{\overline{\rm hc}},\mu)]\left(\frac{-Q^2}{N \mu_{\overline{\rm hc}}^2}\right)^{\!2C_A a_\text{cusp}(\mu_{\overline{\rm hc}},\mu)}\hspace*{-2.2cm}\,,\\
D^\text{B1}_g(-n_+p\omega,\mu^2)&=&\exp[2(C_F-C_A)S(\mu_{D^\text{B1}},\mu)]\left(\frac{-n_+p\omega}{\mu_{D^\text{B1}}^2}\right)^{\!\!-(C_F-C_A) a_\text{cusp}(\mu_{D^\text{B1}},\mu)}\hspace*{-2.2cm}\,,\\
S_\text{sc,sub}^\text{NLP}(\omega,\omega',\mu^2)&=&S_\text{sc,sub}^\text{NLP,LO}(\omega,\omega')\,\exp[-2(C_F-C_A)\,(S(\mu_{sc},\mu)+S(\mu_{sc}^\prime,\mu))]\qquad\notag\\
&&\times \left(\frac{-n_+p\omega}{N\mu_{sc}^2}\right)^{\!(C_F-C_A) a_\text{cusp}(\mu_{sc},\mu)}\,
\left(\frac{-n_+p\omega^\prime}{N\mu_{sc}^{\prime \,2}}\right)^{\!(C_F-C_A) a_\text{cusp}(\mu_{sc}^\prime,\mu)}\hspace*{-1cm},
\label{eq:SscNLPLL}\\
\enp{S_\text{sc,sub}^\text{NLP}(\omega,\omega',\mu^2)}&=&\frac{\as C_F}{\pi}\frac{1}{N}\,\omega\delta(\omega-\omega')\exp[-4(C_F-C_A)S(\mu_{sc},\mu)]\notag\\
&&\times\left(\frac{-n_+p\omega}{N\mu_{sc}^2}\right)^{\!2(C_F-C_A) a_\text{cusp}(\mu_{sc},\mu)}\hspace*{-1cm}.
\label{eq:SscNLPLLdoublebracket}
\end{eqnarray}
Here, we have already used the simplification that in the LL the lowest order initial conditions suffice, which equal 1 for the hard coefficient, anti-hardcollinear jet function, and $D^\text{B1}$ function. The initial condition for the NLP soft function is, in general, non-perturbative, but for the large $\omega^{(\prime)}$-limit $\enp{S_\text{sc,sub}^\text{NLP}(\omega,\omega',\mu^2)}$ the lowest order expression \eqref{eq:diffNLPsctree} in $d=4$ can be used in \eqref{eq:SscNLPLLdoublebracket}. The full LL-resummed, Mellin moment-space hadronic tensor 
can then be read off from \eqref{eq:Wtogetherref} and \eqref{eq:refNLPpdfdef}: 
\begin{align}
    W_\phi^\text{NLP,LL}=\,&\frac{Q^2}{2x^2}\exp\big[4C_A(S(\mu_Q,\mu)-S(\mu_{\overline{\rm hc}},\mu))\big]\notag\\
    &\times\Bigg\{\int_0^Q\frac{d\omega}{\omega}\frac{d\omega'}{\omega'}\exp\big[2(C_F-C_A)(S(\mu_{D^{\text{B1}}},\mu)+S(\mu'_{D^{\text{B1}}},\mu)-2S(\mu_{sc},\mu))\big]\notag\\
    &\quad\times S_\text{sc,sub}^\text{NLP}(\omega,\omega',\mu_{sc}^2)\notag\\
    &+\int_{R_3}\frac{d\omega}{\omega}\frac{d\omega'}{\omega'}\exp\big[2(C_F-C_A)(S(\mu_{D^{\text{B1}}},\mu)+S(\mu'_{D^{\text{B1}}},\mu)-2S(\mu_{sc},\mu))\big]\notag\\
    &\quad\times\left(S_\text{sc,sub}^\text{NLP,LO}(\omega,\omega')-\frac{\as C_F}{\pi}\frac{1}{N}\,\omega\delta(\omega-\omega')\right)\Bigg\}\,f_{q/N}^\text{thr,LP}(\mu)\,.
\label{eq:wphiLL}    
\end{align}
In this expression, the initial scales $\mu_Q$ etc. appearing in \eqref{eq:initialscales} should all be set {\em equal} to their respective canonical values. In this case, it can be shown that the $\mu$ dependence on the right-hand side, including that of the parton distribution, cancels to LL accuracy. 
The exponent in the first line stems from the hard coefficient and the anti-hardcollinear jet function.
We emphasize that, as was the case for the $\msbar$ scheme, the initial condition of the NLP soft-collinear function is non-perturbative. Note, however, that the formula above for the hadronic tensor has been written in a form that is valid beyond perturbation theory.

In the double-logarithmic approximation, we can further approximate the function $S$ in the exponent by  
\begin{equation}
    S(\nu,\mu)=-\frac{\as}{8\pi}\,\ln^2\frac{\mu^2}{\nu^2}\,.
\end{equation}
We now define the perturbative NLP conversion function $\hat V_{gq}^{\circledR}(\Lambda,\mu)$ in the refactorization scheme through
\begin{equation}
    \rf_{g/N}(\Lambda,\mu)=\mathcal{U}_{gq}^{\text{NLP},\circledR}(\Lambda,\mu_0)f_{q/N}^\text{thr,LP}(\mu_0)=\hat V_{gq}^{\circledR}(\Lambda,\mu_0)\,\mathcal{U}_{qq}^{\text{LP}}(\mu_0)f_{q/N}^\text{thr,LP}(\mu_0)\,,
\end{equation}
in analogy with Sec.~\ref{sec:MSbar} for the $\msbar$ scheme. We introduce the lower limit $\mu_0^2/(n_+p)$ in the $\omega^{(\prime)}$ convolution integrals in \eqref{eq:refNLPpdfdef} as the factorization scale between perturbative and non-perturbative PDFs, in the same way as in \eqref{eq:omY}.
Starting from \eqref{eq:Uijdef} and using the refactorization scheme version of \eqref{eq:Ugqfact}, which can be obtained from the explicit formula for the PDF in the refactorization scheme \eqref{eq:refPDFexpl}, we then find for the NLP conversion function 
\begin{equation}
\label{eq:Vgqref}
\begin{aligned}
    \hat V_{gq}^{\circledR}(\Lambda,\mu)=&\;\frac{1}{2N L_N}\frac{C_F}{C_F-C_A}\,\Bigg\{\exp\bigg[\frac\as\pi(C_F-C_A)\left(L_N\ln\frac{\mu^2}{\mu_0^2}+\frac{L_N^2}2\right)\bigg]\\
    &-\exp\bigg[\frac\as\pi(C_F-C_A)\left(L_N \ln\frac{\mu^2}{\Lambda Q}+\frac{L_N^2}2\right)\bigg]\,\Bigg\}\,.
\end{aligned}
\end{equation}
By comparing this result to the $\msbar$ scheme expression \eqref{eq:Vgqmsbar}, it is remarkable that the Bernoulli numbers appear to be specific to $\msbar$ subtractions. The form \eqref{eq:Vgqref} 
of the resummed off-diagonal PDF in the refactorization scheme is indeed more similar to its time-like analogue, resummed ``gluon-thrust'', which is free of collinear divergences, hence no subtractions are needed, and it does not feature the Bernoulli numbers \cite{Beneke:2022obx}.

\subsection{Relationship between the two schemes }

We proceed to derive the relation between the PDFs in the $\msbar$  and the new refactorization scheme and demonstrate that this scheme relation factor is a perturbatively calculable object.

At LP, the PDFs are the same in both schemes. In Mellin space, the partonic PDFs are renormalized multiplicatively
\begin{eqnarray}
    f_{i/N}^\text{thr,LP}(\mu)&=&Z_{ii}^\text{LP}f_{i/N,\text{bare}}^{\text{thr,LP}}\,.
\end{eqnarray}
For better intelligibility in this section, we make the bare quantities explicit, while the renormalized quantities explicitly depend on the renormalization scale $\mu$. The LP renormalization factors $Z_{ii}^{\rm LP}$ have been derived in Sec.~\ref{sec:LPPDFresum} and are given in \eqref{eq:diagZLP}. 

 At NLP, the renormalized PDFs mix under RG evolution because, in addition to soft gluons, soft quarks need to be taken into account. The renormalized NLP gluon PDF in the $\msbar$ scheme is given in all generality by
\begin{equation}
\begin{aligned}\label{eq:msbarren}
    f_{g/N}^\text{thr,NLP}(\mu)=&\;Z_{gg}^\text{LP}f_{g/N,\text{bare}}^{\text{thr,NLP}}+Z_{gg}^\text{NLP}f_{g/N,\text{bare}}^{\text{thr,LP}}\\
    &+Z_{gq}^\text{NLP}f_{q/N,\text{bare}}^{\text{thr,LP}}+Z_{g\bar{q}}^\text{NLP}f_{\bar{q}/N,\text{bare}}^{\text{thr,LP}}.
    \end{aligned}
\end{equation}
In this work, we focus exclusively on off-diagonal NLP terms, disregarding the diagonal ones generated by next-to-soft gluons. We tacitly assume hereby that endpoint divergences in the diagonal part do not mix with off-diagonal contributions. We can therefore simplify the previous equation to
\begin{equation}\label{eq:msbarrenoNLP}
    f_{g/N}^\text{thr,NLP}(\mu)=Z_{gg}^\text{LP}f_{g/N,\text{bare}}^{\text{thr,NLP}}+Z_{gq}^\text{NLP}f_{q/N,\text{bare}}^{\text{thr,LP}} 
   \; \big[+Z_{g\bar{q}}^\text{NLP}f_{\bar{q}/N,\text{bare}}^{\text{thr,LP}} \big],
\end{equation}
and, as before, do not discuss explicitly the identical antiquark term in brackets.
We emphasize again that the subtraction of both UV and endpoint divergences is only performed \textit{after} the convolutions in $\omega^{(\prime)}$ have been carried out. 
The standard $\msbar$ $Z$-factors do not differentiate between UV and endpoint divergences and remove both types of divergences simultaneously.

In the refactorization scheme, the endpoint divergences are removed by  endpoint subtractions, and the NLP PDF is renormalized through the renormalization of its component functions (in the $\overline{\rm MS}$ scheme). Since the $\omega^{(\prime)}$ convolutions do not introduce divergences any longer, the convolution and renormalization operations commute. In order to derive the relation between the two schemes, we may therefore express the refactorization scheme PDF in terms of bare functions where poles are resummed to all orders in $\as$ in $d$ dimensions, and renormalize the UV divergences after the $\omega^{(\prime)}$ integrations. 

We start from the analogue of \eqref{eq:msbarrenoNLP} for the refactorization-scheme PDF,
\begin{equation}\label{eq:refactrenoNLP}
    f_{g/N}^\text{NLP,\circledR}(\Lambda,\mu)=Z_{gg}^\text{LP}f_{g/N,\text{bare}}^{\text{NLP,\circledR}}(\Lambda)+Z_{gq}^\text{NLP,\circledR}(\Lambda)f_{q/N,\text{bare}}^{\text{thr,LP}}\,.
\end{equation}
The bare refactorization-scheme PDF was defined in \eqref{eq:refNLPpdfdef} by 
\begin{equation}
\label{eq:brefgNrelation}
 f_{g/N,\rm bare}^\text{NLP,\circledR}(\Lambda) = 
 f_{g/N,\text{bare}}^{\text{thr,NLP}} - \int_{R_2}d\omega d\omega'\enp{I_\text{bare}(\omega,\omega')} \,f_{q/N,\text{bare}}^{\text{thr,LP}}\,,
 \end{equation}
 where $R_2$ was defined below \eqref{eq:scaleless}, and the integrand of the bare NLP PDF has been abbreviated as
\begin{equation}
    I_{\text{bare}}(\omega,\omega')=\frac{1}{2}\frac1\omega\frac1{\omega'} D^{\text{B1}}_{g,\text{bare}}(\omega)D^{\text{B1}\ast}_{g,\text{bare}}(\omega') S^{\text{NLP}}_\text{sc,sub,bare}(\omega,\omega') \,.   
\end{equation}
Next, we insert \eqref{eq:brefgNrelation} into \eqref{eq:refactrenoNLP} and introduce the renormalized LP quark PDF, 
 \begin{eqnarray}\label{eq:deltaVdefpre}
 f_{g/N}^\text{NLP,\circledR}(\Lambda,\mu)&=&
 Z_{gg}^\text{LP}f_{g/N,\text{bare}}^{\text{thr,NLP}}+Z_{gq}^\text{NLP}f_{q/N,\rm bare}^{\text{thr,LP}} \nonumber\\
 && 
 - Z_{gg}^\text{LP} \int_{R_2}d\omega d\omega'\enp{I_\text{bare}(\omega,\omega')}\, (Z_{qq}^{\rm LP})^{-1}  \,f_{q/N}^{\text{thr,LP}}(\mu)\nonumber\\
 &&  +
\, (Z_{gq}^\text{NLP,\circledR}(\Lambda)-Z_{gq}^{\rm NLP})\, (Z_{qq}^{\rm LP})^{-1} \,f_{q/N}^{\text{thr,LP}}(\mu) \,.\qquad\quad
 \end{eqnarray}
 The first line is the renormalized $\overline{\rm MS}$ NLP gluon PDF. Now we {\em define} $Z_{gq}^\text{NLP,\circledR}(\Lambda)$ such that the third line  cancels the pole part of the second line, leaving 
 \begin{equation}\label{eq:frefreno}
 f_{g/N}^\text{NLP,\circledR}(\Lambda,\mu)=
f_{g/N}^{\text{thr,NLP}}(\mu)
 - \left[Z_{gg}^\text{LP} \int_{R_2}d\omega d\omega'\enp{I_\text{bare}(\omega,\omega')}\, (Z_{qq}^{\rm LP})^{-1}\right]_{\rm finite}  \!\!f_{q/N}^{\text{thr,LP}}(\mu)\,.
 \end{equation}
We then introduce the relation between the refactorization scheme PDF and the $\overline{\text{MS}}$ scheme PDF as
\begin{equation}
    \begin{aligned}\label{eq:deltaZdef}
        \Delta f_g^\text{NLP}&=\rf_{g/N}(\Lambda,\mu)-f_{g/N}^\text{thr,NLP}(\mu)\\
        &\equiv\Delta V_{gq}^\text{NLP}(\Lambda)f_{q/N}^\text{thr,LP}(\mu)\,.
    \end{aligned}
\end{equation}

In the following, we compute the scheme relation explicitly from \eqref{eq:deltaZdef} in the double-logarithmic approximation. 
We take the $d$-dimensional expressions for the resummed $D_g^\text{B1}$ and NLP soft-collinear functions from \eqref{eq:resumDB1} and \eqref{eq:resummSsc}, respectively, use \eqref{eq:diffNLPsctree} as the initial condition for the latter, and perform the remaining $\omega$ convolution with \eqref{eq:omY}, arriving at 
\begin{equation}
    \begin{aligned}
        \Delta V_{gq}^\text{NLP}(\Lambda)=&\;-\frac1{2N}\frac{C_F}{C_F-C_A}\bigg[Z_{gg}^\text{LP}(Z_{qq}^\text{LP})^{-1}\frac{\eps N^\eps}{N^\eps-1}\\
        &\qquad\times\left(\exp\left[\frac{\as(C_F-C_A)}{\pi\eps^2} \,(N^\eps-1)\left(\frac{\mu^2}{\Lambda Q}\right)^{\!\eps}\,\right]-1\right)\bigg]_\text{finite}.
    \end{aligned}
\end{equation}
Without the factor $Z_{gg}^\text{LP}(Z_{qq}^\text{LP})^{-1}$, this has the same form as eq.~(3.50) of \cite{Beneke:2020ibj} under the replacements  $Q^2\to \Lambda Q$, and $C_A\to0$, $C_F\to C_F-C_A$ in this order in the arguments in the exponential functions.
Following the same steps as laid out in this reference, we find, in analogy to eq.~(3.58) in that article,
\begin{equation}
    \begin{aligned}
        \Delta V_{gq}^\text{NLP}(\Lambda)=&\;\frac1{2NL_N}\frac{C_F}{C_F-C_A}\bigg[Z_{gg}^\text{LP}(Z_{qq}^\text{LP})^{-1}\frac{w}{e^w-1}\exp\left(\frac{\as(C_F-C_A)}\pi\frac{L_N}\eps\right)\\
        &\qquad\times\left(e^{a/w}-e^{\hat{S}_{FA}}\right)\bigg]_\text{finite},
    \end{aligned}\label{eq:DZcalc1}
\end{equation}
where $w$ and $a$ are defined as in~\eqref{eq:defwa}. We also introduced the abbreviation
\begin{equation}
        \hat{S}_{FA}=\frac{\as(C_F-C_A)}{\pi}\frac1{\eps^2}\left[\left(\frac{\mu^2}{\Lambda Q}\right)^{\!\eps}(N^\eps-1)-\eps L_N\right]\,.
 \end{equation}
The exponential term in the first line of \eqref{eq:DZcalc1} cancels $Z_{gg}^\text{LP}(Z_{qq}^\text{LP})^{-1}$ exactly, as is readily visible from \eqref{eq:diagZLP}. Moreover, it was already noted in \cite{Beneke:2020ibj} that $w/(e^w-1)$ and $e^{\hat{S}_{FA}}$ are finite for $\eps\to0$. The finite part of $w/(e^w-1)\,e^{a/w}$ is the Bernoulli function appearing in \eqref{eq:Pll_B0}.
Combining everything, we therefore find
\begin{equation}
\begin{aligned}\label{eq:deltaZexp}
    \Delta V_{gq}^\text{NLP}(\Lambda)=&\;\frac{1}{2NL_N}\frac{C_F}{C_F-C_A}\,\Bigg\{\mathcal{B}_0\left(\frac\as\pi(C_F-C_A)L_N^2\right)\\
    &-\exp\bigg[\frac\as\pi(C_F-C_A)\left(L_N \ln\frac{\mu^2}{\Lambda Q}+\frac{L_N^2}{2}\right)\bigg]\,\Bigg\}\,.
    \end{aligned}
\end{equation}

Since the IR physics for the PDFs is the same in both schemes, we can rewrite \eqref{eq:deltaZdef} in terms of the previously introduced NLP conversion functions 
\begin{equation}
    \begin{aligned}
        \Delta f_g^\text{NLP}&=\rf_{g/N}(\Lambda,\mu)-f_{g/N}^\text{thr,NLP}(\mu)\\
        &=[\hat V_{gq}^\circledR(\Lambda,\mu)-\hat V_{gq}(\mu)]\,U_{qq}^\text{LP}(\mu)f_{q/N}^\text{thr,LP}(\mu_0)\,.
    \end{aligned}
\end{equation}
This provides the alternative expression 
\begin{equation}
    \Delta V_{gq}^\text{NLP}(\Lambda)=\hat V_{gq}^\circledR(\Lambda,\mu)-\hat V_{gq}(\mu)
\end{equation}
for the scheme relation factor. Comparing the difference of the explicit expressions \eqref{eq:Vgqmsbar} and \eqref{eq:Vgqref} with the explicit derivation that led to \eqref{eq:deltaZexp}, we find agreement. We note that the dependence on the IR matching scale $\mu_0$ cancels out in the difference of the two conversion factors. This is indeed expected because the two schemes differ in their treatment of the endpoint divergences, which occur in the large $\omega^{(\prime)}$ region. Hence, both schemes should feature the same dependence on the low-energy scale $\mu_0$. 
The two ways of computing $\Delta f_g^\text{NLP}$ serve as a strong cross-check of the derivation.

We close this subsection with the observation that the renormalization condition for the NLP PDF in the refactorization scheme \eqref{eq:refactrenoNLP} can be expressed in the simpler form 
\begin{equation} 
\begin{aligned}\label{eq:refrendef}
    \rf_{g/N}(\Lambda,\mu)=&\;Z_{gg}^\text{LP}\,\Bigg\{\int_{0}^\infty d\omega d\omega'\,I_\text{bare}(\omega,\omega')-\int_{R_2} d\omega d\omega'\,\enp{I_\text{bare}(\omega,\omega')}\Bigg\} \,f^{\rm thr, LP}_{q/N, \text{bare}} \,.
\end{aligned}
\end{equation}
To see this, we start by comparing \eqref{eq:deltaVdefpre} and \eqref{eq:frefreno} and solve for
\begin{eqnarray}
Z_{gq}^{\mathrm{NLP},\circledR}
&=& -\left[
Z_{gg}^{\mathrm{LP}}
\int_{R_2}d\omega\,d\omega'\,
\enp{I_{\mathrm{bare}}(\omega,\omega')}
(Z_{qq}^{\mathrm{LP}})^{-1}
\right]_{\mathrm{finite}}
Z_{qq}^{\mathrm{LP}}
\nonumber\\
&&
+\,Z_{gg}^{\mathrm{LP}}
\int_{R_2}d\omega\,d\omega'\,
\enp{I_{\mathrm{bare}}(\omega,\omega')}
+Z_{gq}^{\mathrm{NLP}}
\nonumber\\[0.15cm]
&=&
\left[
Z_{gg}^{\mathrm{LP}}
\int_{R_2}d\omega\,d\omega'\,
\enp{I_{\mathrm{bare}}(\omega,\omega')}
(Z_{qq}^{\mathrm{LP}})^{-1}
\right]_{\mathrm{poles}}
Z_{qq}^{\mathrm{LP}}
+Z_{gq}^{\mathrm{NLP}}\,.
\label{eq:ZgqR}
\end{eqnarray}
Subsequently, we rewrite the off-diagonal renormalization factor $Z_{gq}^\text{NLP}$ of the $\msbar$ scheme in terms of explicit component functions. Starting from the renormalization condition \eqref{eq:msbarrenoNLP}, we find, after extracting a factor of $f_{q/N}^\text{thr,LP}(\mu)$, that
\begin{equation}
    Z_{gq}^\text{NLP}(Z_{qq}^\text{LP})^{-1}+Z_{gg}^\text{LP}(Z_{qq}^\text{LP})^{-1}\int_0^\infty d\omega d\omega' I_\text{bare}(\omega,\omega')=\text{finite}
\end{equation}
must be fulfilled. In other words,
\begin{equation}
    Z_{gq}^\text{NLP}(Z_{qq}^\text{LP})^{-1}=-\left[Z_{gg}^\text{LP}(Z_{qq}^\text{LP})^{-1}\int_{0}^\infty d\omega d\omega' I_\text{bare}(\omega,\omega')\right]_\text{poles},
\end{equation}
since in the $\msbar$ scheme all $Z$ factors are pure poles.\footnote{In the DLA, this expression reproduces \eqref{eq:cond1_chapter}. In the present case, all poles are UV poles since the non-perturbative soft-collinear function in $I_{\rm bare}$ does not lead to IR singularities at small $\omega^{(\prime)}$. To obtain \eqref{eq:cond1_chapter}, the IR poles have been regularized by $\Lambda_{\rm IR}$, which allows for the direct use of the perturbative quantity 
$\enp{I_\text{bare}}$.}  
Since all poles are either UV poles or endpoint divergences that occur at large $\omega^{(\prime)}$ values, we can replace $I_\text{bare}(\omega,\omega')$ by $\enp{I_\text{bare}(\omega,\omega')}$ and split the integral into the two regions $R_1$ and $R_2$ as defined before,
\begin{eqnarray}
    Z_{gq}^\text{NLP}(Z_{qq}^\text{LP})^{-1}&=&-\,\Bigg[Z_{gg}^\text{LP}(Z_{qq}^\text{LP})^{-1}\,\bigg(\,\int_{R_1} \!d\omega d\omega' \,\enp{I_\text{bare}(\omega,\omega')}
    \nonumber\\
    &&+\,\int_{R_2} d\omega d\omega' \enp{I_\text{bare}(\omega,\omega')}\bigg)\,\Bigg]_\text{poles}\,,
\label{eq:Zgqdiff}
\end{eqnarray}
where now ``$|_{\rm pole}$'' refers to UV poles and 
poles from  large $\omega^{(\prime)}$, but not to the IR poles 
generated by the replacement of $I_\text{bare}$ with 
$\enp{I_\text{bare}}$. 
The first term in this equation is actually finite, which can be understood as follows. For fixed values of $\omega^{(\prime)}$, the A0 contribution of the factorization theorem (the second term in \eqref{eq:Wtogether}) is scale invariant. That implies that the combined renormalization factor of $\enp{I_\text{bare}}\, f_{q/N,\text{bare}}^\text{LP}$ is the inverse of the renormalization factor of $C_g^\text{A0}\mathcal{J}_{\overline{\text{hc}}}^{(g)}$.  The scale invariance of the LP diagonal gluon channel \eqref{eq:WLPgluon} now dictates that the product $C_g^\text{A0}\mathcal{J}_{\overline{\text{hc}}}^{(g)}$ is renormalized by $(Z_{gg}^\text{LP})^{-1}$. Performing the $\omega^{(\prime)}$ integrals does not introduce any further divergences. Thus, $Z_{gg}^\text{LP}(Z_{qq}^\text{LP})^{-1}$ renormalizes $\int_{R_1} d\omega d\omega' \enp{I_\text{bare}(\omega,\omega')}$, and this term does not contribute to the pole part of \eqref{eq:Zgqdiff}. Therefore
\begin{equation}
Z_{gq}^{\mathrm{NLP}}
=-\left[
Z_{gg}^{\mathrm{LP}}
\int_{R_2}d\omega\,d\omega'\,
\enp{I_{\mathrm{bare}}(\omega,\omega')}
(Z_{qq}^{\mathrm{LP}})^{-1}
\right]_{\mathrm{poles}}
Z_{qq}^{\mathrm{LP}}\,,
\end{equation}
and consequently, from \eqref{eq:ZgqR} we find $Z_{gq}^{\text{NLP},\circledR}=0$, hence \eqref{eq:refactrenoNLP} simplifies to \eqref{eq:refrendef}.

\subsection{Juxtaposition of the two schemes}

We briefly summarize our findings for both the $\msbar$ scheme and the proposed re\-factori\-zation-scheme here, as well as their implications for the computation of the hadronic tensor. The factorization theorem for the latter in the two schemes is given in~\eqref{eq:Wtogether} and \eqref{eq:Wtogetherref}, respectively. The main difference is that in the $\msbar$ scheme, the two summands are ill-defined in $d=4$ dimensions due to the occurrence of endpoint divergences. Keeping $d$-dimensional expressions, these divergences cancel only after all contributions have been added, \textit{once} the convolution integrals are performed. The refactorization scheme offers a possibility to work with four-dimensional expressions instead, since the endpoint divergences are removed by refactorization-based rearrangements of the endpoint contributions. This allows us to solve the standard RGEs for the component functions to resum the large threshold logarithms in the full hadronic tensor. 

Furthermore, in the refactorization scheme, the scale $\Lambda$ at which the endpoint subtractions are implemented provides a lever to control which of the two SCET current contributions contains the leading logarithmic corrections. If this scale is chosen to be the hard scale $\Lambda=Q$, all corrections in the leading logarithmic order (LL) are contained in the A0 term, and the hadronic tensor \eqref{eq:Wtogetherref} in Mellin moment-space simplifies to 
\begin{equation}
W_\phi^{\rm NLP, LL} =\frac{Q^2}{x^2}\left|C_g^{\rm A0}(\np p,Q)\right|^2 
\mathcal{J}^{(g)}_{\overline{hc}}\rf_{g/N}(Q)\,,
\label{eq:WNLPLambdaQ}
\end{equation}
with $\rf_{g/N}(Q)$ given by \eqref{eq:refPDFexpl}. In practice, this redistribution of large logarithms is advantageous because the B1 hard and jet functions obey more complicated integro-differential RGEs beyond  leading-logarithmic accuracy \cite{Beneke:2022obx}.


\section{Conclusion}
\label{sec:conclusion}

In this paper, we developed a systematic and fully operator-based formulation of the factorization and definition of parton distribution functions appropriate for the endpoint region $x \to 1$, and explicitly worked out the case of off-diagonal parton scattering in DIS. We established their factorization within the position-space SCET framework at leading and next-to-leading power in $1-x$, always at leading power (``twist'') in the QCD factorization scale $\Lambda_\text{QCD}$. Our analysis resolves several conceptual issues of previous treatments of the threshold region, including the correct identification of the soft-collinear mode, the factorization of collinear virtualities through frozen-collinear fields, and the emergence and cancellation of endpoint divergences, specifically  at subleading power.

At leading power, we demonstrated that the threshold quark and gluon PDFs coincide with the $\xi \to 1$ limit of the ordinary $\overline{\text{MS}}$ PDFs. This follows from the direct operator-level matching in which hardcollinear and collinear fluctuations are integrated out, leaving a soft-collinear function that combines with other factors to reproduce precisely the standard PDF. We further showed that $\overline{\text{MS}}$ renormalization commutes with the endpoint limit and naturally reproduces the DGLAP kernel, confirming that the EFT derivation is fully consistent with the traditional collinear factorization framework.

At next-to-leading power, our results provide a complete EFT description of the two physical mechanisms contributing to the off-diagonal channel, which were only sketched in \cite{Beneke:2020ibj}: the anti-hardcollinear contribution arising from the power-suppressed B1 current, which describes the final-state jet generated from two partons produced in the hard scattering, and the soft-collinear contribution generated by the time-ordered product of the leading-power A0 current with the subleading Lagrangian, which describes the emission of a parton into the proton remnant. This contribution yields a non-local matrix element, which also contains hardcollinear fluctuations captured by a universal matching coefficient, $D^{\rm B1}_g$. The resulting EFT description generalizes the familiar leading-power factorization theorem and makes transparent the origin of endpoint divergences previously observed in \cite{Beneke:2020ibj}. 

The soft-collinear physics of the subleading-power contribution is encoded in the multi-local NLP soft-collinear function. Performing a soft rearrangement, 
we defined a subtracted NLP soft-collinear function whose ultraviolet renormalization leads to a well-defined anomalous dimension. This procedure identifies and clearly separates IR and UV divergences, enabling  standard RG evolution to resum large logarithms in the perturbative functions. Eventually, we recover leading-logarithmic resummed results previously obtained from $d$-dimensional consistency conditions.

A central outcome of this analysis is the identification and cancellation of the aforementioned endpoint divergences. We presented two independent subtraction schemes for implementing this cancellation.
Using $d$-dimensional resummation, the endpoint divergences manifest as poles in the dimensional regulator that eventually cancel between the two SCET contributions, the anti-hardcollinear B1 current and soft-collinear A0 current terms. In the second approach, the  endpoint-divergent convolution integrals are rendered finite by employing refactorization relations for the integrands representing the two aforementioned physical mechanisms in the problematic endpoint regions. Refactorization introduces a scale $\Lambda$ (which cancels in the final result), which enables the use of standard RGEs to resum leading logarithmic corrections since renormalization can be performed before evaluating the convolutions. By choosing $\Lambda$ appropriately, one can redistribute the logarithmically enhanced and resummed terms between the A0 and B1 contributions.

The refactorization framework also suggests a more natural scheme for the parton distributions at large $x$ than the standard $\msbar$ scheme. The difference between the two schemes is encoded in a calculable conversion coefficient, which itself contains large logarithms that can be summed.  
While we do not propose to abandon the conventional $\overline{\rm MS}$-scheme PDFs in phenomenological analyses, the relevance of refactorization-scheme arises from its universal applicability to threshold-sensitive processes with hadrons in the initial state. To go beyond the leading-logarithmic approximation (at next-to-leading power) in the $d$-dimensional framework is difficult. It is simpler to perform the calculation in the refactorization scheme and benefit from solving four-dimensional RGEs for the component functions---and only at the end replace the refactorization-scheme PDF with the conventional $\overline{\rm MS}$-scheme PDF via the universal relation between the two schemes provided (to double-logarithmic accuracy) in the present work. The derivation of this relation inevitably requires a $d$-dimensional treatment of endpoint singularities due to the nature of the $\overline{\rm MS}$-scheme PDFs, but this is then the only place where a $d$-dimensional treatment is ever necessary.

Finally, we computed the perturbative building blocks required for NLP resummation to LL accuracy, including the composite-operator jet function, the radiative jet function, and the NLP soft-collinear function, together with its anomalous dimension. The analysis confirms that soft subtraction is essential for  consistent NLP factorization and highlights its conceptual relation to, as well as its important distinction from, the zero-bin subtraction familiar at leading power. 
Within the framework established here, the step to higher logarithmic accuracy is only a technical one. This extension to next-to-leading logarithmic resummation accuracy will be presented in forthcoming work \cite{NLPDIS2}.

\subsubsection*{Acknowledgements} 
We benefited from discussions with J.~Strohm during the early stages of this project. The work has been supported in part by the Excellence Cluster ORIGINS funded by the Deutsche Forschungsgemeinschaft 
(DFG, German Research Foundation) under Ger\-many's Excellence Strategy --EXC-2094 --390783311. R.S. is supported by the United States Department of Energy under Grant Contract DE-SC0012704. 
Part of the work of M.S. was supported by the Alexander von Humboldt Foundation through a Feodor Lynen Fellowship. 
M.B. wishes to thank Brookhaven National Laboratory for its hospitality in 2024 when part of this work was performed. 
R.S. is grateful to the Erwin-
Schr\"odinger International Institute for Mathematics and Physics for partial support during the 
2026 Programme ``New Paradigms for Harnessing Quantum Field Theory at Colliders”, 
where this article has been finalized.
Figures were drawn with \texttt{Jaxodraw}~\cite{Binosi:2008ig} and FeynGame~\cite{Bundgen:2025utt}. 

\bibliography{NLP}

\end{document}